\documentclass[a4paper,11pt]{article}

\usepackage{jheppub,amsmath, amssymb,amsthm,dsfont,graphicx,mathtools,multirow,rotating,placeins,verbatim} 
\usepackage[normalem]{ulem}

\usepackage{hyperref}
\usepackage{comment}
\usepackage{color}

\usepackage{setspace}
\usepackage{cancel}
\usepackage{graphicx}
\usepackage{enumerate}

\def\be{\begin{equation}}
\def\ee{\end{equation}}
\def\ba{\begin{eqnarray}}
\def\ea{\end{eqnarray}}
\def\bi{\begin{itemize}}
\def\ei{\end{itemize}}

\def\zb{\bar{z}}
\def\wb{\bar{w}}
\def\w{\omega}

\def\I{\mathcal{I}}
\def\Lie{\mathcal{L}}
\def\diff{\text{Diff}}

\def\w{\omega}

\def\hard{\text{hard}}
\def\soft{\text{soft}}

\def\gbms{\text{GBMS}}
\def\Db{\mathbf{D}}
\def\Rb{\mathbf{R}}

\def\as{\text{AS}}

\def\O{\mathcal{O}}

\def\D{\mathcal{D}}
\def\kin{\text{kin}}
\def\phys{\text{phys}}
\def\A{\mathbf{A}}
\def\Ab{\bar{\mathbf{A}}}

\def\Bb{\mathbf{B}}
\def\tb{\mathbf{t}}
\def\extra{\text{extra}}

\def\Lt{\tilde{L}}

\def\bms{\text{BMS}}
\def\conf{\text{Conf}}
\def\Chat{\mathcal{C}}
\def\Ot{\tilde{\mathcal{O}}}
\def\Sb{\mathbf{S}}
\def\Sbt{\tilde{\mathbf{S}}}
\def\ckv{\text{CKV}}
\def\Gzero{\mathcal{G}}

\def\Gb{\mathbf{G}}
\def\Gbt{\tilde{\mathbf{G}}}
\def\Gone{\mathbf{G}}

\def\L{\mathbf{L}}
\def\K{\mathbf{K}}
\def\uinf{\Lambda}
\def\sign{\text{sign}}
\def\Ncal{\mathcal{N}}

\newcommand\None{\overset{\scriptscriptstyle 1}{N}\vphantom{N}}
\newcommand\Gpsi{\overset{\scriptscriptstyle \psi}{\Gamma}\vphantom{\Gamma}}
\newcommand\szero{\overset{\scriptscriptstyle 0}{S}\vphantom{S}}
\newcommand\sone{\overset{\scriptscriptstyle 1}{S}\vphantom{S}}

\newcommand\Ncalzero{\overset{\scriptscriptstyle 0}{\mathcal{N}}\vphantom{N}}
\newcommand\Ncalone{\overset{\scriptscriptstyle 1}{\mathcal{N}}\vphantom{N}}

\newcommand\delone{\overset{\scriptscriptstyle 1}{\delta}\vphantom{\delta}}

\title{Gravitational Poisson brackets  at null infinity compatible  with smooth superrotations}

\author[a]{Miguel Campiglia}
\author[b,c]{Adarsh Sudhakar}

\affiliation[a]{Facultad  de  Ciencias, Universidad  de  la  Rep\'ublica 
Ig\'ua  4225,  Montevideo,  Uruguay}

\affiliation[b]{Institute of Mathematical Sciences, IV Cross Road, CIT Campus, Taramani, Chennai, India 600 113}

\affiliation[c]{Homi Bhabha National Institute,
Training School Complex, Anushakti Nagar,
Mumbai, India 400 085}

\emailAdd{campi@fisica.edu.uy}
\emailAdd{adarshsu@imsc.res.in}

\abstract{

Superrotations are local extensions of the Lorentz group at null infinity that have been argued to be symmetries of  gravitational scattering. In their smooth version, they can be identified with the  group of diffeomorphisms on the celestial sphere. Their canonical realization requires  treating the celestial metric as a variable in the gravitational phase space, along with the news and shear tensors.  In this paper, we derive the resulting Poisson brackets (PB).  The standard PB algebra of the news and shear tensors is augmented by distributional terms at the boundaries of null infinity, including novel PB relations between the celestial metric and the radiative variables.
}

\begin{document}
\maketitle
\flushbottom


\section{Introduction}

Over the past decade, there has been a rich revision on the subject of asymptotically flat spacetimes, largely sparked by Strominger's infrared triangle that connects asymptotic symmetries, soft theorems and memory effects \cite{stromrev}.  An attractive feature of these developments is their wide scope of applicability, ranging from gravitational wave observables \cite{Boersma:2020gxx,Khera:2020mcz,nichols} to flat space holography \cite{raju,Raclariu:2021zjz,Donnay:2023mrd,Pasterski:2023ikd}. Indeed, it  came as a  surprise that new physics was  hiding in a seemingly settled area,   with foundational works dating back to the 60s  \cite{bondi,sachs,penrose1,weinberg}. 

In reality the subject was never closed, and many  landmark  results   slowly accumulated during the  intervening years. To give a few examples that are directly relevant for the  present paper, we can highlight Geroch's conformal description of gravitational radiation \cite{geroch}, Ashtekar and Streubel's (AS) gravitational phase space at null infinity \cite{AS}, and Barnich and Troessaert's extension of the Lorentz group to include holomorphic superrotations (SRs) \cite{BTprl}.

Within the  more recent developments, our starting point is Kapec, Lysov, Pasterski and Strominger's observation \cite{stromvirasoro} that the subleading soft graviton theorem \cite{CS} can be understood as a Ward identity of holomorphic superrotations.  It was  noticed afterwords that one may in fact drop the holomorphicity condition and   interpret the soft theorem in terms of \emph{smooth} superrotations \cite{gbms1}.  Whereas  this may appear to run against the idea of a dual Celestial Conformal Field Theory (CCFT) \cite{Raclariu:2021zjz,Donnay:2023mrd,Pasterski:2023ikd}, 
it has the advantage that  it allows to extend the notion of superrotations (and its relation to the subleading soft theorem) to dimensions grater than four \cite{Colferai:2020rte,Chowdhury:2022gib,CaponeMitra}. The cost to pay, however,  is that   the celestial metric has to be treated dynamically 
in the  gravitational phase space \cite{gbms2}. From a CCFT perspective, one is allowing the boundary theory to live on  arbitrary background metrics.\footnote{See \cite{kapec1,kapec2} for  discussions of background sources in CCFT and their relation to asymptotic symmetries.}

One of the main difficulties one faces in treating the  celestial  metric as a dynamical variable is the appearance of divergences in the gravitational symplectic structure at null infinity. As shown in \cite{compere,cfss,wbms,Freidel:2024tpl}, these can be regularized and renormalized by appropriate boundary counterterms.  These techniques, however, usually leave  undetermined ``corner'' contributions at  the boundaries of null infinity. In \cite{javier} these difficulties were bypassed by requiring consistency of the charge algebra generated by superrotations and supertranslations (ST), leading to a specific proposal for a symplectic structure at null infinity  in which the fundamental variables are the shear tensor (as in the AS case) and the celestial metric.  

 Once a symplectic structure is given, it is natural to ask what  the corresponding Poisson brackets (PB) are. For the case at hand, this turns out to be  quite a challenging problem and  progress was initially made in restricted settings:  In \cite{alok} PBs were obtained in the sector relevant for holomorphic SRs while \cite{adarsh} dealt with smooth SRs at linearized level. Based on the insights gained in those previous studies, in this work we undertake the task of evaluating PBs without  simplifying assumptions. 
 
 \subsection{Summary of results} \label{summarysec}
 Before getting started, let us describe the main result of the paper.  Consider the following gravitational variables at null infinity,
\begin{center}
\begin{tabular}{ l l }
 $\Ncal_{ab}(u,x)$ & news tensor,\\ 
 $C(x)$ &  supertranslation Goldstone mode, \\  
 $q_{ab}(x)$ &   celestial metric  ($\equiv$ smooth superrotation Goldstone mode),
\end{tabular}
\end{center}
where $u$ and $x$ are, respectively, time and angular coordinates at null infinity, while $a,b,\ldots$ are 2d indices on the celestial sphere. $C(x)$   is the ``ST Goldstone mode'' introduced in \cite{stromST}, which, together with the news tensor,  provide a parametrization of the AS phase space. The celestial metric appears as an additional variable,  playing the role of  Goldstone mode for smooth SRs \cite{gbms2}.

The evaluation of PBs requires the introduction of a cut-off $\Lambda$  such that
\be
-\uinf <u <\uinf \quad \text{with} \quad \uinf \to \infty. \nonumber
\ee
The final result of our analysis can be summarized in the following PB algebra:
\ba 
\{ \Ncal_{ab}(u,x),  \Ncal_{cd}(u',x')  \} &=&   \frac{1}{2} \delta_{ab,cd} \,  \delta^{(2)}(x,x')  \, \dot \delta(u-u') \nonumber \\
& &+  \, \delta_{ab,cd} \,  \delta^{(2)}(x,x') \,  \dot \delta_\Lambda(u)   \big(\Lambda \delta_\Lambda(u') - \tfrac{1}{2} \big)  - [u \leftrightarrow u' \nonumber ]\\ 
 & &+  \,  \K_{ab,cd}(x,x') \dot \delta_\Lambda(u) \dot \delta_\Lambda(u')  \label{newsnewsPBintro} \\
& &-   \, \L_{ab,cd}(x,x') {\delta_\Lambda(u) \dot \delta_\Lambda(u') }+ [ (u,x,ab) \leftrightarrow (u',x',cd) ] \nonumber \\
 & &-  \, \frac{1}{2}  q_{cd} \, \Ncal^{mn}(u',x') \Db_{\langle a} \Gb_{b \rangle mn}(x,x') \dot \delta_\Lambda(u) + [(u,x,ab) \leftrightarrow (u',x',cd)  ]\nonumber  \\
 & &  \nonumber\\
\{\Ncal_{ab}(u,x),C(x') \} &=&  - \Gzero_{ab}(x,x')  \big(  C(x)  \dot \delta_\Lambda(u)   + \delta_\Lambda(u)   \big)  \label{newsCPBintro} \\
 & &  \nonumber \\
\{\Ncal_{ab}(u,x) ,q_{cd}(x')  \} &=& - \Db_{\langle a} \Gb_{b \rangle cd}(x,x')  \dot \delta_\Lambda(u) \label{newsqPBintro} \\
 & &  \nonumber\\
 \{C(x),C(x') \} &=&  \{q_{ab}(x),C(x') \}=  \{q_{ab}(x),q_{cd}(x') \}= 0, \label{trivialPBintro}
\ea
where $\delta_{ab,cd} = \tfrac{1}{2}(q_{ac} q_{bd}+q_{ad} q_{bc}- q_{ab}q_{cd})$, $\delta_\Lambda(u)$ is a Dirac delta at $|u|=\Lambda$, and the dot represents a derivative with respect  to time. The  various  bilocal tensors on $x$ and $x'$ will be described in due course. The first line of \eqref{newsnewsPBintro} is the usual news PBs, while the second term in \eqref{newsCPBintro} reproduces the  bracket obtained in \cite{stromST}. All other terms (i.e. the last four lines of  \eqref{newsnewsPBintro}, the  first term in \eqref{newsCPBintro}, and \eqref{newsqPBintro}) are the  new contributions that arise when extending the AS space by $q_{ab}$.   Their role  will become clear during the analysis. 

 \subsection{Organization of the paper}

 The  article is organized as follows. Section \ref{preliminariessec} sets the stage for the paper: In \ref{conventions}, we introduce notation and explain our objective in more detail. In \ref{SRC}, we review the notion of SR covariance, which plays a central role throughout the article. In \ref{hardsoftsplit}, we present a parametrization of the gravitational variables at null infinity that simplifies the symplectic structure. This parametrization suggests a description of the extended phase space as a constrained system, presented in \ref{kinvsphyssec}. The constraints naturally group into two sets, and in \ref{diracsec} we present the strategy for their imposition. In \ref{holocoordsprel}, we comment on the subtleties associated to the use of holomorphic coordinates in a context where the 2d metric is not fixed.

The body of the paper is composed of five sections: In  \ref{kinPBsec} we present the kinematical PBs, while in  \ref{1stconstriantssec} and \ref{2ndconstraintssec} we impose the first and second  set of constraints respectively, resulting in the brackets \eqref{newsnewsPBintro} to \eqref{trivialPBintro}. As an illustration and consistency check, in section  \ref{gbmsactionsec} we use the brackets to evaluate the action of supertranslation and superrotation charges, recovering the expected transformation rules.   
A final discussion is given in section \ref{discussionsec}. 

The article is complemented with eight appendices: In \ref{appA} we review the notions of Weyl scaling at null infinity, (G)BMS group and charges, and establish various properties of the symplectic structure. In  \ref{diffopssec} we present   differential operators and Green's functions that are used  in the analysis. \ref{kinHVFsapp} is a complement of section \ref{kinPBsec} which discusses kinematical HVFs. In \ref{1stDiracapp} we present the evaluation of the first stage Dirac matrix.  \ref{HVF1sec} is a complement of section \ref{2ndconstraintssec} and discusses HVFs on the intermediate kinematical space. In \ref{2nddiracapp} we present details of the evaluation of the second stage Dirac matrix.   \ref{HVF2sec} contains a discussion of HVFs on the physical space and  \ref{nullappendix} discusses null directions of the three symplectic structures appearing in the paper (kinematical, intermediate, and physical).

\section{Preliminaries} \label{preliminariessec}

\subsection{Conventions and statement of the problem} \label{conventions}

We  follow a coordinate-based approach, as in the original BMS treatment \cite{bondi,sachs}, although we shall use  various geometric notions that arise in the coordinate-free 
conformal setting \cite{penrose1,geroch,AS}.

Asymptotically flat space-times in Bondi coordinates  are characterized by the leading $r \to \infty$   angular components of the spacetime metric (see \cite{BT} for a detailed review)
\be 
g_{ab} \stackrel{r\to \infty}{=}  r^2 q_{ab}(x) + r g^{(1)}_{ab}(u,x)+ \cdots \label{gab} ,
\ee
where $a,b,c,\ldots$ are two-dimensional indices on the celestial sphere.  Future (past) null infinity $\I$ is parametrized by sphere coordinates $x^a$ and retarded (advanced) time $u$. 
$q_{ab}$ defines  the 2d metric at null infinity, which is used to raise and lower 2d indices. 

Among Bondi gauge conditions, the one that  plays a key role in our discussion is\footnote{See \cite{wbms,geiller} for  more general  boundary conditions.}
\be \label{detgcond}
 \det g_{ab} =  r^4 \det \mathring{q}_{ab},
\ee
where $\det \mathring{q}_{ab}$ is the square of an area element that is being fixed. 
In terms of the  expansion coefficients in \eqref{gab}, the condition reads
\begin{align}
\det q_{ab} &= \det \mathring{q}_{ab} , \label{detqfixed}  \\
q^{ab}g^{(1)}_{ab} & =0, \label{goneTF}
\end{align}
and can be interpreted as follows. Eq.  \eqref{detqfixed}  freezes  Weyl diffeomorphisms at null infinity (see appendix  \ref{weylapp}) while Eq. \eqref{goneTF} brings down to two the number of independent components of  $g^{(1)}_{ab}$, eventually allowing their identification with the two  polarizations of gravitational waves.  

It has been known since the early works on null infinity asymptotics \cite{geroch} that  gravitational radiation is encoded in a ``renormalized $g^{(1)}_{ab}$'' \cite{compere},
\be \label{defghat}
\Chat_{ab}:= g^{(1)}_{ab} - u T_{ab},
\ee
where $T_{ab}$ is a  symmetric, trace-free (STF)  tensor that is constructed from $q_{ab}$ (see Eq. \eqref{gerochcond} below).  In particular, the energy-flux of gravitational waves is proportional to the square of the time derivative of $\Chat_{ab}$,  
\be
\Ncal_{ab} := \partial_u \Chat_{ab}.
\ee
We refer to $\Chat_{ab}$, $T_{ab}$ and $\Ncal_{ab}$  as the shear,  Geroch, and news tensor respectively. 

Requiring finiteness of  the total energy carried by gravitational waves  implies
\be \label{fallug1ren}
\lim_{|u| \to \infty}\Ncal_{ab} =0. 
\ee
Typical spacetimes of interest satisfy a stronger version of \eqref{fallug1ren}, which is that the spacetime metric becomes flat at early and late times \cite{stromgravscatt}. The condition can be written as 
\be \label{electricbdycond}
\lim_{u \to \pm \infty}  \Chat_{ab}  = (-2 D_{\langle a} D_{b \rangle}  +  T_{ab})\Chat^\pm,
\ee
for certain functions on the celestial sphere $\Chat^\pm$. Here $D_a$ is the covariant derivative of $q_{ab}$ and the angle brackets denote symmetrization and  extraction of the trace.\footnote{Explicitly, given a 2d tensor  $A_{ab}$, we define $A_{\langle ab \rangle} := \frac{1}{2}(A_{ab}+A_{ba}) - \frac{1}{2} q_{ab} q^{cd}A_{cd}$.}

The standard gravitational phase space  at null infinity is defined such that only $\Chat_{ab}$  is allowed to vary, leading to the symplectic structure  introduced by Ashtekar and Streubel  \cite{AS}
\be \label{OmegaAS}
\Omega_{\as} = \int_{\I} \delta  \Ncal^{ab}  \wedge \delta \Chat_{ab}.
\ee
Among other things, this structure allows for a canonical realization of the  asymptotic isometries of the spacetime metric, described by the so-called BMS group. On the other hand, as indicated in the introduction, the subleading soft graviton theorem suggests one should allow for   diffeomorphisms that are not necessarily asympotic isometries. For  spacetime metrics in Bondi gauge, these are generated by smooth superrotations and supertranslations, and span a generalized version of the BMS group.\footnote{From now on, we will omit the adjective ``smooth'' when referring to smooth superrotations. The BMS group and its generalized version are reviewed in appendix \ref{gbmsapp}.}

In general,  SRs change the 2d metric and so their description requires  one to work in 
 the extended space \cite{gbms2}\footnote{The space may appear to depend on the choice of ${\det \mathring{q}_{ab}}$.  Different choices can however be identified thanks to the Weyl invariance of the symplectic form, see appendix \ref{weylapp}.}
\be
\Gamma = \bigcup_{\substack{q_{ab}: \\ \det q_{ab} = \det \mathring{q}_{ab}}} \Gamma^{q_{ab}}_{\as}, \label{defGamma}
\ee
where $\Gamma^{q_{ab}}_{\as}$ is the AS phase space associated to a given 2d metric $q_{ab}$. This space admits a natural symplectic structure that is compatible with both SRs and STs and which takes the form \cite{javier}
\be \label{Omegaintro}
\Omega = \Omega_{\as} + \int_{S^2}\left( \delta p^{ab}_1 \wedge \delta q_{ab} + \delta \Pi^{ab}_1 \wedge \delta T_{ab}\right),
\ee
where $p^{ab}_1$ and $\Pi^{ab}_1$ are  STF tensors that are constructed out of  $\Chat_{ab}$, $\Ncal_{ab}$, $q_{ab}$ and $T_{ab}$ (see subsection \ref{hardsoftsplit} and appendix \ref{simplifiedOmegaapp}).

The aim of this paper is to invert \eqref{Omegaintro}, i.e. to derive the  PBs among elementary variables. The starting point  will be a rewriting of \eqref{Omegaintro}  that  brings $\Omega$ into a simpler form compared to the  original expression of \cite{javier}. A recurring technical tool in our analysis will be the use of  a SR covariant derivative,  which we describe next.

\subsection{SR covariant derivative} \label{SRC}

A $u$-independent 2d tensor $A$ at null infinity is said to be SR covariant if it enjoys the SR transformation rule\footnote{Borrowing CFT terminology, tensors satisfying \eqref{delVA} are also referred to as primary fields \cite{fprsubsub}.}
\be \label{delVA}
\delta_V A = \Lie_V A + \frac{k}{2} D_c V^c A,
\ee
where $\Lie_V$ is the Lie derivative along the 2d vector field $V^a$ generating the SR and $k$ is the Weyl weight of the tensor, see appendix \ref{weylapp}.

 In general, derivatives of SR covariant quantities are not  SR covariant. For instance, the term $-2 D_{\langle a} D_{b\rangle }\Chat^\pm$ in \eqref{electricbdycond} is not SR covariant. In this case, the non-covariance is  ``corrected'' by the addition of the tensor $T_{ab}$ introduced earlier and defined by  the condition \cite{geroch,compere},
\be \label{gerochcond}
D^{b} T_{b a} =- \frac{1}{2} \partial_a R,
\ee
where $R$ is the scalar curvature of $q_{ab}$. More generally, one can always ``SR covariantize'' differential expressions by means of  a SR covariant derivative defined as follows \cite{javier}.\footnote{In the present smooth context, the notion of SR covariant derivative is equivalent \cite{adarsh} to that of  Weyl covariant derivative  (see e.g. \cite{CiambelliLeighWeyl,BarnichRuzziconiWeyl}).  In the holomorphic setting, however, the two notions are independent.}

Let $\psi$ be a potential for $T_{ab}$ such that \cite{compere}\footnote{  $\psi$ can equivalently be defined by the condition  that  $e^{-2 \psi} q_{ab}$ is diffeomorphic to the round sphere metric \cite{compere,javier}. From either definition one finds that $\psi$ is defined modulo an additive term associated to conformal isometries. The tensor $T_{ab}$ is however independent of such ambiguity. In \cite{compere} $\psi$ is interpreted as a Liouville field and  \eqref{defpsi} as the trace-free part of the corresponding stress tensor.}
\be \label{defpsi}
T_{ab}= 2 (D_{\langle a} \psi D_{b \rangle} \psi + D_{\langle a} D_{b \rangle} \psi).
\ee
The SR covariant derivative $\Db_a$ is then  defined as
\be \label{defDb}
\Db_a A := D_a A +  k D_a \psi  A + \Gpsi A,
\ee
where $k$ is the weight of the tensor $A$, as given in \eqref{delVA}, and $\Gpsi A$ represents indices contractions with the Christoffel-like symbols,
\be
\Gpsi_{a b}^{c}= -2 D_{\langle a}\psi \delta_{b\rangle}^c.
\ee
One can then show that the SR covariant derivative of a SR covariant tensor is SR covariant, i.e.
\be
\delta_V A =( \Lie_V + \frac{k}{2} D_c V^c) A \implies \delta_V \Db_a A = (\Lie_V + \frac{k}{2} D_c V^c ) \Db_a A.
\ee
In particular, $\Db_a \Db_b \Chat^\pm$ is SR covariant, and in fact one has the identity
\be
-2 \Db_{\langle a} \Db_{b \rangle} \Chat^\pm = (-2 D_{\langle a} D_{b\rangle }+ T_{ab}) \Chat^\pm .
\ee

Other useful identities to keep in mind are
\ba 
\Db_a q_{bc} &=&0 \\
\Db_a \Rb& =&0,\label{DbRbzero}
\ea
where $\Rb$ is the SR covariant scalar curvature, defined by $[\Db_a, \Db_b] \w_c = \Rb_{abc}^{\phantom{abc}d} \w_d$ with  $\Rb_{abcd} = \Rb q_{a[c} q_{d] b}$. It is related to the ordinary scalar curvature by  $\Rb= R + 2 D_c D^c \psi$.

\subsection{Simplifying the symplectic form through a hard--soft split} \label{hardsoftsplit}
Following \cite{stromST,alok}, it is convenient 
to isolate a $u$-independent part of $\Chat_{ab}$ by writing it as
\be \label{gonehatsigmaC}
\Chat_{ab}(u,x) = \sigma_{ab}(u,x) + C_{ab}(x)
\ee
with $\sigma_{ab}$ satisfying
\be \label{bdycondsigma}
\lim_{u \to \infty} (\sigma_{ab}(-u,x) +  \sigma_{ab}(u,x) ) =0.
\ee
In a slight abuse of language, we shall often  refer to $\sigma_{ab}$ as the shear.  

The splitting \eqref{gonehatsigmaC} translates into a splitting of the AS symplectic structure \eqref{OmegaAS},\footnote{If we had chosen $\Omega_{\as}$ to be $\int_{\I} \delta \Ncal_{ab}  \wedge \delta \Chat^{ab}$ (i.e. with covariant/contravariant indices swapped from \eqref{OmegaAS}), we would still end up with \eqref{OmegaASsplit}. This is because its difference from \eqref{OmegaAS} is proportional to $\delta \sqrt{q}$, as can be seen by writing the difference in holomorphic coordinates.}
\be \label{OmegaASsplit}
\Omega_\as =  \int_{\I} \delta \dot \sigma^{ab} \wedge \delta \sigma_{ab}  + \int_{S^2} \delta \Ncalzero^{ab} \wedge \delta C_{ab},
\ee
where 
\be
\dot \sigma^{ab} = \partial_u \sigma^{ab} \equiv \Ncal^{ab}
\ee
 and
\be \label{defNcalzero}
 \Ncalzero_{ab} := \int^{\infty}_\infty du  \, \dot \sigma_{ab} 
\ee
is the leading soft news.
The first term in \eqref{OmegaASsplit} will be referred to as the hard part of the symplectic structure,
\be \label{defOmegahard}
\Omega^\hard := \int_{\I} \delta \dot \sigma^{ab} \wedge \delta \sigma_{ab}  .
\ee

 It turns out that the second term in \eqref{OmegaASsplit} combines nicely with the second term in \eqref{Omegaintro}, resulting in what we will refer to as the soft part of the symplectic structure (see appendix for \ref{simplifiedOmegaapp} for details)
\be \label{defOmegasoft}
\Omega^\soft  :=  \int_{S^2} \left( \delta (\D N) \wedge \delta C + \delta \Pi^{ab} \wedge \delta T_{ab} + \delta p^{ab} \wedge \delta q_{ab} \right),
\ee
where $C$ and $N$ are such that\footnote{In terms of $\Chat^\pm$ introduced in \eqref{electricbdycond}:  $C=(\Chat^++\Chat^-)/2$ and $N=\Chat^+-\Chat^-$. All these scalars  are to be understood as defined modulo elements in the kernel of $\Db_{\langle a} \Db_{b \rangle} $. These indeterminacies are associated to null directions in the the symplectic structure, see appendix \ref{nullappendix} for further details.} 
\be
\begin{aligned} \label{defCNscalars}
C_{ab} & = -2  \Db_{\langle a} \Db_{b \rangle} C, \\
\Ncalzero_{ab} &  = -2 \Db_{\langle a} \Db_{b \rangle} N,
\end{aligned}
\ee
and $\D$ is the 4th order differential operator defined by
\be \label{defDcalN}
\D N  :=   -2 \Db^{\langle a} \Db^{b \rangle} \Ncalzero_{ab}=  4 \Db^{\langle a} \Db^{b \rangle}  \Db_{\langle a} \Db_{b \rangle}  N.
\ee
The conjugates to $T_{ab}$ and $q_{ab}$ appearing in \eqref{defOmegasoft} are
\ba
\Pi^{ab} & = &   2(\Ncalone^{ab} + C\Ncalzero^{ab}) ,\label{defPiab} \\
p^{ab} & = & \frac{1}{2} \left( D^{\langle a} D_c \Pi^{b \rangle c} - \frac{R}{2} \Pi^{ab} \right) =: \frac{1}{2} \O^{ab}_{\phantom{ab} cd} \Pi^{cd} , \label{defpab}
\ea
where 
\be \label{defNcalone}
\Ncalone_{ab} := \int^\infty_{-\infty} du \, u \dot \sigma_{ab}
\ee
is the subleading soft news.

To summarize, the extended  symplectic structure \eqref{Omegaintro} can be written as
\be
\begin{aligned} 
\Omega & = \Omega^\hard + \Omega^\soft, \\
& = \int_{\I} \delta \dot \sigma^{ab} \wedge \delta \sigma_{ab}   + \int_{S^2} \left( \delta (\D N) \wedge \delta C + \delta \Pi^{ab} \wedge \delta T_{ab} + \delta p^{ab} \wedge \delta q_{ab} \right). \label{Omegahardplussoft}
\end{aligned}
\ee

A few comments are in order:

\begin{enumerate}
\item The above hard--soft splitting does not mean the total phase space  is a product of  hard and soft phase spaces. In particular, $\Omega^\hard$ depends on the  ``soft'' variable $q_{ab}$ through the condition $q^{ab}\sigma_{ab}=0$ while $\Omega^\soft$ depends on the ``hard'' variable $\sigma_{ab}$ through Eqs. \eqref{defDcalN}, \eqref{defPiab}, \eqref{defpab}.   

\item There is certain freedom into how one chooses to write $\Omega$. For instance, we could express the first piece of $\Omega^\soft$ as $\int_{S^2} \delta N \wedge \delta (\D C)$ at the cost of introducing extra terms in the definitions of $\Pi^{ab}$ and $p^{ab}$ \cite{alok,adarsh}. Additionally, one could consider a different prescription in the splitting \eqref{gonehatsigmaC} such that $\alpha \sigma^-_{ab} + (1-\alpha) \sigma^+_{ab} =0$ with $\alpha \neq 1/2$.   The choices we made  simplify the expressions for  $\Pi^{ab}$ and $p^{ab}$ and allow for a relation  between them \eqref{defpab} that does not involve  $C$ nor $N$.  See  appendix \ref{simplifiedOmegaapp} for further details.

\item   In order for \eqref{defNcalone} to be well-defined, we need  $\dot \sigma_{ab}$ to decay faster than $u^{-2}$. We will thus assume
\be \label{phyisicalfallu}
\dot \sigma_{ab} \stackrel{|u| \to \infty}{=} O(1/|u|^{2+\epsilon}) \quad \text{(``physical'' boundary conditions)}
\ee
for some $\epsilon>0$. The qualifier  ``physical'' here is just to distinguish it from the  different  boundary conditions  that are needed at the kinematical level (see Eq. \eqref{kinematicalfallu} below).  It is however well-known that the fall-offs \eqref{phyisicalfallu}  are generally too restrictive, and that one should, in fact,  set $\epsilon=0$ \cite{damour,laddhasen}. See \cite{Agrawal:2023zea,marcalok,chipum,chipum2} for recent progress on how to deal with such situation at the phase space level.
\end{enumerate}

\subsection{Physical phase space as a constrained system} \label{kinvsphyssec}
Following \cite{alok}, we will regard the phase space introduced before as a constrained system  inside an auxiliary kinematical phase space. The latter is defined by regarding all quantities  appearing in \eqref{Omegahardplussoft}  as independent (except for the trace-free conditions on the tensors and the fixed determinant  on the metric):
\be \label{defGammakin0}
\Gamma_{\kin, 0} = \{\sigma_{ab}, C, N, T_{ab},  \Pi^{ab}, q_{ab}, p^{ab} \} .
\ee
In addition, we relax the $u \to \pm \infty$ boundary conditions \eqref{phyisicalfallu} on $\dot \sigma_{ab}$ to allow for an $O(1)$ term:
\be \label{kinematicalfallu}
\dot \sigma_{ab} \stackrel{|u| \to \infty}{=} O(1) + O(1/|u|^{2+\epsilon}) \quad \text{(``kinematical'' boundary conditions)}.
\ee
The vanishing of this $O(1)$ term will  then be treated as part of the constraints.  The reason to include it  is to have a well-defined kinematical action of the subleading news \eqref{defNcalone}, see  section \ref{utoinfsec} for further details. Note that we are assuming the asymptotic values of the kinematical news at $u\to \pm \infty$  coincide.

 Conditions \eqref{fallug1ren}, \eqref{electricbdycond}, \eqref{gerochcond}, \eqref{bdycondsigma}, \eqref{defPiab} and \eqref{defpab} will then be regarded as constraints on  $\Gamma_{\kin, 0}$ that determine the physical space $\Gamma_\phys$. Specifically, we define the  constraint functions:
\be \label{defF1toF6}
\begin{aligned}
F^{ab}_1 & :=  p^{ab} - \frac{1}{2} \O^{ab}_{\phantom{ab} cd} \Pi^{cd} , \\
F_{2 a} & :=  D^{b} T_{b a} + \frac{1}{2} \partial_a R ,\\
F_3^{ab} &:= \Ncalone^{ab} + C N^{ab}  -\frac{1}{2}\Pi^{ab},   \\
F_4^{ab} & := 2\dot \sigma^{+ab} ,\\
F_5^{ab} & := \Ncalzero^{ab} - N^{ab}, \\
F_6^{ab} & := \frac{1}{2}(\sigma^{+ab} + \sigma^{-ab}), 
\end{aligned}
\ee
where the $\pm$ superscripts indicate  $u \to \pm \infty$ limits, $N^{ab} := -2 \Db^{\langle a } \Db^{b\rangle} N$, and $\O^{ab}_{\phantom{ab} cd}$ is the differential operator defined in Eq. \eqref{defpab}.

The origin of $F_2$ and the pair $F_5$, $F_6$ are evident from the definitions of the quantities involved, as given in sections \ref{SRC} and \ref{hardsoftsplit} respectively. The constraints $F_1$ and $F_3$ arose from demanding that the canonical charges associated to the generalised BMS symmetries faithfully represent the GBMS algebra \cite{javier}. Finally, the constraint  $F_4$ demands that the $O(1)$ mode \eqref{kinematicalfallu}  of the news vanishes.\footnote{Unlike the situations studied previously \cite{alok,adarsh},  the explicit inclusion of this constraint  is essential in order to obtain an invertible Dirac matrix.}  For concreteness we impose this condition at $u \to +\infty$, but any condition of the type  $\alpha \dot \sigma^+ + \beta \dot \sigma^- =0$ with $\alpha + \beta \neq 0$  works equally well.

\subsection{Strategy for the imposition of constraints} \label{diracsec}

The constraints \eqref{defF1toF6} naturally split into two sets,
\be
(F_1,F_2) \quad \text{and} \quad  (F_3,F_4,F_5,F_6).
\ee
The first set only involve variables from the ``soft sector''  and can be consistently imposed on its own. This leads to an intermediate kinematical space that we denote by $\Gamma_{\kin, 1}$, 
\be \label{defGammakin1}
\Gamma_{\kin, 1}: = \{F_1=F_2=0\}  \subset  \Gamma_{\kin, 0}.
\ee
As we shall see, the role of these constraints is to make explicit the SR covariance of the PBs. 

The remaining constraints, involve the field $\sigma_{ab}$ and its time derivative, and require dealing with  distributional contributions at $|u| \to \infty$. Their imposition may be regarded as a generalization of the analysis made in \cite{stromST} (which, in our notation, only treated $F_5$ and $F_6$ as there were no SRs), and leads to the  physical phase space  
\be \label{GammaphysfromGammakin1}
\Gamma_{\phys} \equiv \Gamma_{\kin, 2} := \{F_3=F_4 = F_5= F_6=0\}  \subset  \Gamma_{\kin, 1}.
\ee

It  turns out that, in both cases,  the constraints form a second class system. PBs can then be obtained by iterating  Dirac's procedure as follows.    Let 
\be \label{FalphaFalphap}
\{F_\alpha, F_{\alpha'} \}_i  
\ee
be the matrix of constraints'  PBs  at the $i$-th level $(i=0,1)$, where  $\{\cdot , \cdot \}_i$ are the PBs on $\Gamma_{\kin, i}$ and   the index $\alpha$ represents the   constraint labels (possibly including tensor components and points on the celestial sphere). Dirac's procedure requires the  inverse matrix of \eqref{FalphaFalphap} which we denote by $W^{\alpha \beta}_i$,
\be
\{F_\alpha, F_{\alpha'} \}_i  W^{\alpha' \beta}_i = \delta^\alpha_\beta.
\ee
The corrected PBs between two quantities  $\psi$ and $\varphi$ are then given by,
\be \label{abstractcorrectedPB}
\{ \psi, \varphi \}_{i+1} =  \{ \psi, \varphi \}_{i} + \{ \psi , F_\alpha \}_i W^{\alpha \alpha'}_i \{\varphi , F_{\alpha'}\}_i.
\ee

Our  starting point are the  ``zeroth'' kinematical PBs, discussed in section \ref{kinPBsec}. Applying \eqref{abstractcorrectedPB} for $i=0$ yields the ``intermediate'' kinematical PBs (section \ref{1stconstriantssec}) and  applying it again for $i=1$ yields the physical PBs (section \ref{2ndconstraintssec}).

Throughout the process, it will be important to keep in mind the  definition of PBs in terms of Hamiltonian vector fields (HVFs). In a slight abuse of notation, we  denote by $\{ \cdot, \varphi \}$ the HVF of $\varphi$. It is defined as the solution to the equation
\be
\Omega(\delta, \{ \cdot, \varphi \}) = \delta \varphi. \label{defHVF}
\ee
The Poisson brackets between two such functionals is then given by
\be \label{defPB}
\{ \varphi, \psi \} = \Omega(\{ \cdot, \psi \},\{ \cdot, \varphi \}) .
\ee
As implied by the notation, one can  interpret the LHS of  \eqref{defPB}  either as the HVF of $\psi$ acting on $\varphi$ or as (minus) the HVF of $\varphi$ acting on $\psi$. 

Equations \eqref{defHVF} and \eqref{defPB}  hold on each space $\Gamma_{\kin, i}$ separately.
The constraint analysis can in fact be understood as a recipe  to solve  \eqref{defHVF}  on the physical space ($i=2$).  When applied to the kinematical space ($i=0$), Eqs. \eqref{defHVF} and \eqref{defPB} provide the definition of the  ``zeroth'' PBs. This abstract perspective will  help us  deal with the  subtleties associated to the non-compact nature of the time variable $u$, discussed in  subsection \ref{utoinfsec}.

We conclude by introducing  additional notation. In actual computations it will be useful to consider smeared constraints. In the  notation of \eqref{FalphaFalphap} we write them as
\be
F[X]:= F_{\alpha}  X^{\alpha},
\ee
where  $X^{\alpha}$ plays the role of smearing parameter for the constraints. The Dirac matrix can then be thought of as the linear map 
\be \label{Diracmap}
X^{\alpha} \mapsto Y_\alpha  = \{F_\alpha, F_{\alpha'} \}_i X^{\alpha'} \equiv  \{F_\alpha, F[X] \}_i,
\ee
with inverse
\be \label{inverseDiracmap}
X^\alpha = W^{\alpha \alpha'}_i  Y_{\alpha'}.
\ee
Given a function $\varphi$, we can associate a smearing parameter $X_\varphi$ by
\be \label{defXphi}
 X_\varphi^\alpha:=W^{\alpha \beta}_i \{ \varphi, F_\beta\}_i , 
 \ee
in terms of which Eq. \eqref{abstractcorrectedPB} takes the compact form
\be \label{correctedHVFabstract}
\{ \cdot, \varphi \}_{i+1} = \{ \cdot, \varphi\}_i + \{ \cdot, F[X]\}_i\vert_{X=X_\varphi} .
\ee
This version of Dirac's formula will be used in appendices \ref{HVF1sec} and \ref{HVF2sec} to evaluate   HVFs on $\Gamma_{\kin, 1}$ and $\Gamma_{\kin, 2}$ respectively.

\subsection{Comments on the use of holomorphic coordinates}  \label{holocoordsprel} 

In order to simplify computations and expressions, it is  useful to work in holomorphic coordinates $(z,\zb)$ satisfying
\be \label{defholo}
q_{z \zb} > 0 , \quad q_{zz}=q_{\zb \zb}=0.
\ee
In particular, the independent variations of the 2d metric (which are trace-free, since $\det q_{ab}$ is fixed) are simply parametrized by $\delta q_{zz}$ and $\delta q_{\zb \zb}$. That is, we have
\be \label{delqzznotzero}
\delta q_{z \zb} =0 , \quad \text{and} \quad \delta q_{zz},  \delta q_{\zb \zb} \quad \text{unconstrained}.
\ee
As it is evident from  \eqref{delqzznotzero}, when dealing with tensor variations it is important that the evaluation on holomorphic coordinates is done \emph{after} the variation is performed. This in turn requires knowing the expression of the tensor in general coordinates. It will thus be important to be able to go back and forth between expressions in holomorphic  and  in  general coordinates.

STF tensors are naturally  parametrized by their holomorphic/antiholomorphic components. Their variations, however, can include mixed terms. Specifically, let $A^{ab}$ be a  STF tensor and consider  a general variation of its trace:
\be \label{deltraceA}
0= \delta( A^{ab} q_{ab}) = \delta A^{ab} q_{ab} + A^{ab} \delta q_{ab}. 
\ee
Writing \eqref{deltraceA} in holomorphic coordinates and solving for the pure-trace part, one has
\be \label{delAzzb}
\delta A^{z \zb} = - \frac{1}{2}q^{z \zb}(A^{zz} \delta q_{z z}+A^{\zb\zb} \delta q_{\zb \zb}),
\ee
where we emphasize again that the expression is  written in holomorphic coordinates after the variation is performed. The above equation makes it clear that such pure-trace variations are entirely captured by variations of the 2d metric. The truly independent variations of  STF tensors are captured by $\delta A^{zz}$ and $\delta A^{\zb\zb}$.

Similar expressions hold for STF tensors with lower indices. The relation between variations of lower and upper components are
\be \label{delAzzbdown}
\delta A_{z \zb} = -q_{z\zb} q_{z\zb} \delta A^{z \zb}
\ee 
and 
\be
\delta A_{z z}=  q_{z\zb} q_{z\zb}  \delta A^{\zb\zb} , \quad  \quad  \delta A_{\zb \zb}=  q_{z\zb} q_{z\zb}  \delta A^{zz} ,
\ee
where the negative sign in \eqref{delAzzbdown} arises from $\delta q^{zz}= - q^{z \zb} q^{z \zb} \delta q_{\zb \zb}$.

\section{Kinematical brackets} \label{kinPBsec}

As introduced in section \ref{kinvsphyssec}, the  kinematical phase space treats the fields appearing in the symplectic form \eqref{Omegahardplussoft} as independent variables.  In this section we discuss the resulting kinematical PBs. 
We  start in \ref{kinPBsubsec} by presenting PBs in holomorphic coordinates  and restricting to  finite-$u$ values in the shear.   In \ref{utoinfsec} we extend the analysis to include  quantities that are sensitive to  $|u| \to \infty$ asymptotic  values of the shear.  This section is complemented by a  treatment   based on HVFs in general 2d coordinates given in appendix \ref{kinHVFsapp}.

\subsection{Soft variables and finite--$u$ shear} \label{kinPBsubsec}

Let us start with the  PBs associated to what may be referred  as the canonical pairs,\footnote{ We denote points on the celestial sphere by  $z,w,\cdots$, without implying any holomorphic dependence.  Sphere points on   tensor arguments are omitted: 
$\sigma_{zz}(u)\equiv \sigma_{zz}(u,z)$, $\sigma_{\zb\zb}(u)\equiv \sigma_{\zb\zb}(u,z)$, etc.}
\ba \label{sigmadotsigmakinPB}
\{  \sigma_{zz}(u), \dot \sigma^{ww}(u') \}_0 &=&  {\tfrac{1}{2} \delta(u-u')} \delta^{(2)}(z,w) ,\\
\{  C(z),N(w)\}_0 &=& \Gzero_z(w), \label{CNPB0}\\
\{q_{zz}, p^{ww}\}_0 &=& \delta^{(2)}(z,w), \label{qpPB0} \\
 \{T_{zz}, \Pi^{ww}\}_0  &=& \delta^{(2)}(z,w), \label{TPiPB0}  
\ea
where $\Gzero_z(w)$ is the Green's function for the $\D$ operator, see appendix \ref{diffopssec} for details.  Eq. \eqref{sigmadotsigmakinPB} is the well-known news-shear PB, first postulated by Sachs \cite{sachs2}, while the PB relation  \eqref{CNPB0} was  introduced in \cite{stromST}. Eqs. \eqref{qpPB0} and \eqref{TPiPB0} are standard  PB relations for canonical pairs.

The above relations do not yet  exhaust all  PBs, but can be used to generate the remaining ones as follows.

Consider first   PB relations involving  $p^{ab}$. Since $p^{ab}$ has a non-trivial bracket with the 2d metric, it must have non-trivial brackets with all 2d STF tensors. Applying  Eqs. \eqref{delAzzb} and \eqref{delAzzbdown} to a variation generated by $\{\cdot,p^{ww}\}_0$  leads to
\be \label{pwwkinPB}
\begin{aligned}
\{  \sigma_{z\zb}(u),  p^{ww}\}_0 &=  \tfrac{1}{2}q_{z \zb} \sigma^{zz}(u)  \delta^{(2)}(z,w),\\
\{  T_{z\zb},  p^{ww}\}_0 &=  \tfrac{1}{2}q_{z \zb} T^{zz}  \delta^{(2)}(z,w) ,\\
\{  \Pi^{z\zb},  p^{ww}\}_0 &=  - \tfrac{1}{2}q_{z \zb} \Pi^{zz}  \delta^{(2)}(z,w) ,\\
\{  p^{z\zb},  p^{ww}\}_0 &=  - \tfrac{1}{2}q_{z \zb} p^{zz}  \delta^{(2)}(z,w) .\\
\end{aligned}
\ee
These relations in turn imply, upon using  Jacobi identities involving two $p^{ab}$'s, 
\be \label{pzzpwwkinPB}
\{  p_{zz},  p^{ww}\}_0 = \frac{1}{2}\left(  \Pi^{zz} T_{zz} - \Pi^{\zb\zb} T_{\zb\zb} + \int du(\dot \sigma^{zz} \sigma_{zz} - \dot \sigma^{\zb\zb} \sigma_{\zb \zb} ) \right) \delta^{(2)}(z,w) .
\ee

The remaining  non-trivial PBs are those involving  $N$ with $p^{ab}$ and $\Pi^{ab}$. These can be understood from the observation that  $\D N$ Poisson commutes with all variables except $C$, while $\D$ has non-trivial PBs with $p^{ab}$ and $\Pi^{ab}$ (since it depends on $q_{ab}$ and $T_{ab}$ respectively). For $p^{ww}$ this leads to
\ba
\{N(z), p^{ww}\}_0 &= & -\Gzero_z \{\D_z,p^{ww}\}_0 N(z) \label{Np0} \nonumber   \\
&= &   2 \Delta(N(w))\Gzero_{z}^{ww} +2 \Delta(\Gzero_z(w))N^{ww}, \label{NpwwkinPB}
\ea
where in the first line we used an operatorial notation for the Green's function (which leaves implicit the integration, see appendix \ref{greenfnapp}) and the subscript in $\D$ indicates the variable on which the operator acts. In the second line  $\Gzero_{z}^{ww} \equiv -2 \Db^w\Db^w \Gzero_z(w)$ while $\Delta$ is a differential operator that captures the variation of $D_{\langle a} D_{b \rangle}$, see appendices \ref{variationsDDapp} and \ref{holodiffopsapp}. Similarly for $\Pi^{ww}$ one has
\ba
\{N(z), \Pi^{ww}\}_0  &= &  -\Gzero_z \{\D_z,\Pi^{ww}\}_0 N(z) \label{NPi0}\nonumber   \\
&= &   -  (N(w) \Gzero_{z}^{ww}+ \Gzero_{z}(w) N^{ww} ) . \label{NPiwwkinPB}
\ea

One can finally show that Eqs. \eqref{sigmadotsigmakinPB} to \eqref{NPiwwkinPB}, together with their complex conjugates, complete all non-trivial  PBs among the elementary kinematical  variables \eqref{defGammakin0}.  We refer to appendix \ref{kinHVFsapp} for additional details.

\subsection{Prescriptions for brackets when $|u| \to  \infty$}\label{utoinfsec}
In the PB expressions discussed above, the variable $u$ in $\sigma_{ab}(u)$ was  assumed to be bounded. We will however need  PBs involving $|u| \to \infty$ limiting values of the shear, in particular when imposing the constraints $F_3$ to $F_6$ in section \ref{2ndconstraintssec}.  This requires certain prescriptions that we discuss below. From an abstract phase space perspective, the subtleties arise because, in the presence of boundaries, HVFs are not always guaranteed to exist \cite{RT}.

Asymptotic values of the shear  and news tensor can appear either explicitly (as in the constraints $F_4$ and $F_6$), or  through the leading and subleading soft news (as in the constraints $F_3$ and $F_5$).  Let us start   the discussion with the leading soft news \eqref{defNcalzero}. 

One can show this quantity has a perfectly well-defined HVF, with action on the shear given by\footnote{In this subsection we revert to general 2d coordinates an keep implicit the dependance on sphere points.}
\be \label{shearsoftnewsPB}
 \{ \sigma_{ab }(u),   \Ncalzero^{cd} \}_0 = \delta_{\langle ab \rangle}^{\langle c d \rangle} ,
\ee
where $ \delta_{\langle ab \rangle}^{\langle c d \rangle}$ represents the identity kernel on STF tensors. 
As emphasized in  \cite{stromST}, Eq. \eqref{shearsoftnewsPB}  is in conflict with the bracket between the shear and the news, which in the above notation reads,
\be \label{shearnewsPB}
 \{ \sigma_{ab }(u),   \dot \sigma^{ cd }(u')  \}_0 = \tfrac{1}{2}\delta_{\langle ab \rangle}^{\langle c d \rangle} \delta(u-u').
\ee
Indeed, integrating \eqref{shearnewsPB} over $u'$ gives half the expected result from \eqref{shearsoftnewsPB}.  The problematic quantity causing the discontinuity, however,  is not the soft news, but the shear. As reviewed in \ref{utoinfapp}, one can show $\sigma_{ab }$  does not admit a HVF, implying \eqref{shearsoftnewsPB} and \eqref{shearnewsPB} are  not bona fide PBs, in the sense of \eqref{defPB}. Although hidden in the notation, the brackets  \eqref{shearsoftnewsPB} and \eqref{shearnewsPB} are only defined through the HVF action of the  (soft) news. The discontinuity can however be fixed \cite{stromST} when imposing the constraints $F_5, F_6$, as we shall see in section \ref{2ndconstraintssec}.   

To discuss the remaining ``soft" quantities we will make use of a  cutoff  $\uinf$ such that 
\be
-\uinf <u <\uinf \quad \text{with} \quad \uinf \to \infty. 
\ee
 In particular, we will use Dirac deltas supported on the limiting endpoints,  defined  by
\be \label{defdelplusminus}
\delta_{\pm}(u) =  \delta(u \mp \uinf), \quad  \int du \, \delta_\pm(u)  =1.
\ee
We also denote a symmetrized Dirac delta at infinity by
\be \label{defdelLambda}
\delta_\Lambda(u):=\tfrac{1}{2}(\delta_+(u)+\delta_-(u)).
\ee
In terms of these distributions, the constraints $F_4$ and $F_6$ can  be written as\footnote{As mentioned at the end of section \ref{kinvsphyssec}, there is certain freedom in the definition of $F_4$. It may be more natural to keep a symmetrized form so  that   $\delta_\Lambda$, rather than $\delta_+$, appears in \eqref{F4integratedform}. Such is the prescription implicitly  used in the expressions given in  subsection \ref{summarysec}. In the body of the paper, however, we keep the asymmetrical form \eqref{F4integratedform}  to facilitate the identification of the induced distributional terms in the physical PBs. \label{fnotedeltainf}}
\ba
F_4^{ab} & =& \int du \, \delta_+(u) \,\dot \sigma^{ab }(u), \label{F4integratedform} \\
F_6^{ab} & = & \int du \, \delta_\Lambda(u) \, \sigma^{ab }(u),
\ea
with a HVF action of $F_4$ on the shear given by
\be \label{shearF4PB0}
 \{ \sigma_{ab }(u),  F_4^{cd} \}_0 = \delta_{\langle ab \rangle}^{\langle c d \rangle} \delta_+(u).
\ee 
For the reasons discussed earlier,  there is no HVF associated to $F_6$, see appendix \ref{utoinfapp}.  This poses a challenge for defining PBs involving this quantity, needed in the construction of the Dirac matrix. Another difficulty is the appearance of product of  Dirac deltas at coincident points when attempting an evaluation of  brackets involving $F_4$ with itself or with $F_6$. Following \cite{alok} we  take a prescription in which all these quantities have vanishing brackets:\footnote{It is clear that $\{\sigma^{\langle ab \rangle -},  \dot \sigma^{\langle cd \rangle +} \}_0$ should vanish. On the other hand $\{\sigma^{ab +},  \dot \sigma^{cd +} \}_0$ is formally given by $\tfrac{1}{2}\delta_{\langle ab \rangle}^{\langle c d \rangle} \displaystyle{\lim_{U \to \infty, U' \to \infty}} \delta(U-U')$. The prescription in \eqref{F4F4F6} sets this type of terms to zero. A nonzero value for these brackets could impact on the final physical brackets, see footnote \ref{F6F6footnote}. It would be interesting to explore what is  the most general PBs  $F_4$ and $F_6$ could take that is  compatibility with Jacobi identity  and with GBMS covariance.}
\be \label{F4F4F6}
\{F_4,F_4 \}_0 = \{F_4,F_6 \}_0 = \{F_6 , F_6 \}_0=0.
\ee

On the other hand, the brackets between $F_4$, $F_6$ and the soft news can be obtained from the 
HVF action of the latter, leading to
\be \label{F4F6N0}
\{F_4^{ a b } , \Ncalzero^{c d } \}_0=0, \quad  \{F_6^{a b } , \Ncalzero^{ c d } \}_0= \delta^{\langle ab \rangle \langle c d \rangle}.
\ee
Alternatively, the first relation can  be recovered by integrating the time derivative of \eqref{shearF4PB0} and using $\int du \, \dot \delta_+(u)=0$.

Let us finally discuss the subleading soft news. This is yet another quantity not admitting a HVF. As in the case of the shear, the failure for satisfying the HVF condition can be removed at the cost of adding a term proportional to $F_6$, leading to (see appendix \ref{utoinfapp})
\be \label{sigmaNoneminusF6}
\{ \sigma_{ab }(u),  \big( \Ncalone^{ cd } - \Lambda \, F^{ cd  }_6 \big) \}_0 = \tfrac{1}{2} \delta_{\langle ab \rangle}^{\langle c d \rangle} u.
\ee
Using \eqref{sigmaNoneminusF6} in conjunction with \eqref{F4F4F6} and \eqref{F4F6N0} leads to
\ba
\{F_4^{ab} , \Ncalone^{ cd }  \}_0 & =&  \delta^{\langle ab \rangle \langle c d \rangle} ,  \label{F4Ncalone} \\
\{ F_6^{ab},  \Ncalone^{ cd }  \}_0 & =& 0 , \label{F6Ncalone} \\
\{\Ncalzero^{ab} ,  \Ncalone^{ cd }  \}_0 & =& 0, \label{NcalzeroNcalone}\\
\{\Ncalone^{ab} ,  \Ncalone^{ cd }  \}_0 & =& 0. \label{NcaloneNcalone}\
\ea
Note that \eqref{F4Ncalone} is compatible with \eqref{shearF4PB0} (by using $\int du u \, \delta_+(u) = -1$), while \eqref{NcalzeroNcalone} is consistent with the fact that the soft news has a trivial action on the news tensor.

We conclude the section by \emph{postulating} the  bracket between the news tensor and $F_6$ to be\footnote{The prescription \eqref{prescriptionnewsF6} is what eventually solves the analogue of the discontinuity between \eqref{shearsoftnewsPB} and \eqref{shearnewsPB} in the physical space.  The price one pays is that \eqref{prescriptionnewsF6} is discontinuous with what one would get from \eqref{shearnewsPB}. This, however, does not lead to  discontinuities in the physical space.}
 
\be \label{prescriptionnewsF6}
\{ \dot \sigma_{ab }(u),     F^{ cd  }_6  \}_0 := - \delta_\Lambda(u) \delta_{\langle ab \rangle}^{\langle c d \rangle}.
\ee
This definition is consistent with \eqref{F4F6N0}, and leads, through \eqref{sigmaNoneminusF6}, to 
\be \label{newsNcalone}
\{ \dot \sigma_{ab }(u),   \Ncalone^{ cd } \}_0=\big(\tfrac{1}{2}-\Lambda \delta_\Lambda(u) \big)  \delta_{\langle ab \rangle}^{\langle c d \rangle}.
\ee
Eq. \eqref{newsNcalone} is in turn compatible with \eqref{NcalzeroNcalone}, \eqref{NcaloneNcalone}, as well as with \eqref{shearnewsPB}, by appropriately keeping track of boundary terms.

We will return  to the  discussion of this subsection  when imposing the constraints $(F_3, F_4, F_5, F_6)$.  All these subtleties, however, play no role in the imposition of the first pair of constraints to which we turn next.

\section{Intermediate brackets} \label{1stconstriantssec}

In this section we construct the Poisson brackets associated to the  intermediate kinematical space \eqref{defGammakin1},
\be \label{defGammakin1again}
\Gamma_{\kin, 1}: = \{F_1=F_2=0\}  \subset  \Gamma_{\kin, 0},
\ee
where we recall
\be \label{defF1F2}
\begin{aligned}
F^{ab}_1 & =  p^{ab} - \frac{1}{2} \O^{ab}_{\phantom{ab} cd} \Pi^{cd}, \\
F_{2 a} & =  D^{b} T_{b a} + \frac{1}{2} \partial_a R .
\end{aligned}
\ee

Since the constraint $F^{ab}_1$ is symmetric and trace-free, it provides two independent constraints  (per point on the celestial sphere). Together with  $F_{2 a}$, we then have  at this stage  four independent constraint equations. These  can be thought of as  fixing $p^{ab}$ and $T_{ab}$ in terms of $\Pi^{ab}$ and $q_{ab}$. The intermediate kinematical space  \eqref{defGammakin1again} can then be  parametrized by
\be \label{parametrizationGamma1}
\Gamma_{\kin, 1}  =  \{\sigma_{ab}, C, N,   \Pi^{ab}, q_{ab} \} .
\ee

In order to discuss the Dirac matrix associated to \eqref{defF1F2}, we consider the constraints in smeared form,
\be
F_1[X]  := \int_{S^2}   F_1^{ab} X_{ab}, \quad F_2[X]  := \int_{S^2}   F_{2 a} X^a,
\ee
where  $X_{ab}$ and $X^a$ are smearing parameters.\footnote{ $F_1$ identically vanishes if $X^{ab} \propto q^{ab}$, and so one can  restrict $X^{ab}$ to be trace-free.  $F_2$ identically vanish if $D_{\langle a} X_{b \rangle}=0$ and so $X^a$ is  defined modulo CKVs. The resulting Dirac brackets are insensitive to this redundancy. \label{F2smearingCKV}} 
In the abstract notation of section \ref{diracsec}, $X^\alpha$ represents the pair  $(X_{ab},X^a)$ and
\be
F_\alpha X^\alpha  = F_1[X]+F_2[X]  \equiv F[X].
\ee

The Dirac matrix map \eqref{Diracmap} then reads
\be \label{Diracmap1ststage}
 Y^{ab} = \{F_1^{ab}, F[X] \}_0 , \quad  Y_a =  \{F_{2 a}, F[X] \}_0,
\ee
with  $(Y^{ab},Y_a)$ corresponding to $Y_\alpha$.  In appendix \ref{1stDiracapp} we show \eqref{Diracmap1ststage} is given by
\be \label{XtoY0holocoord}
\begin{aligned}
 Y^{\zb \zb} &= \frac{1}{4}\A X^{\zb\zb } + \frac{1}{2} q^{z \zb} q^{z \zb} \Db^3_z X^z ,\\
 Y_{\zb} & = \frac{1}{2}\Db^3_{\zb} X^{\zb\zb},
\end{aligned}
\ee
where $\A$ is a differential operator with $\Pi^{ab}$-dependent coefficients (see appendix \ref{Aoperatorapp} and  Eq. \eqref{Absoft}). The inverse map $(Y^{ab},Y_a) \mapsto (X_{ab},X^a)$  takes the form\footnote{One may worry about the invertibility of the $\Db^{3}_z$ operator.  When seen as a map from tensors to vectors, as in the second line of \eqref{XtoY0holocoord}, it has trivial kernel. The corresponding inverse, appearing in the first line of \eqref{YtoX0holocoord}, is therefore unambiguously defined.  When seen as a map from vector to tensors, as in the first line of  \eqref{XtoY0holocoord}, it has a non-trivial kernel given by global CKVs.  The corresponding inverse, appearing in the second line of \eqref{YtoX0holocoord}, is then defined modulo global CKVs. However, as discussed in footnote \ref{F2smearingCKV}, $X^a$ is itself defined modulo CKVs and hence  there is no ambiguity in the second line of \eqref{YtoX0holocoord}.  See appendix \ref{greenSsec} for further details of the $\Db^{3}_z$ operator and its inverse.} 
\be \label{YtoX0holocoord}
\begin{aligned}
 X_{zz} &= 2 q_{z \zb} q_{z \zb} \Db^{-3}_{\zb}  Y_{\zb} , \\
 X^z & = 2 \Db^{-3}_z Y_{zz} - q_{z \zb} q_{z \zb}   \Db^{-3}_z \A \Db^{-3}_{\zb}   Y_{\zb}.
\end{aligned}
\ee

The  ``coefficients" of $Y$ in \eqref{YtoX0holocoord}  form the elements of the inverse Dirac matrix that go into  Dirac formula \eqref{abstractcorrectedPB} for $i=0$. We present below  the resulting  PBs in holomorphic coordinates.  A complementary discussion in terms of HVFs and in general 2d coordinates is given in appendix \ref{HVF1sec}.\\

We start by noticing that $ \sigma_{zz}$, $C$ and $N$  commute with $F_1$ and $F_2$ and hence their PBs remain unchanged,
\ba
\{  \sigma_{zz}(u), \dot \sigma^{ww}(u') \}_1 &=&  {\frac{1}{2} \delta(u-u')} \delta^{(2)}(z,w),  \label{sigmadotsigma1}\\
\{  C(z),N(w)\}_1 &=& \Gzero_z(w). \label{CN1}
\ea
Likewise, PBs involving asymptotic values of the shear and the news tensor, discussed in subsection \ref{utoinfsec}, also remain unchanged.

For the remaining independent variables in  \eqref{parametrizationGamma1},   the general formula  \eqref{abstractcorrectedPB} leads to the  following non-trivial bracket relations
\ba \label{qzzPiwbwb}
\{q_{zz}, \Pi^{\wb \wb}\}_1& =& -2 q_{z \zb}  \Db^{-3}_{\zb} \Db_z \delta^{(2)}(z,w) \\
\{\Pi^{zz}, \Pi^{\wb \wb}\}_1 &=& \Db_{\zb} \Db^{-3}_z \A \Db^{-3}_{\zb}\Db_z \delta^{(2)}(z,w) \label{PizzPiwbwb} \\
\{  N(z),\Pi^{ww}\}_1 &=& -\Gzero_z \{\D_z,\Pi^{ww}\}_1 N(z)  \label{NPi1} \\
&=& -  \left( N(w) \Gzero^{ww}_{z}+  2 q_{w \wb} \Db_{\wb} \Db^{-3}_w \bar \Bb(N(w)) \Gzero^{\wb \wb}_z  + (N \leftrightarrow \Gzero_{z} )\right)  , \label{NPiwwfinal2}
\ea
where  $\Bb$ is a  differential operator, given in   Eq. \eqref{scalarBb}, that is constructed out of $\O$ and $\Delta$.\footnote{The equivalence between \eqref{NPi1} and \eqref{NPiwwfinal2}  can be shown by writing the first line explicitly as $\{  N(z),\Pi^{ww}\}_1 = -\int  d^2 z' \Gzero_z(z') \{\D',\Pi^{ww}\}_1 N(z')$, and  using the formula \eqref{delD4} for the variation of $\D$, together with \eqref{qzzPiwbwb} and \eqref{TzzPiww1}, \eqref{TzzPiwbwb1}.}

Since $\Pi^{zz}$ has non-trivial brackets with the metric, it induces pure trace deformations on STF tensors. Specializing   Eq. \eqref{delAzzb} to the case where the variation is generated by the PB action of $\Pi^{\wb\wb}$ leads to
\ba
\{\sigma^{z\zb}(u),\Pi^{\wb\wb}\}_1 &=&\sigma^{zz}(u)\Db^{-3}_{\zb} \Db_z \delta^{(2)}(z,w) \label{sigmazzbPiwbwb},\\
\{\Pi^{z\zb},\Pi^{\wb\wb}\}_1 &=&\Pi^{zz}\Db^{-3}_{\zb} \Db_z \delta^{(2)}(z,w) . \label{PizzbPiwbwb}
\ea

Eqs. \eqref{sigmadotsigma1} to \eqref{PizzbPiwbwb}, together with their complex conjugates, complete all non-trivial  PBs among the independent variables on  $\Gamma_{\kin, 1}$.\\

We will also need PB expressions for what appear as    ``non-elementary'' quantities from the perspective of  $\Gamma_{\kin, 1}$. In particular,  for the Geroch tensor one has
\ba
\{T_{zz}, \Pi^{ww}\}_1 & = &\delta^{(2)}(z,w) ,\label{TzzPiww1} \\
\{T_{zz}, \Pi^{\wb\wb}\}_1 &=& q_{z \zb} \bar \O  \Db^{-3}_{\zb} \Db_z \delta^{(2)}(z,w) ,  \label{TzzPiwbwb1} \\
\{T_{z\zb},\Pi^{\wb\wb}\}_1 &=&- T_{\zb\zb}\Db^{-3}_{\zb} \Db_z \delta^{(2)}(z,w) \label{TzzbPiwbwb}.
\ea

Additional non-elementary PBs, such as those involving $\Db^2_z$ and $\Db_z^3$ are discussed in appendix \ref{HVF1sec}.

\newpage

\section{Physical brackets} \label{2ndconstraintssec}

In this section we construct the physical PBs  by imposing the remaining constraints, 
\be 
\Gamma_{\phys} = \{F_3=F_4 = F_5= F_6=0\}  \subset  \Gamma_{\kin, 1} \ ,
\ee
where we recall
\be \label{last4constraints}
\begin{aligned} 
F_3^{ab} &= \Ncalone^{ab} +C N^{ab}  -\frac{1}{2}\Pi^{ab},   \\
F_4^{ab} & = 2\dot \sigma^{ab +}, \\
F_5^{ab} & = \Ncalzero^{ab} - N^{ab}, \\
F_6^{ab} & = \frac{1}{2}(\sigma^{ab +} + \sigma^{ab -}),
\end{aligned}
\ee
and $N^{ab}\equiv -2 \Db^{\langle a }\Db^{b \rangle} N$.

Our first task is to evaluate the brackets between constraints. Given the PBs on  $\Gamma_{\kin, 1}$,  together with the prescriptions discussed in subsection \ref{utoinfsec}, one finds the Dirac matrix takes the following schematic form
\be \label{2ndstageDiracmatrix}
\{F_\alpha,F_\beta \}_1  = 
\begin{pmatrix}
K & -1 & L & 0 \\
1 & 0 & 0 & 0  \\
- \tilde L &  0 & 0 & -1\\
0 &  0 & 1 & 0
\end{pmatrix} ,
\ee
where $\alpha,\beta=3,4,5,6$ enumerate the constraints \eqref{last4constraints}. Each matrix entry is to be understood as an operator in the space of STF tensors, with $1$ representing  the identity (modulo raising of indices). $L$ and $K$ are integro-differential operators discussed in appendix \ref{2nddiracapp}, with   $\tilde L$   the adjoint of $L$.  $K$ is anti-selfadjoint.

The inverse of  \eqref{2ndstageDiracmatrix} is  given by 
\be \label{2ndstageInv_DM}
W^{\alpha \beta}_1  = 
\begin{pmatrix}
0 & 1 & 0 & 0 \\
-1 & K & 0 & L  \\
0 &  0 & 0 & 1\\
0 &  -\tilde L & -1 & 0
\end{pmatrix}.
\ee

We can now apply Dirac's formula \eqref{abstractcorrectedPB} for \(i=1\). For the reasons discussed in subsection \ref{utoinfsec} and appendix \ref{kinHVFsapp}, we will regard the news tensor, rather than the shear, as the fundamental ``hard'' variable. The elementary brackets naturally fall into three categories, according to how many instances of the news tensor appear in them. We discuss each case separately below.

\subsection{Soft--soft brackets}\label{softsoftsec}
We start by considering the independent soft variables in  $\Gamma_{\kin, 1}$, namely $C,N, \Pi^{ab}$ and $q_{ab}$. The four quantities commute with the constraints $F_4$ and $F_6$. On the other hand, the non-zero entries of $W^{\alpha \beta}_1$ are located at positions where at least one of the indices  is  $4$ or $6$ (corresponding to the second and fourth column/raw in \eqref{2ndstageInv_DM}). It then follows that the physical PBs among such  quantities coincide with their brackets on $\Gamma_{\kin, 1}$, 
\be \label{PBunaltered}
\psi, \varphi \in (C,N, \Pi^{ab},q_{ab}) \implies  \{\psi,\varphi \}_2=\{\psi,\varphi \}_1 .
\ee

From the perspective of $\Gamma_{\kin, 2}$, we should also consider $\Ncalzero_{ab}$ and $\Ncalone_{ab}$ among the soft variables. Their brackets  can be obtained either by Dirac's formula,\footnote{Their non-trivial brackets with the constraints are  $\{ \Ncalone_{ zz },  F_4^{ww}  \}_1  = \{ \Ncalzero_{zz}, F_6^{ww} \}_1 = - \delta^2(z,w) $.}  or  by expressing them in terms of $C,N, \Pi^{ab}$ and $q_{ab}$,  
\be \label{F3F5solved}
\begin{aligned}
\Ncalone^{ab} &=  \frac{1}{2}\Pi^{ab} -C N^{ab}, \\
\Ncalzero^{ab} &= N^{ab},
\end{aligned}
\ee
and then  evaluating their brackets  using  Eq. \eqref{PBunaltered}. Regardless the procedure, one finds the following non-trivial  relations
\be \label{softsoftPBs}
\begin{aligned}
\{  C(z),\Ncalzero^{ww}\}_2 &= \Gzero^{ww}_z ,  \\
\{ C(z), \Ncalone^{w w}\}_2 &= - C(w) \Gzero^{ww}_z, \\
\{q_{\zb \zb}, \Ncalone^{w w}\}_2& = - q_{z \zb}  \Db^{-3}_{z} \Db_{\zb} \delta^{(2)}(z,w), \\
\{ \Ncalone_{zz}, \Ncalzero^{ww} \}_2 &=  \L \delta^2(z,w) , \\ 
\{ \Ncalone_{zz}, \Ncalone^{ww} \}_2 &= \K \delta^2(z,w) ,
\end{aligned}
\ee
where  $\K$ and $\L$ are scalar operators that  characterize the Dirac matrix entries $K$ and $L$ in \eqref{2ndstageDiracmatrix},  see  appendix \ref{2nddiracapp} for details. 
In \eqref{softsoftPBs} we focused on the trace-free components. The subleading soft news also induces trace variations on all tensors, similarly to what occurs with  $\Pi^{ab}$ (see discussion before Eq. \eqref{sigmazzbPiwbwb}). In particuar, one has
\be \label{trsubnewsNone}
\begin{aligned}
\{ \Ncalzero_{z\zb}, \Ncalone^{ww} \}_2 &=  - \tfrac{1}{2} \Ncalzero_{zz}\Db^{-3}_{z} \Db_{\zb} \delta^{(2)}(z,w) , \\
\{ \Ncalone_{z\zb}, \Ncalone^{ww} \}_2 &= - \tfrac{1}{2} \Ncalone_{zz}\Db^{-3}_{z} \Db_{\zb} \delta^{(2)}(z,w) .
\end{aligned}
\ee

\subsection{News--soft brackets} \label{newssoftsec}

 According to the prescriptions given in subsection \ref{utoinfsec} (which remain unaltered on $\Gamma_{\kin, 1}$), the non-trivial PBs between the news tensor  and the constraints are 
\be
\label{newsFs}
\begin{aligned}
\{\dot \sigma_{zz}(u), F_3^{ww} \}_1 & =\big(\tfrac{1}{2}-\Lambda\delta_\Lambda(u) \big)  \delta^{(2)}(z,w),\\
\{\dot \sigma_{zz}(u), F_4^{ww} \}_1 &=\dot  \delta_+(u) \delta^{(2)}(z,w), \\
\{\dot \sigma_{zz}(u), F_6^{ww} \}_1 & = -\delta_\Lambda(u) \delta^{(2)}(z,w),
\end{aligned}
\ee
where  $\delta_+$ and $\delta_\Lambda$ are Dirac deltas with support at infinity, according to  Eqs. \eqref{defdelplusminus} and \eqref{defdelLambda}.\footnote{As discussed in footnote \ref{fnotedeltainf}, one could use a definition of $F_4$ such that $\delta_\Lambda$ appears in place of $\delta_+$. All the results of this and the next section remain unchanged if we replace $\delta_+$ by $\delta_\Lambda$.}  Using \eqref{newsFs}, one can show  the PBs between  the news tensor and \emph{any} soft quantity $\varphi_\soft$ take the form\footnote{For $\varphi_\soft= C,N, \Pi^{ab},q_{ab}$, this follows from $\{\varphi_\soft, F_3 \}_1 = \{\Ncalone, \varphi_\soft\}_2$, $\{\varphi_\soft, F_5 \}_1 = \{\Ncalzero, \varphi_\soft\}_2$  and $W^{3 4}_1=W^{5 6}_1=-1$. For $\varphi_\soft=\Ncalzero, \Ncalone$, the Dirac bracket involves other components of $W^{\alpha \beta}_1$. The result can be written as  \eqref{newsgralsoftPB} after using the last two relations in \eqref{softsoftPBs}.}
\be \label{newsgralsoftPB}
\{ \dot \sigma_{zz}(u),  \varphi_\soft\}_2  =  \delta_\Lambda(u) \{ \Ncalzero_{zz}, \varphi_\soft \}_2   -  \dot \delta_+(u)  \{ \Ncalone_{zz}, \varphi_\soft \}_2 .
\ee

The integration  rules
\be \label{integrationrules1}
\begin{aligned}
\int du \, \delta_\Lambda(u) &=1,   & \int du\, \dot \delta_+(u) & =0, \\
 \int du \, u  \delta_\Lambda(u) &=0,  & \int du \,u \dot \delta_+(u) & =-1, 
\end{aligned}
\ee
then imply  \eqref{newsgralsoftPB} is compatible with the leading and subleading soft news brackets,
 \ba
\int du \{ \dot \sigma_{zz}(u),  \varphi_\soft\}_2  &= &   \{ \Ncalzero_{zz},  \varphi_\soft\}_2  , \label{continuity1} \\
\int du u \{ \dot \sigma_{zz}(u),  \varphi_\soft\}_2  &=&   \{ \Ncalone_{zz},  \varphi_\soft\}_2   . \label{continuity2}
\ea
This illustrates the absence of discontinuities in the physical  brackets, in contrast to what occurs at the kinematical level. In particular, the kinematical discontinuity between \eqref{shearsoftnewsPB} and \eqref{shearnewsPB} should be compared with the continuity    \eqref{continuity1} when $\varphi_\soft = C$. See appendix \ref{HVF2sec} for further comments.

In addition to the ``boundary $u$'' brackets \eqref{newsgralsoftPB}, there are also ``bulk  $u$'' brackets involving the trace-part of the news. They can be obtained from \eqref{sigmazzbPiwbwb} (which remains unchanged), leading to
\be \label{newsPI}
\{\dot \sigma_{z\zb}(u),\Pi^{w w}\}_2 = 2 \{\dot \sigma_{z\zb}(u),\Ncalone^{w w}\}_2 = - \dot \sigma_{zz}(u)\Db^{-3}_{z} \Db_{\zb} \delta^{(2)}(z,w) .
\ee
Upon integrating \eqref{newsPI} in $u$ one recovers the analogue bracket involving the trace part of the (sub)leading soft news, Eq. \eqref{trsubnewsNone}.

\subsection{News--news brackets}\label{newsnewssec}
The news bracket with itself can be evaluated by substituting \eqref{newsFs} on Dirac formula \eqref{abstractcorrectedPB}, leading to
\begin{multline} \label{newsnewsPB2}
\{ \dot \sigma_{zz}(u), \dot \sigma^{ww}(u') \}_2 = \left( {\tfrac{1}{2}\dot \delta(u-u')} -\big(\tfrac{1}{2}-\Lambda\delta_\Lambda(u') \big) { \dot \delta_+(u) }+  \big(\tfrac{1}{2}-\Lambda\delta_\Lambda(u) \big){ \dot \delta_+(u') }  \right.  \\ 
\left. + {\dot \delta_+(u) \dot \delta_+(u')} \K+ {\delta_\Lambda(u) \dot \delta_+(u') }\L^{\dagger} - \dot \delta_+(u)\delta_\Lambda(u') \L\right)\delta^{(2)}(z,w).
\end{multline}
We can again verify a continuity condition, this time between \eqref{newsnewsPB2} and \eqref{newsgralsoftPB} with $\varphi_\soft=\Ncalzero, \Ncalone$. Using the integration rules  \eqref{integrationrules1} and 
\be
\int du \, u  =0, \quad \quad \int du \,  \dot \delta(u-u') =0, \quad \quad \int du \, u \dot \delta(u-u') = -1 + 2 \Lambda \delta_\Lambda(u'), 
\ee
one finds
\ba \label{continuityNzero}
\int du'  \{ \dot \sigma_{zz}(u), \dot \sigma^{ww}(u') \}_2  &=&  -  \dot \delta_+(u) \L  \delta^{(2)}(z,w) \nonumber \\
 &\equiv&   \{ \dot \sigma_{zz}(u),  \Ncalzero^{ww} \}_2 ,  \\
 && \nonumber \\
\int du'  u'  \{ \dot \sigma_{zz}(u), \dot \sigma^{ww}(u') \}_2 &  = &  -\left(  \dot \delta_+(u)   \K  + \delta_\Lambda(u)  \L^{\dagger} \right)\delta^{(2)}(z,w) \nonumber \\
& \equiv &     \{ \dot \sigma_{zz}(u),  \Ncalone^{ww} \}_2 .   \label{continuityNone}
\ea

Finally, either from Dirac formula or from consistency with  \eqref{newsgralsoftPB} with $\varphi_\soft=q_{ab}$ (see \eqref{CqnewsPB2} below) we get
\be
\{ \dot \sigma_{z \zb}(u), \dot \sigma^{ww}(u') \}_2=  \tfrac{1}{2}\dot \sigma_{z z}(u) {\dot \delta_+(u') } \Db^{-3}_{z} \Db_{\zb} \delta^{(2)}(z,w). \label{trnewsnewsPB2}
\ee
Integrating \eqref{trnewsnewsPB2} in $u$ or $u'$   leads to news--soft brackets that are consistent with the earlier expressions.

\subsection{Elementary PB algebra} \label{elemPBalgebrasec}
A minimal set of variables parametrizing $\Gamma_\phys$ are: $\dot \sigma_{ab}$, $C$ and $q_{ab}$. The non-trivial PBs among these quantities are given by Eqs. \eqref{newsnewsPB2}, \eqref{trnewsnewsPB2}, together with Eq. \eqref{newsgralsoftPB} for $\varphi_\soft=C,q_{ab}$,
\be \label{CqnewsPB2}
\begin{aligned}
\{C(z), \dot \sigma^{ww}(u) \}_2 &=  \left(  \dot \delta_+(u) C(w)    + \delta_\Lambda(u)   \right) \Gzero^{ww}_z , \\
\{q_{\zb\zb}, \dot \sigma^{w w}(u)  \}_2& =  {\dot \delta_+(u) } q_{z \zb} \Db^{-3}_{z} \Db_{\zb} \delta^{(2)}(z,w).
\end{aligned}
\ee

This ``elementary'' PB algebra was presented in   Eqs. \eqref{newsnewsPBintro} to \eqref{trivialPBintro} at the beginning of the paper. There, we used general 2d coordinates and assumed  a symmetric definition for $F_4$ so that $\delta_+$ is replaced by  $\delta_\Lambda$. The bilocal kernels in those expressions are obtained by applying the corresponding integro-differential operators to the 2d Dirac delta.

Let us conclude the section by emphasizing that the  continuity properties discussed in the previous subsections imply that the elementary PBs  can be used to generate all  physical brackets. These include, in particular, the ``soft'' brackets  discussed in section \ref{1stconstriantssec}, which remain unchanged on $\Gamma_\phys$ according to \eqref{PBunaltered}. In this perspective,  quantities like $\Pi^{ab}$ or $N$ are  regarded as non-elementary/composite. 

A more interesting example of composite quantities is the supertranslation and superrotation charges. We discuss them in the next section.

\section{GBMS Poisson bracket action}\label{gbmsactionsec}

In this section we verify that the PB action of  ST and SR charges reproduce the expected transformation rules,\footnote{In this section we omit the subscript in PBs, with the understanding that they refer to physical brackets.}
\be \label{gbmsPBaction}
\delta_f  = \{\cdot, P_f \}, \quad  \delta_V  = \{\cdot, J_V \}.
\ee

For concreteness, we will evaluate \eqref{gbmsPBaction} on the  following phase space quantities:
\be \label{varphigbmspbaction}
C, N,  q_{ab},  T_{ab},  \Pi^{ab}  \text{ and }   \dot \sigma_{ab}(u) .
\ee

\subsection{Supertranslations}

The supermomentum can be written as $P_f= P_f^\soft+P_f^\hard$ with  (see appendix \ref{gbmschargesapp})
\be \label{PsoftPhard}
P^\soft_f =  \int d^2 z  \, f(z)   \D N(z) , \quad  \quad P^\hard_f =  \int du d^2 z   f(z) \,  \dot \sigma^{ab}(u,z) \dot \sigma_{ab}(u,z)  .
\ee
Let us start with the soft quantities in \eqref{varphigbmspbaction}. From the soft brackets discussed in section \ref{1stconstriantssec} (which remain unchanged according to \eqref{PBunaltered}) one has
\be \label{softPsoft}
\{ C, P^\soft_f\} = f, \quad  \quad \{ \varphi,P_f^\soft \} =0 ,\quad  \varphi= N,  q_{ab},  T_{ab}, \Pi^{ab}.
\ee

On the other hand, the  brackets of the soft quantities with $P^\hard_f$ can be evaluated from the news--soft brackets \eqref{newsgralsoftPB}. It is easy to see that the distributional terms at infinity do not contribute in the $u$ integral  of \eqref{PsoftPhard}  so that
\be \label{softPhard}
\{ \varphi,P_f^\hard \} =0 \quad \text{for } \varphi=C, N,  q_{ab},  T_{ab}, \Pi^{ab}.
\ee
 
 Thus,  \eqref{softPsoft} holds for the total $P_f$, correctly reproducing the ST action on the soft variables.\\

We now discuss the news tensor. Using the general formula \eqref{newsgralsoftPB} together with
\be
\{ \Ncalzero_{zz}, P_f^\soft \}=0, \quad \quad \{ \Ncalone_{zz}, P_f^\soft \}= - f(z) \Ncalzero_{zz}
\ee
one finds
\be \label{newsPsoft}
\{ \dot \sigma_{zz}(u), P_f^\soft \} = \dot \delta_+(u) f(z) \Ncalzero_{zz}.
\ee

To evaluate the bracket with the hard supermomentum, we start with
\be \label{newsPhard1}
 \{ \dot \sigma_{zz}(u),  P_f^\hard \}= 2 \int du' d^2 w   f(w) \{ \dot \sigma_{zz}(u),   \,  \dot \sigma^{ww}(u')\} \dot \sigma_{ww}(u').
\ee
The news--news bracket \eqref{newsnewsPB2} has the form
\be \label{newsnewsGBMS}
 \{  \dot \sigma_{zz}(u), \dot \sigma^{ww}(u')\} =  \left( \tfrac{1}{2}\dot \delta(u-u') -\tfrac{1}{2} \dot \delta_+(u) \right) \delta^{(2)}(z,w) + (\cdots) \delta_\Lambda(u') +   (\cdots) \dot \delta_+(u').
\ee
By the same reasoning given before \eqref{softPhard}, one can check that the distributional terms at $|u'|=\infty$ do not contribute to the $u'$ integral in \eqref{newsPhard1}. The first two terms in  \eqref{newsnewsGBMS} lead to 
\be \label{newsPhard2}
\{ \dot \sigma_{zz}(u),  P_f^\hard \}= f(z) \ddot \sigma_{zz}(u) - f(z) \Ncalzero_{zz} \dot \delta_+(u).
\ee
Adding \eqref{newsPsoft} and \eqref{newsPhard2} we recover the ST action on the news,
\be
\{ \dot \sigma_{zz}(u),  P_f \}= f(z) \ddot \sigma_{zz}(u) .
\ee

\subsection{Superrotations}

The charge can be written as $J_V=J_V^\hard+J_V^\soft$ with
\be
\begin{aligned}
J^\hard_V &=  \int_{S^2} \int du \, \dot \sigma^{ab} \delta_V \sigma_{ab}   ,\\ 
J^\soft_V &=  \int_{S^2} \Big(  \D N \delta_V C -  \Pi^{ab} \Sb_{ab \, c} V^c\Big),
\end{aligned}
\ee
where  (see appendix \ref{gbmschargesapp} for further details)
\ba
\delta_V \sigma_{ab} &=& \left(  \Lie_V  + \frac{1}{2}D_c V^c (u \partial_u -1) \right) \sigma_{ab}, \label{delVshearsec6}  \\
\delta_V C & = &\left(  \Lie_V  - \frac{1}{2}D_a V^a \right) C,\\
 \Sb_{zz \, c} V^c &=& \Db_z^3 V^z .
\ea

We  start with the  observation that  the first four quantities in \eqref{varphigbmspbaction} commute with the hard superrotation charge
\be \label{CNqTcommuteJhard}
\{ \varphi,J_V^\hard \} =0 \quad \text{for } \varphi=C, N,  q_{ab},  T_{ab}.
\ee
This, again, is not entirely trivial, as one needs to check that  distributional terms at infinity do not contribute. The situation is   more subtle than in the supertranslation case, since $J_V^\hard$ depends both on the news and on the shear tensors.\footnote{One thus needs to evaluate  shear--soft brackets. These   can be defined by integrating the  news--soft brackets \eqref{newsgralsoftPB}, with boundary condition  $F_6=0$.  One can  show the resulting distributional terms do not contribute in the bracket \eqref{CNqTcommuteJhard}, by  a similar argument to the one given below in Eq. \eqref{bdyzeroJhard} (with $ \varphi$ playing the role of $\sigma_{zz}(u)$).}

Let us now consider the PBs of the quantities in \eqref{CNqTcommuteJhard} with $J^\soft_V$.  The simplest case is $C$, for which one has
\be
\{C(z),J^\soft_V\}  = \int d^2 w \D_w \{C(z), N(w)\}  \delta^{w}_V C(w) =\delta_{V}C,
\ee
where we used that the only non-trivial PB is the one with $N$, \eqref{CN1}.

$N$, on the other hand, has non-trivial brackets with $C$ and   with $\Pi^{ab}$ \eqref{NPiwwfinal2}, leading to
\ba
\{N(z),J^\soft_V\}  &=& \int d^2 w\left( \D_w N(w)  \delta^{w}_V \{ N(z),C(w)\} -(\{N(z), \Pi^{w w}\} \Db^3_{w} V^{w} +c.c.) \right) \\
&=& \Gzero \delta_V (\D N)  +  \int d^2 w(  \Gzero_z \{\D_z,\Pi^{ww}\} N(z)  \Db^3_{w} V^{w} + c.c) \\
&=& \Gzero \delta_V (\D) N+ \delta_V N - \Gzero \{\D,J^\soft_V \} N =  \delta_V N ,
\ea
where in the last equality we used that $\{\D,J^\soft_V \}=\delta_V \D$. This is a consequence of the SR action of the soft charge on the 2d metric and on the Geroch tensor, which we  verify next.

For the 2d metric  the only non-trivial PB is the one with $\Pi^{ab}$ \eqref{qzzPiwbwb}, leading to 
\ba
\{q_{zz},J^\soft_V\}  &=& -\int d^2 w \{q_{zz}, \Pi^{\wb \wb}\} \Db^3_{\wb} V^{\wb} \\
&=& 2 \Db_{z} V_z  \equiv \delta_V q_{zz},
\ea
where we used  
\be \label{DzD3zbcomm2}
[\Db_z , \Db^3_{\zb}]=0,
\ee
together with the fact that the SR covariant derivative on ($k=0$) SR vector fields coincides with the ordinary covariant derivative, 
\be \label{DbVDV}
\Db_a V^b = D_a V^b.
\ee

For the  Geroch tensor, the non-trivial brackets are those with $\Pi^{ab}$. Using  Eqs. \eqref{TzzPiww1} and \eqref{TzzPiwbwb1}, the trace-free components of the PBs are
\ba
\{T_{zz},J^\soft_V\}  &=& -\int d^2 w \left( \{T_{zz}, \Pi^{w w}\} \Db^3_{w} V^{w}+ \{T_{zz}, \Pi^{\wb \wb}\} \Db^3_{\wb} V^{\wb} \right)\\
&=& -\Db_z^3 V^z -    \bar \O   \Db_z V_z =\delta_V T_{zz},
\ea
where to get the second equality we used  \eqref{DzD3zbcomm2} and in the last equality we used the identity  \eqref{Dz3delVTbarO}. STFs tensors also exhibit  pure-trace variations that compensate for the change in the metric. From \eqref{TzzbPiwbwb} one gets
\ba \label{TzzbJsoft}
\{T_{z\zb},J^\soft_V\}  &=& -\int d^2 w \left( \{T_{z\zb}, \Pi^{w w}\} \Db^3_{w} V^{w}+ c.c \right)  \\
&=&  T_{zz} \Db_{\zb}V^z  + c.c  = \delta_V T_{z \zb},
\ea
where we used \eqref{DzD3zbcomm2}, \eqref{DbVDV}, and the fact that   $\delta_V T_{z \zb}= \Lie_V T_{z \zb} =D_{\zb}V^z T_{zz} + D_{z}V^{\zb} T_{\zb\zb}$.\\

The last soft quantity to analyse is $\Pi^{ab}$. Unlike the other soft variables, there is a non-trivial contribution from the hard SR charge given by (see the end of this section for a derivation)
\be \label{PizzJVhard}
\{\Pi^{zz},J_V^\hard\} = \Db_{\zb}\Db^{-3}_z \A_\hard \Db_z V^{\zb},
\ee
where we recall that
\be
\A_\hard= 2 \int_{-\infty}^\infty du(\dot \sigma^{zz} \sigma_{zz} -  \dot \sigma^{\zb \zb} \sigma_{\zb \zb} ). \label{Abhard2}
\ee
On the other hand, the bracket with the soft charge is a sum of two terms,
\be
\{\Pi^{zz},J_V^\soft\} =  -\int d^2 w \Big( \{\Pi^{zz}, \Pi^{\wb \wb}\} \Db^3_{\wb} V^{\wb}+ \Pi^{ab} \{\Pi^{zz}, \Sb_{ab \, c} \} V^c\Big).
\ee
The first term can be evaluated using \eqref{PizzPiwbwb} leading to 
\be \label{1sttermPiJVsoft}
 -\int d^2 w  \{\Pi^{zz}, \Pi^{\wb \wb}\} \Db^3_{\wb} V^{\wb} =- \Db_{\zb} \Db^{-3}_z \A \Db_z V^{\zb}.
\ee
The second term can be evaluated using the brackets computed in \ref{PiSbsec} and gives
\ba
 -\int d^2 w \Pi^{ab} \{\Pi^{zz}, \Sb_{ab \, c} \} V^c & =& -\{\Pi^{zz}, \Sb[X,V]\}|_{X=\Pi}  \\
 & =&  \delta_V \Pi^{zz} + \Db^{z} \Db^{-3}_z \A_\soft \Db_z V_{z}.  \label{PiJVsoft2}
\ea
Since $\A=\A_\soft+\A_\hard$, the two contributions add up to 
\be \label{PiJsoft}
\{\Pi^{zz},J^\soft_V\} = \delta_V \Pi^{zz} - \Db_{\zb}\Db^{-3}_z \A_\hard \Db_z V^{\zb}.
\ee
The second term in \eqref{PiJsoft} precisely cancels the  hard contribution \eqref{PizzJVhard}, so that we finally  recover the desired result
\be \label{PBdelVPi}
\{\Pi^{zz},J_V\} = \delta_V \Pi^{zz}. 
\ee

The evaluation of $\{\Pi^{z\zb},J_V\}$ is analogous to the one already discussed for $T_{z \zb}$, with the relevant bracket now being \eqref{PizzbPiwbwb}.\\

We finally discuss the news tensor. We start by noticing that the soft brackets found above imply 
\ba
 \{ \Ncalzero_{zz}, J_V^\soft  \} &=& \delta_V \Ncalzero_{zz} \\
\{ \Ncalone_{zz},  J^\soft_V \} &=& \delta_V \Ncalone_{zz} - \tfrac{1}{2} \Db_{\zb}\Db^{-3}_z \A_\hard \Db_z V^{\zb}.
\ea
Therefore, the general news--soft bracket formula \eqref{newsgralsoftPB} leads to 
\be \label{newsJsoft}
\{ \dot \sigma_{zz}(u),  J_V^\soft \}  =  \delta_\Lambda(u) \delta_V \Ncalzero_{zz}   -  \dot \delta_+(u)\big(  \delta_V \Ncalone_{zz} - \tfrac{1}{2} \Db_{\zb}\Db^{-3}_z \A_\hard \Db_z V^{\zb} \big).
\ee

The bracket of the news with $J_V^\hard$ can be written as,
\be \label{newsJhard1}
\{ \dot \sigma_{zz}(u),J_V^\hard\} =\int du' d^2 w \left( \{  \dot \sigma_{zz}(u), \dot \sigma^{ab}\} \delta_V \sigma_{ab} +  \dot \sigma^{ab} \delta_V \{ \dot \sigma_{zz}(u) , \sigma_{ab}\} \right) ,
\ee
where it is understood that $ \sigma_{ab}$ and $\delta_V$ are tensor fields on the $(u',w)$ variables being integrated. The second term in \eqref{newsJhard1} can be shown to be equal to  the first one, after a double integration by parts: One on the sphere for  $\delta_V$,  and one on $u'$. The latter leads to a  boundary piece 
\be \label{bdyzeroJhard}
\left.  \sigma^{ab}(u') \delta_V \{ \dot \sigma_{zz}(u) , \sigma_{ab}(u')\}\right|^{u'=+\infty}_{u'=-\infty}=0,
\ee
that vanishes provided we define the news--shear PBs by integrating  the news--news PBs \eqref{newsnewsPB2} with boundary conditions compatible with $F_6 =0$. 
After the double integration by parts, one gets
\be \label{newsJhard2}
\{ \dot \sigma_{zz}(u),J_V^\hard\} =  2 \int du' d^2 w \delta_V \sigma_{ab}(u')  \{  \dot \sigma_{zz}(u), \dot \sigma^{ab}(u')\} .
\ee

Let us now  split \eqref{newsJhard2} into trace and trace-free  components of $ \sigma_{ab}$,
\be \label{newsJhardtrace}
\{ \dot \sigma_{zz}(u),J_V^\hard \}|_{\text{trace part}} =  4 \int du' d^2 w  \delta_V \sigma_{w \wb}(u')  \{  \dot \sigma_{zz}(u), \dot \sigma^{w \wb}(u')\},
\ee
and
\be \label{newsJhardtracefree}
\{ \dot \sigma_{zz}(u),J_V^\hard\}|_\text{trace-free} =  2 \int du' d^2 w  \delta_V \sigma_{ww}(u')  \{  \dot \sigma_{zz}(u), \dot \sigma^{ww}(u')\} .
\ee
The evaluation of \eqref{newsJhardtrace} follows the same pattern of  \eqref{PizzJVhard} (see the end of the section) and gives
\be \label{newsJhardtrace2}
\{ \dot \sigma_{zz}(u),J_V^\hard \}|_{\text{trace part}} = -  \tfrac{1}{2} \dot \delta_+(u)  \Db_{\zb}\Db^{-3}_z \A_\hard \Db_z V^{\zb}.
\ee
To evaluate \eqref{newsJhardtracefree}, we write the news--news bracket as we did in \eqref{newsnewsGBMS}
\be \label{newsnewsGBMS2}
 \{  \dot \sigma_{zz}(u), \dot \sigma^{ww}(u')\} =  \left( \tfrac{1}{2}\dot \delta(u-u') -\tfrac{1}{2} \dot \delta_+(u) \right) \delta^{(2)}(z,w) + (\cdots) \delta_\Lambda(u') +   (\cdots) \dot \delta_+(u').
\ee
As before, one can show that the boundary distributions do not contribute to the $u'$ integral (unlike the supertranslation case, however, we need to  make explicit use of the condition $F_6=0$).  The integration in $u'$ of the first two terms in \eqref{newsnewsGBMS2} requires keeping track of the cutoff $\Lambda$ according to 
\be \label{integrationrule3}
\begin{aligned}
\int du' \gamma(u') \dot \delta(u-u') & = \dot \gamma(u) - 2 \gamma^+ \delta_\Lambda(u), \\
\int du' \gamma(u')  &= - \int du u' \dot \gamma(u')
\end{aligned}
\quad \quad \text{for any  $\gamma(u')$  s.t.  $\gamma^++\gamma^- =0$,}
\ee
where $ \gamma^\pm = \lim_{\Lambda \to \pm \infty} \gamma(u)$.  
Applying  \eqref{integrationrule3} in \eqref{newsJhardtracefree} and using the fact that $\Ncalzero_{zz} = 2 \sigma^+_{zz}$ we  get
\be \label{newsJtracefree2}
\{ \dot \sigma_{zz}(u),J_V^\hard\}|_\text{trace-free} = \delta_V \dot \sigma_{zz}(u) - \delta_\Lambda(u) \delta_V \Ncalzero_{zz}  + \dot \delta_+(u)  \delta_V \Ncalone_{zz}.
\ee

Adding the three contributions \eqref{newsJsoft}, \eqref{newsJhardtrace2} and \eqref{newsJtracefree2} we recover the expected result
\be \label{finalnewsJ}
\{ \dot \sigma_{zz}(u),J_V \} = \delta_V \dot \sigma_{zz}(u)
\ee
where $\delta_V \dot \sigma_{zz}=\Lie_V \dot \sigma_{zz} + \tfrac{1}{2}D_a V^a u \ddot \sigma_{zz}$.\\
 
The above discussion dealt with the trace-free components of the news--$J_V$ bracket. As for the other tensors, there is also a non-trivial bracket for the trace part. The computation parallels the one given in \eqref{TzzbJsoft}, with the relevant elementary bracket now being \eqref{newsPI},
\ba
\{\dot \sigma_{z\zb}(u),J_V\}  &=& \{\dot \sigma_{z\zb}(u),J^\soft_V\}\\
&=& -\int d^2 w \left( \{\dot \sigma_{z\zb}(u), \Pi^{w w}\} \Db^3_{w} V^{w}+ c.c \right)  \\
&=&  \dot \sigma_{zz }(u) \Db_{\zb}V^z  + c.c.  = \delta_V \dot \sigma_{z\zb}(u).  \label{newszzbJsoft}
\ea
(One can show that the contribution from $J_V^\hard$ vanishes,  by the same reason that the boundary terms in \eqref{newsnewsGBMS2} do not contribute to $\{\dot \sigma_{zz},J^\hard_V\}$.)

 Let us comment on the fact, that, if one assumes  $|u| < \infty$, so that no distributions at infinity ever appear, the derivation of \eqref{finalnewsJ} would be  identical to a PB computation in the AS phase space \cite{gbms1,fprsubsub}.   From this perspective, the real novelty is the  non-trivial action on the trace-part of the news, Eq. \eqref{newszzbJsoft}. We notice that, without such trace variations, the commutator between two smooth superrotations would not close \cite{gbms1,schwarz}.

\subsection*{Derivation of Eq. \eqref{PizzJVhard}}

The PB of $\Pi^{zz}$ with the hard superrotation charge can be written as a sum of two terms,
\be \label{PiJVhard}
\{\Pi^{zz},J_V^\hard\} =\int du d^2 w \left(  \{ \Pi^{zz}, \dot \sigma^{ab}\} \delta_V \sigma_{ab} +  \dot \sigma^{ab}\{\Pi^{zz},\delta_V \sigma_{ab}\} \right) .
\ee
The only  brackets that  contribute are those involving the trace components of the shear/news tensor, Eq. \eqref{sigmazzbPiwbwb}. For the first term in \eqref{PiJVhard} this leads to
\ba
\int du d^2 w  \, \{ \Pi^{zz}, \dot \sigma^{ab}\} \delta_V \sigma_{ab} &=& 2 \int du d^2 w  \{ \Pi^{zz}, \dot \sigma^{w \wb}\} \delta_V \sigma_{w \wb} \\
&=& - 2 \int du \Db_{\zb}\Db^{-3}_z ( \dot \sigma^{\zb \zb} (D_{\zb} V^{z}  \sigma_{zz}+ D_z V^{\zb}  \sigma_{\zb\zb} )) .\label{PiJVhardone}
\ea
The second term in \eqref{PiJVhard} may seem to vanish, since $\delta_V \sigma_{w w}$ and $\delta_V \sigma_{w \wb}$ appear to depend solely on the trace-free part of the shear according, see e.g. \eqref{finalnewsJ} and \eqref{newszzbJsoft}. There is however a  hidden trace term in  $\delta_V \sigma_{w w}$ coming from the Lie derivative in \eqref{delVshearsec6},
\be
\Lie_V \sigma_{ww} = (V^c D_c+ 2 D_w V^w )\sigma_{ww} + 2 D_w V^{\wb} \sigma_{w \wb},
\ee
leading to 
\ba
\int du d^2 w \, \dot \sigma^{ab}\{\Pi^{zz},\delta_V \sigma_{ab}\} &=&  \int du d^2 w \left(  \dot \sigma^{w w}\{\Pi^{zz},\delta_V \sigma_{w w}\} +  (w \leftrightarrow \wb) \right)  \\
&=&   2 \int du d^2 w \left( D_w V^{\wb}  \dot \sigma^{w w}\{\Pi^{zz}, \sigma_{w \wb}\} + (w \leftrightarrow \wb)\right)   \\
&=& 2 \int du \Db_{\zb}\Db^{-3}_z (D_z V^{\zb}  \dot \sigma^{zz} \sigma_{zz} + D_{\zb} V^{z}  \dot \sigma^{\zb\zb} \sigma_{zz}). \label{PiJVhardtwo}
\ea
Adding \eqref{PiJVhardone} and \eqref{PiJVhardtwo} and comparing with Eq. \eqref{Abhard2} leads to Eq. \eqref{PizzJVhard}.

\section{Discussion}\label{discussionsec}

In this paper, we obtained Poisson brackets at null infinity that are compatible with a generalized BMS symmetry group, spanned by supertranslations and smooth superrotations. The presence of smooth superrotations requires treating the celestial metric as a phase space variable, along with the  news tensor. Following previous work \cite{alok,adarsh}, we  realized the phase space as a constrained system, reducing the problem to one of evaluating Dirac brackets.

The analysis involved two types of challenges related to the dependence of the fields on the sphere and on the time variable. These were addressed separately by dividing the constraints into two sets. The first set is independent of the time variable \( u \) and ensures SR covariance of Poisson brackets. The second set is sensitive to  \( |u| \to \infty \) values of the   news tensor. Their imposition leads to distributional terms at \( |u| = \infty \), whose role can be understood as follows.

It has long been appreciated \cite{stromvirasoro} that the subleading soft news is canonically conjugated to the Geroch tensor, and hence (modulo a differential operator) to the celestial metric \cite{gbms2,compere}. However, the news tensor at finite \( u \) should Poisson commute with the celestial metric. These two conditions are reconciled precisely by the distributional terms described above.

Several points deserve a more thorough understanding. Regarding the sphere dependence, there appears to be a rich geometrical structure that merits further study. For instance, the sector spanned by \((q, p, T, \Pi)\), along with the constraints \(F_1, F_2\), supports a canonical action of superrotations that may have applications beyond the present  context \cite{Donnelly:2020xgu,Freidel:2024jyf}. As for the time dependence, our treatment of distributional terms at infinity was rather formal, and a more rigorous approach would be desirable \cite{Riello:2022din, Riello:2023ptb, barnichspeziale}.  In particular, it would be important to understand  the potential ambiguities in the  prescriptions we have given.

When restricted to a single helicity sector, supertranslations and superrotations are just the first two rungs of an infinite ladder of higher 2d-spin symmetries \cite{Guevara:2021abz,stromwinf,Freidel:2021ytz,geillerwinfty}. An open question is whether there exists a generalization  of such a tower that treats both helicities on equal footing. This would presumably require the inclusion of higher spin ``edge modes'' \cite{Donnelly:2016auv,Geiller:2017xad,Speranza:2017gxd}, see \cite{javierym,javierrn,javiernagy} for similar discussion in the context of gauge theory.

There are situations of interest in which the 2d celestial space exhibits a  non-trivial topology \cite{BarnichRuzziconiWeyl,foster,Strominger:2016wns,Adjei:2019tuj,Atanasov:2021oyu}. In such cases, the Geroch tensor is no longer uniquely determined by the 2d metric and contains new, independent degrees of freedom. This leads to an extension of the phase space, even if the celestial metric is fixed (see \cite{alok} for the case of a celestial plane). A natural research direction would be to attempt a general description that is valid for celestial spaces with dynamical metrics and arbitrary topologies. 

Our treatment  only dealt with the gravitational field at null infinity, but we expect no obstacles in including other massless fields. Massive fields are however more challenging and will require additional considerations. Relatedly, what we have referred to as ``charges'' are in fact ``fluxes" over all null infinity, i.e.  difference of \emph{surface} charges \cite{zoupas,Barnich:2011mi,geiller2,Rignon-Bret:2024gcx,Ashtekar:2024stm}. A canonical realization of GBMS at time-like (and spatial) infinity would require control of such surface charges at $u=\pm \infty$. Whereas this is understood for supertranslations \cite{Campiglia:2015kxa,Troessaert:2017jcm,Henneaux:2018cst,Prabhu:2019fsp,Chakraborty:2021sbc,Capone:2022gme,Compere:2023qoa,Ti}, the case of  superrotations remains an open problem (see \cite{Fiorucci:2024ndw} for recent progress at  spatial infinity).

What may be the most serious limitation of our analysis is the too-strong \( u \)-fall-off assumption needed for a finite subleading soft news. Generic gravitational scattering exhibits \( 1/u \) tails in the shear that spoil this assumption \cite{damour,laddhasen}. A divergent subleading news implies a divergent canonical conjugate to the celestial metric. The most conservative conclusion would be that the celestial metric cannot be a dynamical variable, thus reverting us to the standard radiative space. It is, however, difficult to believe this is the end of the story, as non-trivial conservation laws can be derived from   divergent superrotation charges \cite{Agrawal:2023zea,chipum,marcalok,chipum2}.  We hope  the results of this paper  will   be relevant in  more complete treatments that incorporate these tail effects.


\acknowledgments

We would like to thank Alok Laddha for suggesting the problem, for his feedback and for encouragement. We  thank Gianni Boschetti, Federico Capone, Luca Ciambelli, Laurent Freidel, Nishant Gupta, Guzman Hernandez, Javier Peraza, Michael Reisenberger, Aldo Riello and Amit Suthar for insightful discussions. MC thanks the organizers and participants of the 2024 Workshop on Celestial Holography at Santiago de Chile for stimulating discussions. AS  thanks  the Instituto de Fisica at Facultad de Ciencias, Universidad de la Republica, for hospitality during the beginning of this project. MC is supported by PEDECIBA and ANII grant FCE-1-2023-1-175902.

\appendix

\section{Background material}\label{appA}

\subsection{Weyl scaling at null infinity} \label{weylapp}

In  Geroch's conformal approach \cite{geroch}, a central role is played by ``residual" Weyl rescalings at null infinity (see e.g. \cite{geometryandphysics}). In our notation, these correspond to 
\be \label{weylscaling}
\begin{aligned}
&q_{ab} \rightarrow e^{2 \w}q_{ab}, & \quad  & \partial_u \rightarrow e^{-\omega} \partial_u  \\
&\sigma_{ab} \rightarrow e^{ \w}\sigma_{ab}, &\quad & \partial_a \rightarrow \partial_a
\end{aligned}
\ee
where $\w=\w(x)$ is independent of $u$. We emphasize that even though $\partial_u$ has a non-trivial scaling, this version of Weyl  transformations does not move points in $\I$: The field $\sigma_{ab}$, before and after the transformation, is evaluated at the same location.  This is to be contrasted with the Weyl diffeomorphisms of the BMSW group, see subsection \ref{weyldiffapp}.

Under Weyl rescalings \eqref{weylscaling} the  fields at null infinity exhibit different transformation rules.  Weyl-covariant quantities  transform homogeneously as
\be
A \rightarrow e^{- k \omega} A , \quad \text{(Weyl-covariant tensor of weight $k$)},
\ee
where we have defined the Weyl weight in a way that matches the one given in Eq. \eqref{delVA}. Non-covariant quantities such as $T_{ab}$ and $p^{ab}$ exhibit additional pieces. Below we give a table summarizing the weights of various fields of interest.\footnote{Since $\sqrt{q}$ is a density, the action of the Lie derivative is $\Lie_V \sqrt{q} = D_a V^a \sqrt{q}$ and so $\delta_V \sqrt{q} =0$. }
\be\label{weightlist}
\begin{array}{| l | c | c | c | c | c | c | c | c | c | c | c |c | c | }
\hline 
\text{Field} & q_{ab} & q^{ab} & \sqrt{q}& \sigma_{ab}  & \dot \sigma_{ab} & C & N & \D N & \Pi^{ab}    &  T_{ab}  & p^{ab} & f & V^a  \\
\hline 
\text{weight} \, k & -2 & 2&-2& -1& 0& -1 & -1 & 3 &  2  & "0" & "4" & -1 & 0 \\
\hline 
\end{array}
\ee
The numbers in quotation marks  account for the covariant part of an otherwise non-covariant  transformation rule (see below). We also included the symmetry parameters $f$ and $V^a$ of the GBMS group. Notice that the corresponding vector fields $f \partial_u$ and $V^a \partial_a$ are Weyl-invariant.

We now discuss the Weyl invariance of  the symplectic form \eqref{Omegahardplussoft}. For the hard part, the argument is just a repeat of the AS case \cite{geometryandphysics}. The volume form changes as\footnote{The  $\sqrt{q}$ factor is kept implicit in all the integrals of the paper. It  has non-trivial contributions under Weyl scalings and when integrating by parts  Lie derivatives on the sphere. These two contributions cancel out when  integrating by parts  SR variations of SR-covariant quantities.}
\be
du d^2 x \sqrt{q} \rightarrow e^{3 \w}  du d^2 x \sqrt{q} 
\ee
i.e. it has $k=-3$. This cancels the $k=3$ weight of the integrand.

For the soft part there is again a cancellation of weights between the integrand and the area form. There are however non-covariant terms, coming from $T_{ab}$ and $p^{ab}$. As we now show, these also cancel out, ensuring the symplectic form is Weyl invariant.

Let us first discuss the transformation rules for $T_{ab}$ and $p^{ab}$. For  the Geroch tensor it is useful to use the representation \eqref{defpsi} in terms of the potential $\psi$, which transforms  as
\be \label{weylpsi}
\psi \rightarrow \psi+ \w.
\ee
Using \eqref{weylpsi}, together with the transformation rule for the covariant derivative 
\be
D_a V_b \rightarrow D_a V_b - 2 D_{\langle a} \w V_{b \rangle}
\ee
one gets\footnote{Notice that if one takes $\w =-\psi$, the transformed Geroch tensor vanishes, in accordance with the fact that $e^{-2\psi}q_{ab}$ is a round sphere metric.}
\be \label{weyltransformedgeroch}
T_{ab} \rightarrow T_{ab} - 2 ( D_{\langle a} \omega D_{b \rangle} \omega- D_{\langle a} D_{b \rangle} \omega).
\ee

To see the transformation rule for $p^{ab}$, consider its definition in terms of $\Pi^{ab}$, Eq.\eqref{defPiab}. Using
\ba
D_a V^b & \rightarrow & D_a V^b+ \delta^a_b V^c D_c \w+ D_a \w V^b-V_a D^b \w ,\\
R & \rightarrow & e^{-2 \w}(R-2 D^c D_c \w),
\ea
together with the scaling properties of the metric and $\Pi^{ab}$, one gets
\be
p^{ab} \rightarrow e^{-4 \w}\left( p^{ab} + \Delta_{\w} p^{ab} \right),
\ee
where
\be
\Delta_\w p^{ab}= D^{\langle a } \Pi^{b \rangle c}D_c \w -D_c \Pi^{c \langle a} D^{b \rangle} \w -2 \Pi^{c \langle a} D^{b \rangle} \w D_c \w + \Pi^{c \langle a} D^{b \rangle}  D_c \w + \frac{1}{2}\Pi^{ab}D^c D_c \w .
\ee

We now have all the ingredients to study the Weyl transformation of the symplectic form. It is actually simpler to look at the symplectic potential, and we focus on the   $(\Pi,T,p,q)$ part that exhibits the non-covariant terms.
Under Weyl scaling, the respective contributions from the  $(\Pi,T)$ and $(p,q)$ pairs are
\ba
\int_{S^2}   \Pi^{ab} \delta T_{ab} &\rightarrow  &\int_{S^2} \left(  \Pi^{ab} \delta T_{ab}  + \Pi^{ab } D^c \w D_c \w\delta q_{ab} + 2  \delta q_{ab} \Delta^{ab}_{\phantom{ab} cd}(\w)\Pi^{cd} \right), \label{weylPidelT}\\
\int_{S^2}   p^{ab} \delta q_{ab} &\rightarrow & \int_{S^2} \left(  p^{ab} \delta q_{ab}  + \Delta_\w p^{ab} \delta q_{ab} \right), \label{weylpdelq}
\ea
where the variation of the inhomogeneous term in \eqref{weyltransformedgeroch} were evaluated with the help of the formula \eqref{XabdeltaDaDbf} together with $- 2 \Pi^{ab}\delta( D_{\langle a} \omega D_{b \rangle} \w)=\Pi^{ab}\delta q_{ab} D^c \w D_c \w$. By writing \eqref{weylPidelT} and \eqref{weylpdelq} in holomorphic coordinates, one can show the non-covariant terms exactly cancel upon addition, thus leaving invariant the symplectic potential.

\subsubsection*{Weyl diffeomorphims}\label{weyldiffapp}
As shown in \cite{BT,wbms}, Weyl rescaling at null infinity can  be realized by large diffeomorphisms in Bondi coordinates if ones imposes  the determinant condition \eqref{detgcond} in the weaker form
\be \label{detgcondweyl}
 \partial_r (r^{-4}\det g_{ab} )=  0.
\ee
This augments the GBMS group by ``Weyl diffeomorphisms'', generated by vector field with asymptotic form 
\be
\xi_\w = 2 \w (u \partial_u - r \partial_r )+ \cdots
\ee
where $\w=\w(x)$ is an infinitesimal $u$-independent  Weyl rescaling. The analogue of \eqref{weylscaling} for the corresponding finite diffeomorphisms is
\be \label{finiteweydiff}
\begin{aligned}
&q_{ab} \rightarrow e^{2 \w}q_{ab}, & \quad  & \partial_u \rightarrow  \partial_u , \\
&\sigma_{ab}(u) \rightarrow e^{ \w}\sigma_{ab}(e^{-\omega} u), &\quad & \partial_a \rightarrow \partial_a,
\end{aligned}
\ee
where we kept implicit the sphere label $x$ as it is left unchanged. 

One can show that the symplectic structure is invariant under this version of Weyl scalings. For the soft part the argument is identical to the one given above, while for the hard part the intermediate steps are slightly different as they require a change of integration variable in the time direction. It is this version of Weyl transformations that is present in superrotations, since the latter can be thought of as a sphere diffeomorphisms followed by a Weyl scaling \eqref{finiteweydiff}  that ensures the area element on the sphere is unchanged.

\subsection{GBMS group} \label{gbmsapp}

The asymptotic diffeomorphisms preserving the Bondi form of the metric \eqref{gab} 
are generated by vector fields with an $r \to \infty$ asymptotic form
\be \label{xifxiV}
\xi_{f} =  f \partial_u + \cdots ,\quad \text{and} \quad \xi_{V} =  V^a \partial_a + \frac{1}{2} D_c V^c( u \partial_u - r \partial_r) + \cdots.
\ee
These are referred to as supertranslations (ST) and superrotations (SR) respectively. Here  $f=f(x)$ and $V^a=V^a(x)$ are arbitrary functions on the sphere. The vector field Lie brackets relations of  \eqref{xifxiV} are\footnote{Eq. \eqref{liebracketalgebragbms} is to be interpreted as valid modulo ``small" diffeomorphisms that decay to zero at null infinity. Alternatively, one can use the modified bracket of \cite{BT} for which the analogue of \eqref{liebracketalgebragbms} is valid for all values of $r$,  not just asymptotically.}
\be \label{liebracketalgebragbms}
[\xi_{f}, \xi_{f'} ]=0, \quad [\xi_V,\xi_f ] = \xi_{V(f)}, \quad   [\xi_V, \xi_{V'} ] = \xi_{[V, V']},
\ee
where  $[V,V']$ is the 2d vector field Lie bracket and
\be
V(f) := V^a \partial_a f - \frac{1}{2}D_a V^a f.
\ee
Relations \eqref{liebracketalgebragbms} define the (infinitesimal) generalized BMS group (GBMS),
\be
\gbms = \diff(S^2) \ltimes ST,
\ee
where the $\diff(S^2)$ factor is generated by $\xi_V$ and $ST$  is the Abelian group of supertranslations, generated by $\xi_f$.

The action of \eqref{xifxiV}  in the asymptotic metric \eqref{gab} implies the transformations rules
\be
\begin{aligned} \label{eq:gBMS_var}
\delta_V q_{ab} & = \Lie_V q_{ab} - D_c V^c q_{ab} = 2 D_{\langle a} V_{b \rangle }, &  \delta_f q_{ab} & =0  \\
\delta_V  T_{ab} & = \Lie_V T_{ab} - D_{\langle a} D_{b \rangle } D_c V^c ,    &  \delta_f T_{ab} & =0 \\
 \delta_V \Chat_{ab} & =  \Lie_V \Chat_{ab}  + \frac{1}{2}D_c V^c (u \partial_u -1) \Chat_{ab} ,   &  \delta_f \Chat_{ab} & =  f \partial_u \Chat_{ab} + (-2 D_{\langle a} D_{b \rangle}  +  T_{ab})f ,
\end{aligned}
\ee
with the corresponding algebra of variations reproducing the algebra \eqref{liebracketalgebragbms} \cite{gbms2}.

Given \eqref{eq:gBMS_var},  one can derive the transformation rules for the different quantities that were defined in section \ref{hardsoftsplit}. In particular, under supertranslations one finds
\be
\begin{aligned}
\delta_f \sigma_{ab} & = f \dot \sigma_{ab}, & \delta_f C &= f, & \delta_f N & =0 \\
\delta_f \Ncalone_{ab} &= - f \Ncalzero_{ab}, & \delta_f \Pi^{ab}& =0, & \delta_f p^{ab}& =0.
\end{aligned}
\ee

For superrotations, $\sigma_{ab}$ transforms in the same way as $\Chat_{ab}$,
\be \label{eq:delVsigma}
\delta_V \sigma_{ab} = \left(  \Lie_V  + \frac{1}{2}D_c V^c (u \partial_u -1) \right) \sigma_{ab},   
\ee
while the remaining $u$-independent quantities transform as in \eqref{delVA} (except for $T_{ab}$ and $p^{ab}$ which exhibit  non-covariant pieces), with weights given by \eqref{weightlist}.

The standard  BMS group arises as the subgroup of GBMS  that leaves invariant a given 2d metric $q_{ab}$,
\be
\bms(q_{ab}) = \conf(q_{ab}) \ltimes ST,
\ee
where $\conf(q_{ab})$ is the group of conformal isometries of $q_{ab}$,  generated by 2d vector fields satisfying $\delta_V q_{ab} =0$. Finally, the translation subgroup of BMS is generated by ST vector fields $\xi_f$ satisfying
\be \label{deftranslationsqab}
\text{Translations}(q_{ab}) := \{ \xi_f :  (-2 D_{\langle a} D_{b \rangle}  +  T_{ab})f =0\} \subset ST \subset \bms(q_{ab}) .
\ee

\subsection{Symplectic form} \label{simplifiedOmegaapp}
In this section we review the symplectic structure \eqref{Omegaintro} introduced in \cite{javier} and show how it can be brought into the form \eqref{Omegahardplussoft} upon isolating the constant-in-$u$ mode of the shear. 

We will  phrase the discussion in terms of the symplectic potential.  After isolating a constant-in-$u$ mode of the shear as in Eq. \eqref{gonehatsigmaC} the  potential splits as in \eqref{Omegahardplussoft}, 
\be
\theta= \theta^\hard+\theta^\soft
\ee
where $\theta^\hard$  is the potential for $\Omega^{\hard}$ (given by the first term of  \eqref{Omegahardplussoft}) and $\theta^\soft $ is a sum of two terms as follows.
\be
\theta^\soft = \theta_0+ \theta_1,
\ee
where
\be
\theta_0 := - \int_{S^2} C^{ab} \delta N_{ab} ,
\ee
is the ``soft" contribution of the AS symplectic potential, after isolating the zero mode of the shear (that is, the potential for the second term in  \eqref{OmegaASsplit}). $\theta_1$ is the ``extra'' contribution found in \cite{javier}, which, in the notation of that reference reads\footnote{Using $N=\hat C^+-\hat C^-$ and $C=(\hat C^++\hat C^-)/2$ we recover the expression from \cite{javier}. If one uses a definition of $C$ that is not symmetric under $\hat C^+ \leftrightarrow \hat C^-$ (corresponding to an asymmetric boundary condition in  \eqref{bdycondsigma}), one gets an extra term in \eqref{deftheta1} that is quadratic in  $N$. This  leads to additional terms in \eqref{defPi1} and \eqref{defp1} that modify the relation Eq. \eqref{defpab} 
by an additive piece that is quadratic in $N$.} 
\be \label{deftheta1}
\theta_1: =  \int_{S^2}\left( \None^{ab} \sone_{ab}(\delta) - \frac{1}{2} \left[ N^{ab} (\delta \szero(C)_{ab} -\szero(\delta C)_{ab}-   C \sone_{ab}(\delta)) + (N \leftrightarrow C) \right] \right)
\ee
with
\be \label{defszero}
\szero(C)_{ab}  = -2 D_{\langle a} D_{b \rangle} C  + C T_{ab} 
\ee
\be
\sone_{ab}(\delta) =   2 \delta T_{ab} +  D_{(a} D^c \delta q_{b)c}- \frac{R}{2}\delta q_{ab} 
\ee
\be  \label{defCNscalarsapp}
C_{ab} = \szero(C)_{ab}, \quad N_{ab}= \szero(N)_{ab}.
\ee
Here  $C_{ab}$ is the zero mode of the shear defined in \eqref{gonehatsigmaC} while $N_{ab} \equiv  \Ncalzero_{ab}$ , $\None^{ab} \equiv \Ncalone^{ab}$ are the leading and  subleading soft news, defined in Eqs. \eqref{defNcalzero} and  \eqref{defNcalone} respectively. The scalars $C$ and $N$ are defined exactly as in \eqref{defCNscalars}, but in the notation of \cite{javier} we write them  as in \eqref{defCNscalarsapp}. Note that $\theta_{\as}=\theta^\hard+\theta_0$ is the potential for the AS symplectic form.

We would like to  express $\theta^\soft = \theta_0+ \theta_1$ in a way that the terms proportional to  $\delta T_{ab}$ and $\delta q_{ab}$ appear with no derivatives.  

For $\theta_1$ this leads to 
\be
\theta_1 = \int_{S^2}\left( \Pi^{ab}_1 \delta T_{ab} + p^{ab}_1 \delta q_{ab} \right)
\ee
with\footnote{These are the momenta appearing in Eq. \eqref{Omegaintro}}
\ba
\Pi^{ab}_1 &=& 2 \None^{ab} + \frac{1}{2} N C^{ab}+ \frac{1}{2} C N^{ab},\label{defPi1} \\
p^{ab}_1 &=& \O^{ab}_{\phantom{ab}cd} (\None^{cd} + \frac{1}{2} N C^{cd}+ \frac{1}{2} C N^{cd}) + \Delta^{ab}_{\phantom{ab} cd}(C)N^{cd} + \Delta^{ab}_{\phantom{ab} cd}(N)C^{cd}, \label{defp1}
\ea
where $ \O^{ab}_{\phantom{ab}cd}$ and $\Delta^{ab}_{\phantom{ab} cd}$ are the differential operators  defined in Eqs. \eqref{defOop} and \eqref{defDelta} respectively.

For $\theta_0$ there is certain freedom in how one chooses to write derivatives acting on  $C$ or $N$, see for example Eqs. (4.38) and (4.39) in \cite{alok}. We choose the form corresponding to \eqref{defOmegasoft}, which yields, up to total variation terms:
\be
\theta_0 = \int_{S^2}\left(- C \delta (\D N) +   \Pi^{ab}_0 \delta T_{ab} + p^{ab}_0 \delta q_{ab} \right),
\ee
with
\be
\Pi^{ab}_0 = C N^{ab}, \quad  p^{ab}_0 = - 2 \Delta^{ab}_{\phantom{ab} cd}(C)N^{cd}.
\ee

The total coefficient of $\delta T_{ab}$ in $\theta^\soft$ can then be written as:
\be  \label{Pi1Pi2}
 \Pi^{ab}_0 + \Pi^{ab}_1  =  2(\None^{ab} + CN^{ab}) + D^{\langle a} X^{b \rangle } 
\ee
where
\be
X^a :=  C D^a N- N D^a C.
\ee
The first term in \eqref{Pi1Pi2} is invariant under supertranslations. The second term can be moved into a contribution to $p^{ab}$ as follows: 
\ba
D^{\langle b} X^{a \rangle } \delta T_{ab} & = & - X^a D^b \delta T_{ab} - \frac{1}{2}D \cdot X q^{ab} \delta T_{ab} + (\text{total derivative})\\
& = &- X^a \delta( D^{b} T_{ab}) + X^a \delta(D^b) T_{ab} -  \frac{1}{2}D \cdot X T^{ab} \delta q_{ab}  \label{XDivT}
\ea
Using the equality  $D^b  T_{ab}= -\partial_a R/2$, the first term in \eqref{XDivT} can be written as
\be
- X^a \delta( D^{b} T_{ab})  =  \frac{1}{2}  X^a \delta \partial_a R   =  - \frac{1}{2} D \cdot X \delta R   +  (\text{total derivative})
\ee
Collecting all terms, one then has
\be
D^{\langle b} X^{a \rangle} \delta T_{ab} = \tilde{p}^{ab} \delta q_{ab},
\ee
where 
\be
\tilde{p}^{zz} = -\frac{1}{2}D^{z}D^z D \cdot X + \frac{1}{2}(D_{\zb}X^{\zb})T^{zz} + \frac{1}{2}X^a D_a T^{zz}.
\ee

We finally discuss the coefficient of $\delta q_{ab}$ in $\theta^\soft$. The contributions from $\theta_0$ and $\theta_1$ can be written as
\be \label{p0p1}
p_0^{ab}+p_1^{ab}= \O^{ab}_{\phantom{ab}cd} (\None^{cd} + CN^{cd}) + \left[ \frac{1}{2} \O^{ab}_{\phantom{ab}cd} (N C^{cd}) +  \Delta^{ab}_{\phantom{ab} cd}(N)C^{cd}  - (C \leftrightarrow N) \right]
\ee
where again we have isolated a supertranslation invariant part. The claim now is that the reminder, in square brackets, is nothing but $-\tilde{p}^{ab}$ so that it goes away once we add that contribution. In other words, if write
\be
p^{ab}=p_0^{ab}+p_1^{ab} + \tilde{p}^{ab} = \O^{ab}_{\phantom{ab}cd} (\None^{cd} + CN^{cd}) + \beta(C,N),
\ee
where $\beta(C,N)$ is the bilinear obtained by adding the square bracket in \eqref{p0p1} to $\tilde{p}^{ab}$, then $\beta(C,N)=0$. This vanishing can be verified by explicit computation. Alternatively, we can provide an abstract argument  as follows.

Since $\Omega^\soft(\delta, \delta_f)= \delta \int_{S^2}  f  \D N$ (see appendix \ref{gbmschargesapp}) one should have
\be \label{delfPidelfp}
\int_{S^2}( \delta_f \Pi^{ab} \delta T_{ab} + \delta_f p^{ab} \delta q_{ab})=0.
\ee
For $\Pi^{ab}$ defined by the first term in \eqref{Pi1Pi2} one has $\delta_f \Pi^{ab}=0$. Then the corresponding $p^{ab}$ should satisfy $\delta_f p^{ab}=0$ for \eqref{delfPidelfp} to hold. But
\be
\delta_f p^{ab} = \beta(f,N),
\ee
so that $\delta_f p^{ab}=0 \implies \beta=0$.

Summarizing, the soft symplectic potential can be written as
\be
\theta^\soft = \int_{S^2}\left(- C \delta (\D N)+ \Pi^{ab} \delta T_{ab} + p^{ab} \delta q_{ab} \right)
\ee
with
\ba
\Pi^{ab} & = & 2(\None^{ab} + CN^{ab}), \label{defPiapp} , \\
 p^{ab} & = & \O^{ab}_{\phantom{ab}cd} (\None^{cd} + CN^{cd}) \label{defpapp}.
\ea
This precisely leads to the soft symplectic form  presented in Eq. \eqref{defOmegasoft}. 
The total symplectic potential $\theta= \theta^\hard+\theta^\soft$ then leads to the total symplectic form \eqref{Omegahardplussoft}.

\subsection{GBMS charges} \label{gbmschargesapp}
In this appendix we review the form of supertranslation  and superrotation  charges.
We start by recalling the argument  \cite{javier} as to why STs and SRs act canonically with respect to  the symplectic form \eqref{Omegahardplussoft}.  The idea is to work with a symplectic potential that is invariant under STs and SRs, i.e. that it satisfies (see e.g. \cite{abr})
\be
\begin{aligned}
\delta_f \theta(\delta)+ \theta([\delta,\delta_f]) &=0 ,\\
  \delta_V \theta(\delta)+ \theta([\delta,\delta_V]) &=0.
\end{aligned}
\ee
A symplectic potential for \eqref{Omegahardplussoft} satisfying this condition is given by\footnote{Invariance under SRs follows from the SR invariance of the integrands (for which one needs to take into account the cancellation of non-covariant terms from $\delta T_{ab}$ and $p^{ab}$  discussed in section \ref{weylapp}).  ST invariance is verified as follows: The hard term yields a total derivative that vanishes while for the soft part  the only potential contribution comes from  $\delta C$. This leads  a term proportional to $\delta f$ which vanishes, as long as we keep the symmetry parameters to be field independent.}
\be
\theta = \int_{\I}  \dot \sigma^{ab}  \delta \sigma_{ab}   + \int_{S^2} \left( \D N \delta C +  \Pi^{ab} \delta T_{ab} +  p^{ab} \delta q_{ab} \right). \label{thetagbms}
\ee
The corresponding canonical charges are then simply given by
\be
P_f = \theta(\delta_f), \quad  J_V = \theta(\delta_V) .
\ee
Using the GBMS transformation rules given in section \ref{gbmsapp}, this leads to 
\ba
P_f& = & \int_{\I} f \dot \sigma^{ab} \dot \sigma_{ab} +  \int_{S^2} f \D N \\
J_V &=&  \int_{\I} \dot \sigma^{ab} \delta_V \sigma_{ab} +  \int_{S^2} \left( \D N \delta_V C - \Pi^{ab} \Sb_{ab \, c} V^c  \right) ,
\ea
where for $J_V$ we did an integration by parts and used Eq. \eqref{defSboneVv2} to get the following simplification:
\ba
\int_{S^2} \left( \Pi^{ab} \delta_V T_{ab} +  p^{ab} \delta_V q_{ab} \right) &=& \int_{S^2}  \Pi^{ab}(\delta_V T_{ab}+\frac{1}{2}  \O_{ab}^{\phantom{ab} cd} \delta_V q_{cd}) \\
&=&  - \int_{S^2} \ \Pi^{ab} \Sb_{ab \, c} V^c  .
\ea

\section{Differential operators and Green's functions} \label{diffopssec}
In this section we describe  differential operators, Green's functions, and identities among them, that are used throughout the paper.

\subsection{$\Delta^{ab}_{\phantom{ab} cd}$ and variations of $\Db_{\langle a}\Db_{b \rangle} $ and $\D$ } \label{variationsDDapp}
Given a scalar $f$ and a STF tensor $X^{ab}$, we define the differential operator $\Delta^{ab}_{\phantom{ab} cd}$
\be 
(f,X^{ab}) \mapsto \Delta^{ab}_{\phantom{ab} cd}(f)X^{cd}
\ee
 by the condition
\be \label{XabdeltaDaDbf}
\int_{S^2} X^{ab} \delta( D_{\langle a} D_{b \rangle} )f =  \int_{S^2} \delta q_{ab} \Delta^{ab}_{\phantom{ab} cd}(f)X^{cd} , 
\ee
or equivalently by,\footnote{Here and in the following we keep implicit a $1/\sqrt{q}$ factor in front of functional derivatives.}
\be
\Delta^{ab}_{\phantom{ab} cd}(f)X^{cd}= \frac{\delta}{\delta q_{\langle ab \rangle}}\int_{S^2} X^{cd} D_{\langle c} D_{d \rangle} f,
\ee
(provided  $X^{ab}$ and $f$ are  independent of $q_{ab}$, save for a trace-free condition on $X^{ab}$). Its explicit form is given by
\be \label{defDelta}
\Delta^{ab}_{\phantom{ab} cd}(f)X^{cd} = D_c D^{\langle a} f X^{b\rangle c}+D^{\langle a} f D_c X^{b \rangle c } - \frac{1}{2}D^c f D_c X^{ab} - D^2 f X^{ab} .
\ee
One can check that it is a symmetric (or self-adjoint) operator, in the sense of\footnote{This property can be  understood by writing $D^{\langle a} D^{b \rangle}f = \delta/\delta q_{\langle ab \rangle}  I[f]$ with $I[f]=-\tfrac{1}{2}\int D^a f D_a f$, thus realising  $\Delta_{ab cd}(f)$ as a second functional derivative. }
\be \label{symDelta}
\int_{S^2} Y^{ab} \Delta_{ab cd}(f)X^{cd} = \int_{S^2} X^{ab} \Delta_{ab cd}(f)Y^{cd} ,
\ee
where we recall that indices are raised and lowered with the 2d metric $q_{ab}$. 

The above formulas can be used to evaluate the variations of the  SR-covariant operators
\begin{align}  \label{minus2Db2}
f \mapsto f_{ab} & :=-2 \Db_{\langle a}\Db_{b \rangle} f ,\\
f \mapsto \D f & := 4 \Db^{\langle a} \Db^{b \rangle}  \Db_{\langle a} \Db_{b \rangle} f, \label{defDcalf}
\end{align}
where  $f$ and $X^{ab}$ are now assumed  to have Weyl weights $-1$ and $3$ respectively. The variation of $\Db_{\langle a}\Db_{b \rangle}$ is just  given by \eqref{XabdeltaDaDbf} plus a term proportional to $\delta T_{ab}$. Concatenating this result, one finds the variation of $\D f$ can be expressed as
\be \label{delD4}
\int_{S^2} f \delta(\D) g =  \int_{S^2} \left( -2 ( \Delta^{ab}_{\phantom{ab} cd}(g)f^{cd} + \Delta^{ab}_{\phantom{ab} cd}(f)g^{cd} )\delta q_{ab}  + (g f^{ab}+ f g^{ab}) \delta T_{ab}\right),
\ee
where  $f_{ab} \equiv -2 \Db_{\langle a}\Db_{b \rangle}f$ and $g_{ab} \equiv -2 \Db_{\langle a}\Db_{b \rangle} g$.

\subsection{$\O^{ab}_{\phantom{ab} cd} $}

We now turn attention to the differential operator introduced in Eq.  \eqref{defpab}. Given a STF  tensor $X^{ab}$, we define
\be \label{defOop}
\O^{ab}_{\phantom{ab} cd} X^{cd} = D^{\langle a} D_c X^{b \rangle c} - \frac{R}{2} X^{\langle ab \rangle} .
\ee
Its adjoint $\Ot$, defined by the condition,
\be
\int_{S^2} \gamma_{ab}\O^{ab}_{\phantom{ab} cd}  X^{cd} =  \int_{S^2} X^{ab}\Ot^{\phantom{ab} cd}_{ab}\gamma_{cd}
\ee
is given  by
\be \label{defOtop}
\Ot^{\phantom{ab} cd}_{ab} \gamma_{cd} = D_{\langle a} D^c \gamma_{b \rangle c}- \frac{R}{2}\gamma_{\langle ab \rangle} ,
\ee
and satisfies,
\be \label{Oselfadj}
\Ot^{\phantom{ab} cd}_{ab} = \O_{ab}^{\phantom{ab} cd} ,
\ee
i.e., $\O$ is self-adjoint.

\subsection{$\Sb_{ab \, c}$} \label{Ssec}

Consider the differential operator \cite{gbms2} 
\be \label{defSoneV}
S_{ab \, c}V^c  :=  D_{\langle a} D_{b \rangle} D_c V^c-  \O_{ab}^{\phantom{ab} cd} D_c V_d ,
\ee
which, in holomorphic coordinates takes the familiar form,
\be \label{SzzcVz}
S_{zz \, c}V^c = D_z^3 V^z,
\ee
that  features in the soft contribution of the SR charge \cite{Barnich:2011mi,stromvirasoro}.  Its SR covariant version (acting on $k=0$ vectors) is  given by \cite{javier}
\be \label{defSboneV}
\Sb_{ab \, c}V^c  =   S_{ab \, c}V^c -\Lie_V T_{\langle ab \rangle},
\ee
where the trace-free part is being taken after the Lie derivative. Using the SR transformation rules for $T_{ab}$ and $q_{ab}$ (see \eqref{eq:gBMS_var}), the  operator can alternatively be written as \cite{javier}
\be \label{defSboneVv2}
\Sb_{ab \, c}V^c  =  -\delta_V T_{\langle ab\rangle}-\frac{1}{2}  \O_{ab}^{\phantom{ab} cd} \delta_V q_{cd}.
\ee
When expressed in holomorphic coordinates, it  satisfies  the SR covariant version of \eqref{SzzcVz}
\be \label{Szz}
\Sb_{zz \, c}V^c  = \Db^3_z V^z.
\ee

We denote by $\Sbt_{a \, b  c}$ the adjoint of \eqref{defSboneV}. It maps $k=2$ STF tensors to $k=2$ covectors. In holomorphic coordinates, it is simply given by
\be
\Sbt_{z \, b  c} X^{bc} = -\Db^3_z X^{zz}.
\ee

\subsection{Green's function for $\D$} \label{greenfnapp}
We denote by  $\Gzero_x(x')$ the Green's function of $\D$, that is, 
\be \label{defGzero}
g = \D f \implies  f(x) = \int d^2 x' \Gzero_x(x') g(x')  \mod \ker \D,
\ee
where we have indicated that the inverse of $\D$ is  defined modulo its kernel.\footnote{It is easy to verify that this kernel coincides with that of $\Db_{\langle a } \Db_{b \rangle}$, the latter being given by ``pure translations'', Eq. \eqref{deftranslationsqab}.}

Since $\tilde{\D} = \D$ it follows that $\Gzero_x(x')=\Gzero_{x'}(x)$.  We shall occasionally write \eqref{defGzero} in the compact form
\be \label{defGzeroabstract}
g = \D f \implies  f = \Gzero g  \mod \ker \D,
\ee
where $\Gzero$ is now regarded as an  operator. 

As in \eqref{minus2Db2} and \eqref{delD4} we use the notation
\be \label{defGab}
\Gzero^{ab}_{x'}(x) =-  2 \Db^{\langle a} \Db^{ b \rangle}  \Gzero_{x'}(x),
\ee
where derivatives are taken with respect  to the $x$ variable. The expression can be interpreted as the Green's function for the differential operator $- 2 \Db^{\langle a} \Db^{ b \rangle} $.

The above expressions display certain simplifications when written in holomorphic coordinates. We first note that $ \Db^2_{z}$ commutes with $\Db^2_{\zb}$ (thanks to Eq. \eqref{DbRbzero}). The operator $\D$  can then be written in the following two ways
\be
\D = 8 q^{z \zb}q^{z \zb} \Db^{2}_{\zb} \Db^2_{z} = 8 q^{z \zb}q^{z \zb} \Db^{2}_{z} \Db^2_{\zb}.
\ee
Likewise,   $\Gzero$ can be written as
\be\label{Gzerohol}
\Gzero=\frac{1}{8}q_{z \zb}q_{z \zb} \Db^{-2}_{\zb} \Db^{-2}_{z}  = \frac{1}{8}q_{z \zb}q_{z \zb} \Db^{-2}_{z} \Db^{-2}_{\zb} .
\ee
This in turn allows for the following compact form for $\Gzero^{zz}$,
\be \label{Gzerozz}
\Gzero^{zz}\equiv -2 q^{z \zb}q^{z \zb} \Db^{2}_{\zb}\Gzero =-\frac{1}{4} \Db^{-2}_{z}.
\ee

\subsection{Green's function for $\Sb_{ab \, c}$} \label{greenSsec}

We now discuss the  Green's function for the operator $\Sb_{ab \, c}$: 
\be \label{defGone}
Y_{ab} =  \Sb_{ab \, c}V^c   \implies V^a(x) = \int d^2 x' \, \Gone^{a \, b'c'}(x,x') Y_{b' c'}(x') \mod \ckv ,
\ee
where we used the fact that the kernel of $\Sb_{ab \, c}$ is given by global conformal Killing vector fields (CKVs) on the sphere \cite{Barnich:2011mi,stromvirasoro}, that is, $V^c$  satisfying $D_{\langle a} V_{ b \rangle}=0$.
For the adjoint operator $\Sbt_{a \, bc}$ we have,
\be \label{Gonetilde}
Y_a =  \Sbt_{a \, bc} X^{bc} \implies X^{ab}(x)= \int d^2 x' \, \Gbt^{ab \, c'}(x,x') Y_{a'}(x').
\ee
In this case there is no indeterminacy in inverting the differential operator since its  kernel is trivial.\footnote{This can be seen by performing a spherical harmonic decomposition of the electric and magnetic part of the tensor $X^{ab}$.} The two Green's functions are related by
\be
 \Gbt^{ab \, c'}(x,x') =  \Gone^{a' \, ab}(x',x) .
\ee

We  write \eqref{defGone} and \eqref{Gonetilde} in abstract operator notation as
\be \label{operatorGone}
\begin{aligned}
Y_{ab}&  =  \Sb_{ab \, c}V^c   \implies  V^a = \Gb^{a \, bc} Y_{bc}   \quad \mod \ckv , \\
Y_a & =  \Sbt_{a \, bc} X^{bc} \implies  X^{ab} =  \Gbt^{ab \, c} Y_{c}  .
\end{aligned}
\ee

In holomorphic coordinates,  the operators \eqref{operatorGone} can be written as
\be
\begin{aligned}
\Gb^{z \, bc} Y_{bc} = \Db^{-3}_z Y_{zz}  , \\
 \Gbt^{zz \, c} Y_{c}   =-\Db^{-3}_z Y_{z}  .
\end{aligned}
\ee

\subsection{$\A^{abcd}$} \label{Aoperatorapp}
Given three $k=-2$ STF tensors $X_{ab}, Y_{ab}$ and $Z_{ab}$, let us consider the operator associated to an antisymmetric variation of $\O$,
\be \label{abstractdefAsoft1}
 A_\soft[Z,Y,X]:=  2 \int_{S^2} X_{ab}  \frac{\delta}{\delta q_{\langle a b \rangle}}  \O[Y,Z] - (Z \leftrightarrow X) 
\ee
where 
\be
\O[Y,X]  = \int_{S^2} Y_{ab} \O^{ab cd}X_{cd} .
\ee
The symmetry property  $\O[Y,X]=\O[X,Y]$ \eqref{Oselfadj}, implies $A_\soft$ satisfies a Jacobi-type identity
\be \label{jacobiAsoft}
A_\soft[X,Y,Z]+A_\soft[Y,Z,X] + A_\soft[Z,X,Y]=0.
\ee

By factoring out $Z$ (or $X$)  we can define a $Y$-dependent differential operator $A_\soft^{abcd}(Y)$ by the condition
\be \label{abstractdefAsoft2}
 A_\soft[Z,Y,X]=\int_{S^2} Z_{ab} A_\soft^{abcd}(Y) X_{cd}.
\ee
Its explicit expression in holomorphic coordinates is computed in Eq. \eqref{defAsoft} for $Y_{ab}=\Pi_{ab}$. 

It turns out this operator can be made SR covariant by a simple additive term:
\be \label{abstractdefAsoftcov}
\A_\soft^{abcd}(Y)X_{cd} = A_\soft^{abcd}(Y) X_{cd}  - 2 T^{cd} Y_{cd}  X^{ a b }  +    2 T^{ a b }  Y^{cd}  X_{cd} .
\ee

The operator $\A^{abcd}$ is finally defined by the addition of a ``hard'' piece
\be \label{abstractdefAtotal}
\A^{abcd}(Y) = \A^{abcd}_\soft(Y) + 2 \int du  (\sigma^{ab} \dot \sigma^{cd}-  \dot \sigma^{ab} \sigma^{cd}).
\ee

The kinematical PBs of the constraint $F_1$ with itself features this operator with $Y_{ab}=\Pi_{ab}$, in which case we omit the $\Pi$ label. We notice the terms added to $A_\soft$ in \eqref{abstractdefAsoftcov} and \eqref{abstractdefAtotal} are compatible with \eqref{jacobiAsoft} so that  $\A_\soft$ and $\A$ also satisfy an   identity of the type \eqref{jacobiAsoft}.

\subsection{$\Bb^{ab}_{\phantom{ab} cd}$}\label{Bopappendix}

Given a $k=-1$ scalar $f$ and a $k=3$ STF tensor $X^{ab}$, we define the map
\be \label{defBabst}
\Bb^{ab}_{\phantom{ab} cd}(f)X^{cd}:= 2 \Delta^{ab}_{\phantom{ab} cd}(f)X^{cd} + \frac{1}{2} \O^{ab}_{\phantom{ab} cd}( f X^{cd} ).
\ee
As indicated by the notation, one can show the resulting expression is SR covariant. The map is related to variations of the differential operator $\Db_{\langle a} \Db_{b \rangle}$ on the physical phase space  and features in the kinematical PB between $N$ and $F_1$ (see  footnote \ref{defBbfootnote}).

A useful identity that can be proven by going to holomorphic coordinates is 
\be \label{antisymB3rdderivativecov}
\Bb^{ab}_{\phantom{ab} cd}(f) g^{cd}- \Bb^{ab}_{\phantom{ab} cd}(g) f^{cd} = \Sb^{ab}_{\phantom{ab } c}(  g \Db^c f - f \Db^c g ),
\ee
where  $f_{ab} \equiv -2 \Db_{\langle a}\Db_{b \rangle}f$ and $g_{ab} \equiv -2 \Db_{\langle a}\Db_{b \rangle} g$.

\subsection{Simplifications in holomorphic coordinates}\label{holodiffopsapp}
One of the simplifications that  occur in holomorphic coordinates is the decoupling of holomorphic/antiholomorphic components of (integro) differential operators,  as already seen in Eq. \eqref{Szz} for  $\Sb_{ab \, c}$. In particular, for the differential operators with four indices discussed before, one has:
\be  \label{scalarBb}
\begin{aligned}
\Delta^{zz}_{\phantom{zz} cd}(f)X^{cd} & = \Delta^{zz}_{\phantom{zz} zz}(f)X^{zz} &=: &  \ \Delta(f) X^{zz} \\
\O^{zz}_{\phantom{ab} cd} X^{zz}  & = \O^{zz}_{\phantom{zz} zz} X^{zz}    &=: & \  \O X^{zz} \\
\A^{\zb \zb cd}(Y) X_{cd} &= \A^{\zb \zb zz}(Y) X_{zz} &=: & \ \A(Y)X^{\zb\zb}   \\
\Bb^{zz}_{\phantom{zz}cd}(f)X^{cd} & = \Bb^{zz}_{\phantom{zz}zz}(f)X^{zz} &=: & \ \Bb(f) X^{zz} , 
\end{aligned}
\ee
where  $\Delta(f)$,  $\O$ ,  $\A(Y)$ and $\Bb(f)$ are scalar complex operators. In terms of these, \eqref{defBabst} reads
\be
\Bb(f) X^{zz} = 2 \Delta(f) X^{zz} + \frac{1}{2}\O(f X^{zz}),
\ee
while Eq. \eqref{defSboneVv2} becomes
\be \label{Dz3delVTbarO}
\Db_z^3 V^z  =  -\delta_V T_{zz}- \bar  \O D_z V_z.
\ee
Similar decoupling occurs for the four-index operators  $\K^{abcd}$ and $\L^{abcd}$ that appear in the second stage Dirac matrix,
\be \label{AKL}
\begin{aligned}
\K^{\zb \zb cd} X_{cd} &=:  \K X^{\zb\zb}   ,\\
\L^{\zb \zb cd} X_{cd} &=: \L X^{\zb\zb}   ,
\end{aligned}
\ee
 which are therefore captured by complex scalar operators  $\K$ and $\L$ as in \eqref{scalarBb}.


Additional simplifications occur with   holomorphic/antiholomorphic components of the SR covariant derivative: From \eqref{DbRbzero} one can show
\be  \label{DzD3zbcomm}
[\Db^2_z , \Db^2_{\zb}]  = 0, \quad \quad [\Db_z , \Db^3_{\zb}]  =  0.
\ee

We conclude by describing how  the adjointness properties of the tensorial operators get reflected in adjointness properties of their complex scalar counterparts. We have noted  that $\Delta^{ab}_{\phantom{ab} cd}$ and $\O^{ab}_{\phantom{ab} cd}$ are self-adjoint, whereas $\A^{ab cd}$ and $\K^{ab cd}$ are anti-selfadjoint. For the complex scalar operators, the notion of adjoint includes, in addition to an integration by parts (indicated with a tilde in  \eqref{Oselfadj} and elsewhere in the text), a complex conjugation. Denoting the combined action by a dagger, we then have
\be 
\Delta^\dagger = \Delta, \quad \O^\dagger = \O, \quad  \A^\dagger = -\A, \quad \K^\dagger  = - \K .
\ee

\section{Kinematical HVFs}\label{kinHVFsapp}
Let us denote by  $\Omega_{\kin, 0}$  the symplectic form on $\Gamma_{\kin, 0}$,
\be \label{Omegakinzero}
\Omega_{\kin, 0} = \int_{\I} \delta \dot \sigma^{ab} \wedge \delta \sigma_{ab}   + \int_{S^2} \left( \delta (\D N) \wedge \delta C + \delta \Pi^{ab} \wedge \delta T_{ab} + \delta p^{ab} \wedge \delta q_{ab} \right).
\ee
It has the same form as $\Omega$ \eqref{Omegahardplussoft}, except  that the constraints \eqref{defF1toF6}  are not being imposed. The HVF $ \{ \cdot, \varphi \}_0$ of a phase space function $\varphi$ is defined by the relation
\be
\Omega_{\kin, 0}(\delta, \{ \cdot, \varphi \}_0) = \delta \varphi. \label{defHVF0}
\ee
We present below the solution to \eqref{defHVF0} for all the variables appearing in \eqref{Omegakinzero}.\footnote{We regard  $\dot \sigma^{ab} $, rather than $ \sigma^{ab} $ as the ``fundamental" hard variable since the latter does not admit a HVF.  See appendix A of \cite{gbms1} and subsection \ref{utoinfapp} for further details.} 
\begin{align}
\{ \cdot, q_{ab} \}_0 & = -\frac{\delta}{\delta p^{\langle a b \rangle}} \\
\{ \cdot, T_{ab} \}_0 &= - \frac{\delta}{\delta \Pi^{\langle a b \rangle}} - \frac{1}{2}q_{ab} T^{cd} \frac{\delta}{\delta p^{ c d }} \label{hvfTabzero}\\
\{\cdot , \Pi^{ab} \}_0  &=  \frac{\delta}{\delta T_{\langle a b \rangle}} +\frac{1}{2}q^{ab} \Pi^{cd} \frac{\delta}{\delta p^{ c d }}    -\int d^2 x'  (N \Gzero_{x'}^{ab}+ \Gzero_{x'} N^{ab}) \frac{\delta}{\delta N(x')} \label{hvfPiabzero} \\
\{ \cdot, \dot \sigma^{ab} \}_0 & = \frac{1}{2}\frac{\delta}{\delta \sigma_{\langle a b \rangle}} + \frac{1}{2}q^{ab} \dot \sigma^{cd} \frac{\delta}{\delta p^{ c d }} \label{hvfsigmadotzero} 
\end{align}
\begin{multline} \label{hvfpabzero}
\{\cdot , p^{ab} \}_0  =  \frac{\delta}{\delta q_{\langle a b \rangle}} +\frac{1}{2}T^{ab} q_{cd} \frac{\delta}{\delta T_{ c d }}  -\frac{1}{2}\Pi^{ab} q^{cd} \frac{\delta}{\delta \Pi^{ c d }} +\frac{1}{2} \int du \sigma^{ab} q_{cd}\frac{\delta}{\delta \sigma_{ c d }} \\
+\frac{1}{2}\left( (q^{ab} p^{cd} - p^{ab}q^{cd}) + (\Pi^{ab} T^{cd} - T^{ab}\Pi^{cd}) + \int du(\dot \sigma^{ab} \sigma^{cd} - \sigma^{ab} \dot \sigma^{cd}) \right)\frac{\delta}{\delta p^{ c d }} \\
+  2 \int d^2 x' (   \Delta^{ab}_{\phantom{ab} cd}(N)\Gzero_{x'}^{cd} + \Delta^{ab}_{\phantom{ab} cd}(\Gzero_{x'})N^{cd} )\frac{\delta}{\delta N(x')} 
 \end{multline}

\begin{multline}  \label{hvfNzero}
\{ \cdot,   N(x) \}_0=  \int d^2 x' \Gzero_x(x')  \frac{\delta}{\delta C(x')}  \\
+\int d^2 x'  (N \Gzero_x^{ab}+ \Gzero_x N^{ab})(x') \frac{\delta}{\delta \Pi^{ab}(x')} \\
-2 \int d^2 x' (   \Delta^{ab}_{\phantom{ab}cd}(N)\Gzero_x^{cd} + \Delta^{ab}_{\phantom{ab}cd}(\Gzero_x)N^{cd} )(x')\frac{\delta}{\delta p^{ab}(x')}  
\end{multline}

\be
\{ \cdot , C(x) \}_0 = -\int d^2 x' \Gzero_x(x')  \frac{\delta}{\delta N(x')} \label{hvfCzero} 
\ee

The main source of difficulties in solving \eqref{defHVF0} is the implicit dependence of $q_{ab}$ in the various tensors and in the differential operator $\D$ (which in addition depends on $T_{ab}$).  These are responsible for what appears as ``correction terms" to the standard canonical expressions, as explained in section \ref{kinPBsec} and  below.\\

\noindent \emph{Comments}
\begin{itemize}

\item  The extra terms in the first line of \eqref{hvfpabzero} (together with the $p^{ab}q^{cd}$ term in the second line), induce a pure trace deformation on the various tensors in order to keep them traceless under the deformation of the metric induced by the first term in \eqref{hvfpabzero}.
\item The aforementioned  terms have their counterparts in the terms proportional to $\delta/ \delta p^{ c d }$ in  Eqs. \eqref{hvfTabzero} to \eqref{hvfsigmadotzero} (as well as the $ q^{ab} p^{cd}$ term in \eqref{hvfpabzero}). These terms can be understood from the following variation formula of STF tensors:\footnote{Together with analogous formula for contravariant tensors, $\delta A^{ab}  = \delta A^{\langle ab \rangle} - \frac{1}{2}q^{ab} A^{cd} \delta q_{cd}$.  We also notice the  identity  $\delta A^{\langle ab \rangle}=q^{ac}q^{bd}\delta A_{\langle bd \rangle}$, where the trace-free part is being extracted after the variation.}
\be \label{variationformulaSTF}
\delta A_{ab} = \delta A_{\langle ab \rangle} + \frac{1}{2}q_{ab} A^{cd} \delta q_{cd} \implies  \{ \cdot, A_{ab} \}_0 =  \{ \cdot, A_{\langle ab \rangle} \}_0+ \frac{1}{2}q_{ab} A^{cd} \{ \cdot ,  q_{cd} \}_0
\ee
where the trace-free part is being taken \emph{after} the variation/evaluation of PBs.

\item Jacobi identity between two $p^{ab}$'s and one of these tensors then requires the second line in \eqref{hvfpabzero}. Alternatively, the second line in \eqref{hvfpabzero} is  needed when solving \eqref{defHVF0} in order to cancel $\delta q_{ab}$ pieces induced by the extra terms in the first line of \eqref{hvfpabzero}.
\item The second and third line of \eqref{hvfNzero} can be understood by expanding the HVF relation  $\{ \cdot, \D N\}_{0} = \delta/\delta C$ and using the formula for the variation of $\D$ given in Eq. \eqref{delD4}.
\item The aforementioned  terms have their counterparts in the terms proportional to $\delta/ \delta N$ in   \eqref{hvfPiabzero} and \eqref{hvfpabzero}. Equivalently, these terms can be understood by requiring that $\Pi^{ab}$ and $p^{ab}$ Poisson commute with $ \D N$. This is most easily seen if we consider smeared versions.  Using \eqref{delD4},  one can show 
the smeared versions of \eqref{hvfPiabzero} and \eqref{hvfpabzero} take the form (we only display terms relevant for this discussion)
\begin{align}
\{\cdot , \int_{S^2} \tau_{ab} \Pi^{ab} \}_0  &= \int_{S^2} \left( \delta_\tau T_{ab}  \frac{\delta}{\delta T_{ a b}} - \Gzero \delta_\tau \D N \frac{\delta}{\delta N}  \right) + \cdots  \label{smearedhvfp0}\\ 
\{\cdot , \int_{S^2} X_{ab}  p^{ab} \}_0 &  =  \int_{S^2} \left( \delta_{X}q_{ab} \frac{\delta}{\delta q_{a b }} - \Gzero \delta_X \D N \frac{\delta}{\delta N}   \right) + \cdots  \label{smearedhvfPi0}
\end{align}
where
\be
\delta_\tau T_{ab}=  \tau_{\langle a b \rangle} , \quad \delta_X q_{ab}=  X_{\langle a b \rangle} ,\label{defdelX0}
\ee
and $\delta_\tau \D$, $\delta_X \D$ represents the infinitesimal variation of the operator $\D$ under  \eqref{defdelX0}. The terms proportional to $\delta/\delta N$ in \eqref{smearedhvfp0} and \eqref{smearedhvfPi0} can be understood as being determined by the condition that $ \Pi^{ab}$ and $p^{ab}$ Poisson commute with $\D N$. For instance:
\be
0 = \delta_X (\D N)=  \delta_X \D N + \D \delta_X N \implies \delta_X N = - \Gzero \delta_X \D N.
\ee

\item The appearance of $\Gzero$ in some of the HVFs lead to ambiguities in their expressions, as $\Gzero$ is defined modulo the kernel of $\D$ \eqref{defGzeroabstract}. As discussed in appendix \ref{nullappendix}, such ambiguities are associated to  null directions of the symplectic structure and should therefore be understood  as intrinsic  redundancies of $\Gamma_{\kin, 0}$. Similar null directions are present in $\Gamma_{\kin, 1}$ and $\Gamma_{\kin, 2}$ (see appendix \ref{nullappendix}). All physical quantities should of course be independent of such redundancies.

\end{itemize}

\subsection{Obstructions for a shear and subleading news HVF}\label{utoinfapp}
In this subsection we explain why the shear and the subleading news do not admit HVFs.\footnote{These are specific instances of more general quantities that can be expressed through  $u$-smearings of the shear/news tensor, such that their smearings do not satisfy  appropriate antisymmetric or symmetric boundary conditions, respectively.  See Appendix A of \cite{gbms1} for further discussion.}

A natural candidate for a shear HVF  may be obtained by integrating that of the news tensor. Assuming boundary conditions compatible with $F_6=0$ leads to
\be \label{naiveHVFshear}
`` \{ \cdot, \sigma^{\langle ab \rangle }(u) \}_0 "= \frac{1}{4} \int du' \sign(u'-u) \frac{\delta}{\delta \sigma_{\langle a b \rangle}(u')}  ,
\ee
where, for simplicity we only display the trace-free part components  (the trace-part is determined by the general expression \eqref{variationformulaSTF} and is insensitive to the issues we are now discussing). Plugging \eqref{naiveHVFshear} in the symplectic structure leads to
\be \label{OmeganaiveHVFshear}
\Omega_{\kin, 0}( \delta, `` \{ \cdot, \sigma^{\langle ab \rangle }(u) \}_0 " ) = \delta \sigma^{\langle ab \rangle }(u) - \tfrac{1}{2} \delta F^{\langle ab \rangle }_6.
\ee
The second term in the RHS of  \eqref{OmeganaiveHVFshear}, which results from an integration by parts when evaluating the LHS, appears as an obstruction for finding a HVF for the shear. Although this obstruction is absent when the constraint $F_6=0$ is satisfied, it cannot be ignored when working in the kinematical space. Indeed, Eq. \eqref{OmeganaiveHVFshear}  tell us that, the quantity whose HVF is well defined on the kinematical space is the combination $\big( \sigma^{\langle ab \rangle }(u) - \tfrac{1}{2} F^{\langle ab \rangle }_6 \big)$
\be \label{HVFcorrectedshear}
 \{ \cdot,\big( \sigma^{\langle ab \rangle }(u) - \tfrac{1}{2} F^{\langle ab \rangle }_6 \big)\}_0 = \frac{1}{4} \int du' \sign(u'-u) \frac{\delta}{\delta \sigma_{\langle a b \rangle}(u')}  .
\ee
The above discussion also shows that $F_6$ lacks a HVF. This represents an obstruction for evaluating the Dirac matrix, which we overcome with the prescriptions given in subsection \ref{utoinfsec}.

For the subleading news, the candidate HVF obtained by integrating that of the news is
\be
 `` \{ \cdot, \Ncalone^{\langle a b \rangle} \}_0  "=    \frac{1}{2}  \int du u  \frac{\delta}{\delta \sigma_{\langle a b \rangle}(u)} .
\ee
The evaluation of the symplectic structure along this direction can be written as
\ba
\Omega_{\kin, 0}( \delta,  `` \{ \cdot, \Ncalone^{\langle a b \rangle} \}_0  " ) &=& \delta \Ncalone^{\langle ab \rangle } - \tfrac{1}{2} \big[ u \delta \sigma^{\langle ab \rangle } \big]^{+\Lambda}_{-\Lambda}  \\
&=& \delta \Ncalone^{\langle ab \rangle } - \Lambda\delta  F^{\langle ab \rangle }_6,
\ea
where, as in \eqref{OmeganaiveHVFshear} we kept track of boundary terms when integrating by parts. This time, however, there remains an explicit dependence on the cutoff $\Lambda$.

We formally write this result as 
\be
\{ \cdot, \Ncalone^{\langle a b \rangle} \}_0 =  \frac{1}{2}  \int du u  \frac{\delta}{\delta \sigma_{\langle a b \rangle}(u)} +  \Lambda\{ \cdot,  F^{\langle ab \rangle }_6 \}_0. \label{correctedHVFNcalone}
\ee

\section{First stage Dirac matrix} \label{1stDiracapp}
Let us start by considering the PB of $F_1$ with itself. From the kinematical HVFs described in appendix \ref{kinHVFsapp} we have
\ba \label{F1F1PB}
\{F_1[X'],F_1[X]\}_0 & =&  \frac{1}{2}\int_{S^2} X'_{ab} \frac{\delta}{\delta q_{\langle a b \rangle}} \int_{S^2} X_{cd} \O^{cd}_{\phantom{ab} mn} \Pi^{mn} - \frac{1}{4} \int_{S^2}X'_{ab} \Pi^{ab} D^{\langle c} D^{d \rangle} X_{cd} \nonumber \\
&&\quad +  \frac{1}{2}\int_{S^2}X'_{ab} \left( p^{ab} q^{cd}  +T^{ab} \Pi^{cd}  + \int du  \sigma^{ab} \dot \sigma^{cd} \right) X_{cd} \nonumber \\ 
&& \quad \quad \quad - \left(X' \leftrightarrow X \right).   
\ea
Below, we will work out the explicit expression for this bracket in holomorphic coordinates. Before doing so, let us make a few remarks about \eqref{F1F1PB} and its relation to the computation in holomorphic coordinates: 
\begin{itemize}
\item One can show \eqref{F1F1PB}  depends on $X$ an $X'$ only through their trace-free parts (as it should). The way this happen is that the first term in the second line of \eqref{F1F1PB} cancels, upon using $F_1=0$, an equal an opposite  term coming from the variations of $\O$. This is what will allow us to focus on the $X_{zz} \O^{zz}_{cd} \Pi^{cd}$ piece (and its complex conjugate) when working with holomorphic coordinates
\item As written, the metric functional derivative in $\eqref{F1F1PB}$ does not act on $\Pi^{ab}$ even though  $\Pi^{ab}q_{ab}=0$. Such implicit dependance of $\Pi^{ab}$ on $q_{ab}$ is instead captured in the second term  in the first line of \eqref{F1F1PB}, which comes from the ``correction terms'' in the kinematical brackets discussed in section \ref{kinPBsec}. 

Alternatively, if we allow  the metric functional derivative to act on $\Pi^{ab}$, we do not need to explicitly include the second term  in the first line of \eqref{F1F1PB}. This is how we will setup the computation in section \ref{F1F2app}  (see footnote \ref{footnotetracefreeholo}).


\item  We will  express the result of \eqref{F1F1PB} as
\be \label{defAitoF1PB}
\{F_1[X'],F_1[X]\}_0 = \frac{1}{4}  \int_{S^2}X'_{ab} \A^{abcd} X_{cd},
\ee
or,
\be
\A^{abcd}X_{cd} :=  4 \{F_1^{ab},F_1[X]\}_0,
\ee
with
\be
\A^{abcd} = \A^{abcd}_\soft + 2 \int du  (\sigma^{ab} \dot \sigma^{cd}-  \dot \sigma^{ab} \sigma^{cd}).
\ee
  $\A^{abcd}_\soft $ is thus constructed from the first line of \eqref{F1F1PB} plus the the term proportional to $T^{ab} \Pi^{cd} $ in \eqref{F1F1PB}.  As indicated by the notation, the resulting expression is SR covariant. 
\item As in the case of other differential operators introduced before, $\A^{abcd}$ exhibits a holomorphic/antiholomorphic splitting. Specifically, when working in holomorphic coordinates, one has
\be \label{Azbzbcd}
\A^{\zb \zb cd} X_{cd} =  (\A_\soft+\A_\hard) X^{\zb\zb}   
\ee
where 
\ba
\A_\soft & = & -3 \Db_z^2 \Pi^{zz} + \Db^2_{\zb} \Pi^{\zb \zb}-3 \Db_z \Pi^{zz}\Db_z - \Db_{\zb} \Pi^{\zb \zb} \Db_{\zb} - \Pi^{zz} \Db^2_z + \Pi^{\zb \zb}\Db^2_{\zb} , \label{Absoft}\\
\A_\hard & = &  2 \int_{-\infty}^\infty du(\dot \sigma^{zz} \sigma_{zz} -  \dot \sigma^{\zb \zb} \sigma_{\zb \zb} ) \label{Abhard}. 
\ea
Further discussion of this operator is given in \ref{Aoperatorapp} and \ref{JacobiAsec}.

\end{itemize}

The PB between $F_1$ and $F_2$ will be computed below  in holomorphic coordinates, resulting in the differential operator whose general coordinate expression is discussed in section 
\ref{Ssec}. The PB of $F_2$ with itself vanishes.

\subsection{Constraint variations and PBs involving $F_1$} \label{F1F2app}
We start by considering variations of $F_1^{zz}$ \eqref{defF1F2}
\be
\delta F_1^{zz} = \delta p^{zz} - \frac{1}{2} \O^{zz}_{zz} \delta \Pi^{zz} - \frac{1}{2}  \frac{\partial \O^{zz}_{cd} \Pi^{cd}}{\partial q_{ab}} \delta q_{ab} 
\ee
where $\partial \O^{zz}_{cd} \Pi^{cd}/ \partial q_{ab}$ denotes  the differential operators acting in $\delta q_{ab}$  when evaluating $\delta(\O^{zz}_{cd} \Pi^{cd})$ and in holomorphic coordinates, take the following form:\footnote{Note that the $\delta q_{ab}$ contribution is obtained by computing not only the variational derivative w.r.t $q_{ab}$, but also the derivatives with respect to the trace-free tensors $W^{ab}$, i.e, $\delta W^{ab} = \delta W^{\langle ab \rangle}- \frac{1}{2} q^{ab} W^{cd} \delta q_{cd}$. \label{footnotetracefreeholo}}
\ba
\frac{\partial \O^{zz}_{cd} \Pi^{cd}}{\partial q_{zz}} & = & - \frac{1}{2} q^{z \zb} q^{z \zb}( D^2_{\zb} \Pi^{zz} +3 D_{\zb} \Pi^{zz} D_{\zb} + 3 \Pi^{zz} D^2_{\zb} )\\
 \frac{\partial \O^{zz}_{cd} \Pi^{cd}}{\partial q_{\zb\zb}} & = &   - \frac{1}{2} q^{z \zb} q^{z \zb}( 2 D^2_z \Pi^{zz} + D^2_{\zb} \Pi^{\zb \zb} - D_z \Pi^{zz} D_z + \Pi^{zz}D^2_z) 
\ea

We also need  expressions for the  variations of  the integrated version of $\O^{zz}_{cd} \Pi^{cd}$.  Given a smearing parameter $X_{zz}$, let
\be
\O(\Pi,X) := \int_{S^2} X_{zz} \O^{zz}_{cd} \Pi^{cd} 
\ee
be the corresponding integrated operator.  From the condition
\be
\int_{S^2} X_{zz} \frac{\partial \O^{zz}_{cd} \Pi^{cd}}{ \partial q_{ab} } \delta q_{ab} =  \int_{S^2} \delta q_{ab} \frac{\delta}{\delta q_{ab}} \O(\Pi,X) .
\ee
one finds
\ba
\frac{\delta}{\delta q_{zz}} \O(\Pi,X) &=&  - \frac{1}{2} q^{z \zb} q^{z \zb} ( D^2_{\zb} \Pi^{zz}+ 3  D_{\zb}\Pi^{zz}  D_{\zb} +3  \Pi^{zz} D^2_{\zb} )X_{zz}\\ 
 \frac{\delta}{\delta q_{\zb\zb}} \O(\Pi,X)  &=& - \frac{1}{2} q^{z \zb} q^{z \zb} (4 D^2_z \Pi^{zz} + D^2_{\zb} \Pi^{\zb\zb} + 3 D_z \Pi^{zz} D_z   + \Pi^{zz} D^2_z) X_{zz}.
\ea

The HFV of the smeared $F_1^{zz}$ constraint then takes the form\footnote{Here and in the rest of the section we omit the zero subscript in the PBs.}
\begin{multline} \label{hvfF1X}
\{ \cdot ,  \int X_{zz} F_1^{zz}  \}  =  \int  X_{zz} \{ \cdot, p^{zz} \} 
- \frac{1}{2} \int \frac{\delta \O(\Pi,X)}{\delta q_{ab}} \{ \cdot , q_{ab}\} - \frac{1}{2}\int  \O^{zz}_{zz} X_{zz} \{ \cdot , \Pi^{zz} \}  .
\end{multline}
To evaluate the PB of $F_1$ with itself, consider the unsmeared version of \eqref{hvfF1X} 
\be \label{Fzz1PB}
\{ F^{zz}_1, \cdot \} = \{ p^{zz}, \cdot \}  - \frac{1}{2} \frac{\partial \O^{zz}_{cd} \Pi^{cd}}{\partial q_{ab}} \{q_{ab} , \cdot \} - \frac{1}{2} \O^{zz}_{zz} \{ \Pi^{zz}, \cdot \} .
\ee
Using the kinematical PBs and the expressions given above one finds
\be
\{ F^{zz}_1,  \int X_{z'z'} F_1^{z'z'}   \}  =  \frac{1}{2} \frac{\delta \O(\Pi,X)}{\delta q_{zz}} -\frac{1}{2} \frac{\partial \O^{zz}_{cd} \Pi^{cd}}{\partial q_{zz}}X_{zz} =0
\ee
and
\ba
\{ F^{\zb \zb}_1,  \int X_{z'z'} F_1^{z'z'}   \}  & = &  \frac{1}{2} \frac{\delta \O(\Pi,X)}{\delta q_{\zb\zb}} -\frac{1}{2} \frac{\partial \O^{\zb\zb}_{cd} \Pi^{cd}}{\partial q_{zz}}X_{zz}   +  \int d^2 z' X_{z'z'} \{p^{\zb \zb}, p^{z'z'} \}\\  
& = & \frac{1}{4} q^{z \zb} q^{z \zb}A^\soft X_{zz} + \frac{1}{2} q^{z \zb} q^{z \zb} (\Pi^{zz} T_{zz} - \Pi^{\zb\zb} T_{\zb \zb} +\frac{1}{2}\A^\hard)  X_{zz}
\ea
where
\begin{multline} \label{defAsoft}
A_\soft X_{zz} := 
 2 q_{z \zb} q_{z \zb} \left( \frac{\delta \O(\Pi,X)}{\delta q_{\zb\zb}} -\frac{\partial \O^{\zb\zb}_{cd} \Pi^{cd}}{\partial q_{zz}}X_{zz} \right) = \\
 (-3 D_z^2 \Pi^{zz} + D^2_{\zb} \Pi^{\zb \zb}-3 D_z \Pi^{zz}D_z - D_{\zb} \Pi^{\zb \zb} D_{\zb} - \Pi^{zz} D^2_z + \Pi^{\zb \zb}D^2_{\zb}) X_{zz}.
\end{multline}
Remarkably, as can be shown by a slightly lengthy but straightforward computation, the structure of $A^\soft$ is such that
\be
\A_\soft = A_\soft + 2(\Pi^{zz} T_{zz} - \Pi^{\zb\zb} T_{\zb \zb} ),
\ee
where $\A_\soft$ is the superrotation covariant operator obtained by doing the replacement $D_a \to \Db_a$ in \eqref{defAsoft}. 
One thus arrives at 
\be
\{ F^{\zb \zb}_1,  \int X_{z'z'} F_1^{z'z'}   \}  =   \frac{1}{4} q^{z \zb} q^{z \zb}(\A^\soft +\A^\hard) X_{zz} .
\ee

\subsection{Constraint variations and PBs involving $F_2$} \label{appF2}
We start by considering
\be \label{deltaF2z}
\delta F_{2 z} = D^{z} \delta T_{z z} + \frac{\partial F_{2 z}}{\partial q_{ab}} \delta q_{ab}
\ee
where
\ba
\frac{\partial F_{2 z}}{\partial q_{zz}} &=& \frac{1}{2} q^{z \zb} q^{z \zb}  (D_z T_{\zb \zb} + D_z D^2_{\zb}) \\
&\stackrel{F_{2 \zb}=0}{=}&   \frac{1}{2} q^{z \zb} D_{\zb} (D_z D^z - R/2) \\
\frac{\partial F_{2 z}}{\partial q_{\zb\zb}} &=&   \frac{1}{2} q^{z \zb} q^{z \zb} (-D_z T_{zz} - 2 T_{zz} D_z + D^3_z ) \\
&=& \frac{1}{2} q^{z \zb} q^{z \zb} \Db^3_z.
\ea
In addition to the above SR covariant expressions, we note that the first term in \eqref{deltaF2z} can be written as
\be
D^{z} \delta T_{z z} =  \Db^{z} \delta T_{z z} .
\ee

Since $F_2$ only depends on $q_{ab}$ and $T_{ab}$,  one has
\be
\{F_{2 z} , F_{2 z'} \} = \{F_{2 \zb} , F_{2 z'} \} =0.
\ee

As before, let us consider the smeared version of the constraint
\be
F_2(X) := \int_{S^2}  X^z F_{2 z} .
\ee
The corresponding HVF takes the form
\be \label{hvfF2}
\{ \cdot ,  F_2(X)\} =  \int \frac{\delta  F_2(X)}{\delta T_{zz}} \{ \cdot, T_{zz} \} +  \int \frac{\delta  F_2(X)}{\delta q_{ab}} \{ \cdot , q_{ab}\}
\ee
where
\ba
\frac{\delta F_2(X)}{\delta T_{zz}}  &= & - D^z X^z  = -\Db^z X^z\\
\frac{\delta F_2(X)}{\delta q_{zz}} &=&   \frac{1}{2} q^{z \zb} q^{z \zb}  (D_z T_{\zb \zb} - D^2_{\zb} D_z ) X^z\\ 
&\stackrel{F_{2 \zb}=0}{=}&   -\frac{1}{2} \O^{zz}_{zz} D^z X^z =   -\frac{1}{2} \O^{zz}_{zz} \Db^z X^z \\
 \frac{\delta F_2(X)}{\delta q_{\zb\zb}}   &=&   \frac{1}{2} q^{z \zb} q^{z \zb} (D_z T_{zz}  +2 T_{zz} D_z - D^3_z ) X^z = -\frac{1}{2} q^{z \zb} q^{z \zb} \Db^3_z X^z,
\ea
where in the last line one can express the differential operator in a superrotation covariant manner, as can be checked by direct evaluation.

We finally compute the PBs involving $F_1$ and $F_2$.  The simplest one is
\be
\{F_1^{\zb\zb} , F_2(X) \} = - \frac{\delta  F_2(X)}{\delta q_{\zb \zb}}  = \frac{1}{2} q^{z \zb} q^{z \zb}  \Db_z^3 X^z.
\ee
 For the bracket involving $F_1^{zz}$ one has:
\ba
\{F_1^{zz} , F_2(X) \}& = & \frac{1}{2} \O^{zz}_{zz} \frac{\delta  F_2(X)}{\delta T_{zz}}  - \frac{\delta  F_2(X)}{\delta q_{zz}}  \\
& = & -\frac{1}{2} D^z D_z D^z X^z + \frac{1}{4} R D^z X^z - \frac{1}{2} q^{z \zb} q^{z \zb}  (D_z T_{\zb \zb} - D^2_{\zb} D_z ) X^z \\
& = & -\frac{1}{2} q^{z \zb}( D_{\zb}D_z D^z X^z - \frac{R}{2}D_{\zb} X^z + D^{\zb} T_{\zb \zb} - D_{\zb} D^{z}D_z X^z \\
& = & -\frac{1}{2} q^{z \zb} X^z F_{2 \zb},
\ea
where we used that $[D_z,D^z]X^z = \frac{1}{2}R X^z$. Thus, $F_1^{zz}$ commutes with $F_{2 z}$ on the constraint surface.

\subsection{Bracket relations between two $F_1$'s and one $F_2$}  \label{JacobiAsec}
Let us introduce the notation
\ba
 \Sb[X,V] &:=& \int_{S^2} X^{ab} \Sb_{ab \, c}  V^c \\
 &:=& \int X^{ab} D_{\langle a} D_{b \rangle} D_c V^c -\frac{1}{2}  \O[X,\delta_V q] - \int X^{\langle a b \rangle} \Lie_V T_{ab} .\label{SXVexpanded}
 \ea
\be
\A[Y,\Pi,X] =  \int_{S^2}Y_{ab} \A^{abcd}(\Pi) X_{cd},
\ee
where we have explicitly included the $\Pi^{ab}$ dependence  of the operator $\A$. The brackets evaluated in the previous subsections can then be written as
\ba
  \{F_2[V],F_1[X] \} & =& \tfrac{1}{2} \Sb[X,V], \label{F2F1Sb} \\
  \{F_1[Y],F_1[X] \} & =& \tfrac{1}{4} \A[Y,\Pi,X]. \label{F1YF1X}
\ea
We will now evaluate the bracket between $F_1$ and \eqref{F2F1Sb}. The HVF of the smeared $F_1$ can be written as
\be \label{HVFF1Y}
\{ \cdot, F_1[Y] \} = \int Y_{ab}   \frac{\delta}{\delta q_{\langle a b \rangle}}  - \frac{1}{2}(\O Y)_{ab} \frac{\delta}{\delta T_{\langle a b \rangle}}  +  \frac{1}{2} (Y^{cd}  T_{cd}) q_{ab}\frac{\delta}{\delta T_{a b }}+ \cdots
\ee
where the dots are terms that do not act on $q_{ab}$ and $T_{ab}$ and hence have no impact in the bracket we are interested in. Acting with \eqref{HVFF1Y} on \eqref{SXVexpanded} gives

\begin{multline} \label{SXVF1pb}
\{ \Sb[X,V], F_1[Y] \} =  \int Y_{ab} \Delta^{ab cd}(D \cdot V) X_{cd}  \\
-\frac{1}{2}  \int Y_{ab}   \frac{\delta}{\delta q_{\langle a b \rangle}}  \O[X,\delta_V q] -\frac{1}{2} \O[X,\delta_V Y]\\
+\frac{1}{2}\int \left( X^{ab}\Lie_V (\O Y_{ab})   - Y^{cd}  T_{cd} X^{ a b } \delta_V q_{ab} +    Y^{cd}  X_{cd} T^{ a b } \delta_V q_{ab} \right),
\end{multline}
where each line corresponds to each of the terms in \eqref{SXVexpanded}. Using the fact that $ \Delta^{ab cd}$ is a symmetric operator, the term in the first line can be combined with the first one of the last line to  give a SR variation of $\O Y_{ab}$ ,
\be
\delta_V(\O Y_{ab}) =  \Lie_V (\O Y_{ab}) + 2 \Delta_{ab}^{\phantom{ab}cd}(D \cdot V) Y_{cd},
\ee
as  can be deduced from the (non-covariant) Weyl scaling properties  discussed in \ref{weylapp}. Adding the resulting expression to the second term in the second line gives,
\ba
 \frac{1}{2}\int X^{ab}\delta_V (\O Y_{ab})  -\frac{1}{2} \O[X,\delta_V Y] &=& \frac{1}{2}\int X^{ab}\delta_V (\O_{ab}^{\phantom{ab} cd}) Y_{cd}\\
&=& \frac{1}{2}\int \delta_V q_{ab}  \frac{\delta}{\delta q_{\langle a b \rangle}}  \O[X,Y].  \label{2nd2ndplus3rd}
\ea
When \eqref{2nd2ndplus3rd} is added to the first term of the second line of \eqref{SXVF1pb} one gets the operator $A_\soft$ \eqref{abstractdefAsoft1}. Furthermore, the last two terms of \eqref{SXVF1pb} are precisely the Geroch tensor factors needed to make the expression SR covariant. The end result is then given by
\be 
\{ \Sb[X,V], F_1[Y] \} =  \tfrac{1}{4} \A_\soft[Y,X,\delta_V q] .\label{aim}
\ee

We conclude by noting that the Jacobi-type identity satisfied by $\A_\soft$ \eqref{jacobiAsoft} can be understood from the Jacobi identity involving two $F_1$'s and one $F_2$:
\be \label{jacobi112}
\{ \{F_2[V],F_1[X] \} , F_1[Y] \}  - (X \leftrightarrow Y) =  \{ \{F_1[X],F_1[Y] \},F_2[V]\} .
\ee
From  \eqref{aim}, the LHS of \eqref{jacobi112} is given by
\be
\tfrac{1}{8} \A_\soft[Y,X, \delta_V q ] - (X \leftrightarrow Y).
\ee
From \eqref{F1YF1X} and the HVF action of $F_2$ it is easy to check that the RHS of \eqref{jacobi112} is given by
\be
\tfrac{1}{8} \A_\soft[X, \delta_V q , Y].
\ee
Writing $\delta_V q_{ab} =Z_{ab}$  the Jacobi identity \eqref{jacobi112} can then be seen to be equivalent to
\be \label{jacobiA}
\A_\soft[X,Y,Z]+\A_\soft[Y,Z,X] + \A_\soft[Z,X,Y]=0.
\ee
It is easy to see that a similar identity trivially holds for $ \A_\hard$, and hence for the full $\A$.

\section{Intermediate HVFs} \label{HVF1sec}
In this appendix we discuss the HVFs on $\Gamma_{\kin, 1}$. Our starting point is  the Dirac map \eqref{XtoY0holocoord} written in general coordinates:
\be \label{XtoY0gralcoord}
\begin{aligned}
 Y^{ab} &= \frac{1}{4}\A^{abcd}X_{ c d } + \frac{1}{2} \Sb^{ab}_{\phantom{ab} c} X^c ,\\
 Y_a & = -\frac{1}{2} \Sbt_{a}^{\phantom{a} bc}X_{ b c },
\end{aligned}
\ee
where $\Sb_{ab \, c}$ and  $\A^{abcd}$  are the  differential operators introduced in  \ref{Ssec} and   \ref{Aoperatorapp} respectively and $\Sbt_{a \, b  c}$ is the adjoint of $\Sb_{ab \, c}$ ($\A^{abcd}$ is anti-selfadjoint). 

The inverse map $(Y^{ab},Y_a) \mapsto (X_{ab},X^a)$ can  be written with the help of the Green's functions  introduced in section \ref{greenSsec} and reads
\be \label{YtoX0gralcoord}
\begin{aligned}
 X_{ab} &= -2 \Gbt_{ab}^{\phantom{ab} c} Y_c , \\
 X^a & = 2 \Gb^a_{\phantom{a} bc} Y^{bc} + \Gb^a_{\phantom{a} bc} \A^{bcmn} \Gbt_{mn}^{\phantom{ab} d} Y_d.
\end{aligned}
\ee

We next consider the smeared constraint's HVF. Using the expressions of section \ref{kinHVFsapp} one finds
\begin{multline} \label{hvfFX}
 \{ \cdot , F[X] \}_0 =  \int_{S^2}  \left(     X_{\langle ab \rangle} \frac{\delta}{\delta q_{ab}} +  \big(D^{\langle a} X^{b \rangle} - \frac{1}{2} \Pi^{cd}X_{cd}q^{ab} \big) \frac{\delta}{\delta \Pi^{ab}}    \right.\\
+ \frac{1}{2}\big(- \O_{ab}^{\phantom{ab} cd} X_{\langle cd \rangle} + T^{cd}X_{cd} q_{ab} \big)\frac{\delta}{\delta T_{ab}}     +  \frac{1}{2} \int du \sigma^{cd}X_{cd}q_{ab}  \frac{\delta}{\delta \sigma_{ab}}  \\
 \left. + X_{ab} \int d^2 x' \left( \Bb^{ab}_{\phantom{ab} cd}(N) \Gzero^{cd}_{x'}+  \Bb^{ab}_{\phantom{ab} cd}(\Gzero_{x'}) N^{cd} \right)  \frac{\delta}{\delta N(x')} \right) +   \propto \frac{\delta}{\delta p^{ab}}      ,
\end{multline}
where $\O_{ab}^{\phantom{ab} cd}$ and $\Bb^{ab}_{\phantom{ab} cd}$ are given in Eqs. \eqref{defOop} and \eqref{defBabst}. We do not explicitly write the term proportional to $\delta/\delta p^{ab}$, as it is not an independent direction on $\Gamma_{\text{kin}, 1}$. We do display the term proportional to $\delta/\delta T_{ab}$ (even though it is neither an independent direction) because it facilitates the analysis of variations involving $T_{ab}$. Note that the term proportional to $\delta/\delta N$ includes an additional 2d integral compared to the other terms; this arises from the intrinsic non-locality of $\{\cdot, N\}_0$, see Eq. \eqref{hvfNzero}.

This completes the ingredients needed to use the general formula \eqref{correctedHVFabstract} for $i=0$.  We present below the resulting  HVFs. We first discuss those associated to the ``elementary'' fields \eqref{parametrizationGamma1} parametrizing $\Gamma_{\kin, 1}$ and then discuss some of the ``derived'' HVFs.

\subsection{Elementary HVFs}  \label{HVF1}


We start by discussing the independent tensorial quantities: $q_{ab}$,  $\Pi^{ab}$ and $\dot \sigma^{ab}$. It will be convenient to work with their smeared versions, defined by:\footnote{We assume $\gamma_{ab} \stackrel{|u| \to \infty}{\to} 0$. The case of non-trivial asymptotic values should be treated separately, according to the discussions given in \ref{utoinfsec} and \ref{utoinfapp}.} 
\be \label{smearedkin1}
q[\rho]  := \int_{S^2}  \rho^{ab} q_{ab}    , \quad  \Pi[\tau]  := \int_{S^2}\tau_{ab} \, \Pi^{ab} , \quad \dot \sigma[\gamma] := \int_{\I}  \gamma_{ab}\, \dot  \sigma^{ab} .
\ee

Evaluating \eqref{correctedHVFabstract} for $i=0$ and $\varphi= q[\rho]$, $\sigma[\gamma] $, $\Pi[\tau]$, one finds:
\ba
\{ \cdot, q[\rho] \}_1 & =&  \int_{S^2} 2 \Db^{a} \big( \Gb^{b }_{\phantom{b}cd} \rho^{cd} \big)\frac{\delta}{\delta \Pi^{\langle ab \rangle}} \label{HVF1qab} \\
\{ \cdot,\dot \sigma[\gamma] \}_1  &=  & \int_{\I} \left( \frac{1}{2} \gamma_{\langle a b \rangle} \frac{\delta}{\delta \sigma_{ab}} - \Db^{a }\big( \Gb^{b}_{\phantom{b} cd}(q^{mn}\gamma_{mn} \dot \sigma^{cd}) \big) \frac{\delta}{\delta \Pi^{\langle ab \rangle}} \right) \label{HVF1dotsigmagamma}\\
\{ \cdot, \Pi[\tau] \}_1 &= & \int_{S^2} \left(  \delta_\tau q_{ab} \frac{\delta}{\delta q_{ab}}  +      \delta_\tau T_{\langle ab\rangle} \frac{\delta}{\delta T_{ab}} + \delta_\tau \Pi^{\langle ab \rangle} \frac{\delta}{\delta \Pi^{ab}}  +  \delta_\tau N \frac{\delta}{\delta N} \right)\nonumber \\
& &  \quad \quad \quad + \frac{1}{2} \int_{S^2} \delta_\tau q_{cd}  \left(T^{cd} q_{ab}  \frac{\delta}{\delta T_{ab}} - \Pi^{cd} q^{ab} \frac{\delta}{\delta \Pi^{ab}}  +  \int du \, \sigma^{cd} q_{ab} \frac{\delta}{\delta \sigma_{ab}}  \right) , \label{hvfPiabone}
\ea
where in \eqref{hvfPiabone} we isolated the trace and  trace-free variations of the tensorial variables and
\be \label{deltatau1}
\begin{aligned}
\delta_\tau q_{ab}  &= 2 \Gbt_{ab}^{\phantom{ab} c}\Db^d \tau_{\langle c d \rangle} \\
\delta_\tau T_{\langle ab \rangle}  &=  \tau_{\langle a b \rangle}  - \frac{1}{2} \O_{ab}^{\phantom{ab} cd} \delta_\tau q_{cd}   \\
\delta_\tau \Pi^{\langle ab \rangle} &= -\frac{1}{2}\Db^{\langle a}  \Gb^{b\rangle}_{\phantom{b}cd} \A^{cd mn} \delta_\tau q_{mn} - \Db^{\langle a} \Gb^{b\rangle }_{\phantom{b} cd}(q^{mn}\tau_{mn} \Pi^{cd}) \\
\delta_\tau N & = -\Gzero \delta_\tau \D N. 
\end{aligned}
\ee

For the (unsmeared) scalar quantities $C$ and $N$ one finds
\be
\{ \cdot, C(x_0) \}_1  = - \int_{S^2}  \Gzero_{x_0} \frac{\delta}{\delta N}  
\ee
\be
\{ \cdot, N(x_0) \}_1 =  \int_{S^2} \left( \Gzero_{x_0} \frac{\delta}{\delta C}  +  \left( N \Gzero^{ab}_{x_0}+  2 \Db^{a}  \Gb^{b }_{\phantom{b}cd} \Bb^{cd}_{\phantom{cd}mn}(N) \Gzero^{mn}_{x_0}  + (N \leftrightarrow \Gzero_{x_0} )\right)    \frac{\delta}{\delta \Pi^{\langle ab \rangle}} \right) \label{hvfNone}
\ee

Let us  make a  few comments about these expressions.
\begin{itemize}

\item As a consistency check, one can verify  the above expressions  satisfy the  defining condition
\be
\Omega_{\kin, 1}(\delta, \{ \cdot, \varphi \}_1) = \delta \varphi. \label{defHVF1}
\ee
where $\Omega_{\kin, 1}$ is the symplectic structure on $\Gamma_{\kin, 1}$. 

\item The appearance of Green's functions could potentially introduce ambiguities in the HVFs. We have already noted that  $\Gbt^{ab \, c}$ is free from ambiguities while $\Gb^{a \, bc}$ is defined modulo CKVs. The latter however only appears in  the form $\Db^{\langle a}  \Gb^{b\rangle}_{\phantom{b}cd} $,  thus eliminating such ambiguity. The situation with the Green's function $\Gzero$ is resolved in the same way as  in the kinematical case: As discussed in appendix \ref{nullappendix},  the potential ambiguities are in one-to-one  correspondence with the null directions of $\Omega_{\kin, 1}$.

\item The expression for $\delta_\tau N$ in \eqref{deltatau1} can be obtained from \eqref{correctedHVFabstract}, after using the variation formula \eqref{defDelta} for $\D$, together with the expressions for $\delta_\tau q_{ab}$ and  $\delta_\tau T_{\langle ab \rangle} $.

\item The ``pure-trace'' variations of the various tensors given in the second line of \eqref{hvfPiabone} are analogous to similar terms in $\{ \cdot, p_{ab}\}_0$. They ensure the preservation of trace-free condition of the different tensors given the variation in the metric.

\item The terms involving  the traces of the smearing parameters 
in  $\{ \cdot,\dot \sigma[\gamma] \}_1$ and $\{ \cdot, \Pi[\tau] \}_1 $
can be understood from the general formula \eqref{variationformulaSTF}, applied to the $\{ , \}_1$ brackets.

\item  One way of understanding why the operators of the Dirac matrix given in \eqref{XtoY0gralcoord} are SR covariant (when acted upon smearing parameters of appropriate weight) is the Weyl invariance of the symplectic form. Recall that SR covariance of a tensor is the property of being a primary under the pure Weyl scaling.  The smearing are chosen such that the quantities $q[\rho]$ and $\Pi[\tau]$ are of SR weight zero (keeping in mind that the implicit $\sqrt{q}$ term  in the measure has weight $-2$). Since the pure Weyl scaling leaves the symplectic form invariant, the brackets we compute of any SR covariant quantity with $q[\rho]$ and $\Pi[\tau]$ must then scale with the same weight as that of the said quantity. This lets us conclude that $\Gbt_{ab}^{\phantom{ab} c}$ acts on weight $+2$ vectors $Y_{c}$ to give a weight $-2$ tensor and that $\Gb^{b}_{\phantom{b}cd}$ , $\A^{cd mn}$ act on tensors $Y^{ab}$, $Y'_{ab}$ of weight $+4$ and $-2$ to give a vector and tensor of weight $0$ and $+4$ respectively. Since these operators are inverses of entries from the Dirac matrix, we see that weights claimed for the smearing parameters are the correct ones.

\end{itemize}

There are several other quantities of interest that, from the point of view of $\Gamma_{\kin, 1}$, are not elementary, since they can be written in terms of those discussed above. Of particular interest are those involving  second and third holomorphic SR covariant derivatives, which we discuss below.

\subsection{$\Db^2_z$} \label{Nzzsec}
Let us first consider\footnote{Unlike previous instances, the trace-free parts in \eqref{HVF1XabDaDbf} and \eqref{HVF1DaDb} are  taken before the evaluation of PBs.}
\begin{multline} \label{HVF1XabDaDbf}
\{\cdot, \int_{S^2} X^{ab} \Db_{\langle a}\Db_{b \rangle}f \}_1  = \int_{S^2} X^{ab}  \{\cdot, \Db_{\langle a}\Db_{b \rangle} \}_1 f + \\  \int_{S^2}  \Db_{\langle a}\Db_{b \rangle}f \{\cdot,  X^{ab} \}_1  + \int_{S^2} \Db_{\langle a}\Db_{b \rangle} X^{ab} \{\cdot,  f \}_1  ,
\end{multline}
where $X^{ab}$ and $f$ are  unspecified functions that may be phase-space dependent. Our main interest is  $N_{ab}$, for which we will take   $f=-2 N$ and $X^{ab}$ a (phase-space independent) smearing parameter.

Let us  focus on  the first line of \eqref{HVF1XabDaDbf}. Since $\Db_{\langle a}\Db_{b \rangle}=D_{\langle a}D_{b \rangle}- \tfrac{1}{2} T_{ab}$, this  is a sum of two contributions. The contribution from the first term can be obtained from the variation formula for $D_{\langle a}D_{b \rangle}$ \eqref{XabdeltaDaDbf}, together with the HVFs for $q_{ab}$ \eqref{HVF1qab}. The second contribution requires the HVF of $ T_{ab}$. In smeared form, this is  given by
\be
\{ \cdot,T[\pi] \}_1  =   \int_{S^2} \left( - \pi^{ab} +\Db^{a } \Gb^{b}_{\phantom{b} cd}(q_{mn}\pi^{mn} T^{cd}- \O^{cd}_{\phantom{ab} mn} \pi^{\langle  m n \rangle} \right) \frac{\delta}{\delta \Pi^{\langle ab \rangle}} ,
\ee
where $\pi^{ab}$ is the smearing parameter.  Combining both contributions and using the  definition of the operator $\Bb^{ab}_{\phantom{ab}cd}$ \eqref{defBabst}, one  finds
\begin{multline} \label{HVF1DaDb}
 \int_{S^2} X^{ab}  \{\cdot, \Db_{\langle a}\Db_{b \rangle} \}_1 f = 
  \int_{S^2} \left(\Db^{a } \Gb^{b}_{\phantom{b} cd} \Bb^{cd}_{\phantom{cd}mn}(f) X^{\langle m n \rangle} + \frac{1}{2}f X^{\langle a b \rangle} \right. \\
\left. + \Db^{a } \Gb^{b}_{\phantom{b} cd} (q_{mn }X^{mn}\Db^{\langle c}\Db^{d \rangle} f  )\right) \frac{\delta}{\delta \Pi^{\langle ab \rangle}} .
\end{multline}

From the above, one finds the  HVF of (smeared) $N^{ab}$ is: 
\begin{multline}  \label{HVF1Nab}
 \int_{S^2}  X_{\langle ab \rangle} \{ \cdot, N^{ab} \}_1 =  \int_{S^2} \left(-2 \Db^{a } \Gb^{b}_{\phantom{b} cd} \Bb^{cd}_{\phantom{cd}mn}(N) X^{\langle m n \rangle} - N X^{\langle a b \rangle} \right)  \frac{\delta}{\delta \Pi^{\langle ab \rangle}} \\
 -2 \int_{S^2} \Db^{\langle a}\Db^{b \rangle} X_{ab} \{\cdot,  N \}_1   ,
\end{multline}
where for simplicity we only displayed the contribution from the  trace-free part of the smearing parameter.

We finally show how \eqref{HVF1Nab} may be simplified by going to  holomorphic coordinates. We  phrase the discussion in terms of the quantity 
\be \label{XzzNzzHVF}
 \int X^{zz} \{ \cdot, N_{zz} \}_1 \equiv   \int X_{\zb\zb} \{ \cdot, N^{\zb\zb} \}_1 .
\ee

Expanding \eqref{XzzNzzHVF} as in  \eqref{HVF1XabDaDbf} we have
\be
 \int X^{zz} \{ \cdot, N_{zz} \}_1 = - 2 \int X^{zz}\{ \cdot, \Db^2_z \}_1 N -2 \int \Db^2_z X^{zz} \{\cdot, N\}_1   \label{hvfNzerozz1b}.
\ee
The first term can be read off from the general formula \eqref{HVF1DaDb}, leading to
\be \label{1stNzz}
- 2 \int X^{zz}\{ \cdot, \Db^2_z \}_1 N = \int \left( -2 q_{z \zb}\Db_z \Db^{-3}_{\zb} \Bb(N) X^{zz} \frac{\delta}{\delta \Pi^{\zb \zb}} - N X^{zz} \frac{\delta}{\delta \Pi^{z z}} \right),
\ee
where  we recall that $\Bb(N) \equiv \Bb^{zz}_{\phantom{zz}zz}(N)$.

In order to simplify the subsequent discussion  we now introduce a few definitions. 
Let
\be \label{defg}
g :=  -2 \Gzero \Db^2_z X^{zz}  
\ee
and
\be \label{gzzgzbzb}
g^{zz} = -2 \Db^{z}\Db^{z}g , \quad  g^{\zb\zb} = -2 \Db^{\zb}\Db^{\zb}g .
\ee
Using the expressions for $\Gzero$ and its derivatives  given in Eqs. \eqref{Gzerohol} and \eqref{Gzerozz}, one can show \eqref{defg} and \eqref{gzzgzbzb} can be written as
\be \label{defghol}
\begin{aligned}
g &= -\frac{1}{4}  \Db^{-2}_{\zb} X_{\zb\zb} \\
g^{zz} & =   \frac{1}{2}X^{zz} \\
g^{\zb\zb} & =   \frac{1}{2}  \Db^{-2}_{\zb} \Db^2_{z} X^{zz}.
\end{aligned}
\ee
Let us also define\footnote{This is the combination that appears in the PBs between $N$ and $F_1$: $\{N(w),F_1^{zz}\}_0= \Bb^{zz}(N,\Gzero_w)$. \label{defBbfootnote}}
\be \label{defBb}
 \Bb^{zz}(N,g):= \Bb(N) g^{zz}+ \Bb(g) N^{zz} .
\ee
Given this notation and Eq. \eqref{hvfNone}, the second term in \eqref{hvfNzerozz1b} reads
\begin{multline} 
 -2 \int \Db^2_z X^{zz} \{\cdot, N\}_1 =  \int \left( g  \frac{\delta}{\delta C}  +    (N g^{zz}+ g N^{zz}  + 2 q_{z \zb} \Db_{\zb}   \Db^{-3}_{z} \Bb^{\zb \zb}(N,g)) \frac{\delta}{\delta \Pi^{z z}} \right. \\
+ \left.   (N g^{\zb\zb}+ g N^{\zb\zb}  + 2 q_{z \zb} \Db_z   \Db^{-3}_{\zb} \Bb^{z z}(N,g)) \frac{\delta}{\delta \Pi^{\zb \zb}}  \right). \label{2ndNzz}
\end{multline}

In order to combine  \eqref{2ndNzz} and \eqref{1stNzz}, we will use the identity \eqref{antisymB3rdderivativecov} for  the difference between $\Bb(N) g^{zz}$ and  $\Bb(g) N^{zz}$:
\be \label{antisymB3rdderivative}
\Bb(N) g^{zz}- \Bb(g) N^{zz} = -q^{z \zb}q^{z \zb}q^{z \zb}\Db^3_{\zb}( N \Db_z g - g \Db_z N).
\ee

The   $\Pi^{z z}$ components add up to:
\ba
 \int d^2 w X^{ww} \{ \Pi^{zz}, N_{ww} \}_1 & = &  -N g^{zz}+ g N^{zz}  + 2 q_{z \zb} \Db_{\zb}   \Db^{-3}_{z} \Bb^{\zb \zb}(N,g)  \label{unsimplifiedPzzNzz}\\
& = & 2 \Db^{z} \left( N \Db^z g - g \Db^z N + \Db^{-3}_{z} \Bb_{z z}(N,g) \right) \\
& = & 2 \Db^{z}  \Db^{-3}_{z} \left(\Bb(g) N_{zz}- \Bb(N) g_{zz} + \Bb_{z z}(N,g) \right) \\
& = & 4 \Db^{z} \Db^{-3}_{z} \Bb(g) N_{zz}.
\ea
While for the  $\Pi^{\zb \zb}$ terms one gets
\ba
 \int d^2 w X^{ww} \{ \Pi^{\zb\zb}, N_{ww} \}_1  & = & N g^{\zb\zb}+ g N^{\zb\zb}  + 2 q_{z \zb} \Db_z   \Db^{-3}_{\zb}\left( \Bb^{zz}(N,g)  - \Bb(N) X^{zz}\right) \\
& = & 2 g N^{\zb\zb}  
\ea

In summary,  the HVF  \eqref{XzzNzzHVF} takes the form
\be \label{HVF1XzzNzz}
 \int X^{zz} \{ \cdot, N_{zz} \}_1 =  \int \left( g  \frac{\delta}{\delta C}  + 4  \Db^{z} \Db^{-3}_{z} \Bb(g) N_{zz} \frac{\delta}{\delta \Pi^{z z}} + 2 g N^{\zb\zb} \frac{\delta}{\delta \Pi^{\zb \zb}}  \right).
\ee

\subsection{$\Db_z^3 $} \label{PiSbsec}
We now discuss the quantity
\be
 \Sb[X,V] = \int_{S^2} X^{ab} \Sb_{ab \, c}  V^c = \int d^2 z  X^{zz} \Db_z^3 V^z + c.c.
 \ee
The only non-trivial directions of its HVF are  along $\Pi^{ab}$, and so we just focus on evaluating the bracket
\be
\{\Pi^{zz}, \Sb[X,V]\}_1= \{\Pi^{zz}, \Sb[X,V]\}_0+\{\Pi^{zz}, \Sb[X,V]\}_\extra.
\ee
The kinematical piece is given by
\be
\{\Pi^{zz}, \Sb[X,V]\}_0=- \delta_V X^{zz},
\ee
while the correction term takes the schematic form,
\be 
\{\Pi^{zz}, \Sb[X,V]\}_\extra = \{\Pi^{zz}, F_2 \}_0 W^{21} \{\Sb[X,V],F_1\}_0.
\ee
Using \eqref{aim} and \eqref{YtoX0holocoord} one finds
\be 
\{\Pi^{zz}, \Sb[X,V]\}_\extra =- \Db^{z} \Db^{-3}_z \A_\soft(X) \Db_z V_{z},
\ee
where $\A_\soft(X)$ is the operator $\A_\soft$ after doing the replacement $\Pi^{ab} \to X^{ab}$.

\section{Second stage Dirac matrix} \label{2nddiracapp}
The non-trivial entries in the Dirac matrix \eqref{2ndstageDiracmatrix} are those involving the constraints $F_3$ and $F_5$.\footnote{Non-trivial in the sense of being neither vanishing nor proportional to the identity operator. We emphasize that our considerations rest on the prescriptions presented in subsection \ref{utoinfsec}. In particular, any putative non-trivial bracket of $F_6$ with itself would introduce additional terms in the physical HVFs of  $\Pi^{ab}$ and $C$.  To see this, notice that changing the entry at the bottom right corner of the matrix \eqref{2ndstageDiracmatrix} potentially changes the entries in the ``central" $2 \times 2$ block in the inverse Dirac matrix \eqref{2ndstageInv_DM}. The physical fields having brackets with the constraints associated to these entries are precisely $\Pi^{ab}$ and $C$. \label{F6F6footnote}}  We write the corresponding smeared PBs as
\ba 
\{F_3[X'],F_5[X]\}_1 & = &   \int_{S^2}X'_{ab} \L^{abcd} X_{cd}, \label{defLabcd} \\
\{F_3[X'],F_3[X]\}_1 & = &   \int_{S^2}X'_{ab} \K^{abcd} X_{cd}, \label{defKabcd}
\ea
where the weight of the smearing tensors is $k=0$ for $F_3$ and $k=-1$ for $F_5$. We regard \eqref{defLabcd} and \eqref{defKabcd} as definitions for $\L^{abcd}$ and $\K^{abcd}$ respectively. The main objective of this section is to provide explicit expressions for these operators. 

Since $\Ncalzero^{ab}$ and $\Ncalone^{ab}$ Poisson commute on  $\Gamma_{\kin, 1}$,    the calculation only involves PBs between ``soft" variables.\footnote{Additionally, since PBs involving soft variables on  $\Gamma_{\kin, 1}$ remain uncorrected on $\Gamma_{\phys}$ (see section \ref{2ndconstraintssec}), one has   $\{F_3,F_5\}_1=\{\Ncalone,\Ncalzero\}_2$ and $\{F_3,F_3\}_1=\{\Ncalone,\Ncalone\}_2$. From this perspective, the computations \eqref{LabcdXcd} and \eqref{F3F31} can  be interpreted as the evaluation of the physical (sub)leading soft news brackets. See also the discussion around Eq. \eqref{F3F5solved}.}  
For \eqref{defLabcd}, the relevant PB is
\ba
\L^{abcd} X_{cd} &=& \int_{S^2}  \{C N^{\langle ab\rangle}  -\frac{1}{2}\Pi^{\langle ab\rangle } , - N^{\langle cd \rangle} \}_1 X_{ cd }  \nonumber \\
&=& \int_{S^2}\left( - N^{\langle ab\rangle} \{C  ,  N^{\langle cd \rangle} \}_1+  \frac{1}{2} \{\Pi^{\langle ab\rangle } ,  N^{\langle cd \rangle} \}_1 \right) X_{ cd } , \label{LabcdXcd}
\ea
where we used that $N^{ab}$  commutes with itself. In the next subsection we evaluate \eqref{LabcdXcd} in holomorphic coordinates, and show the operator enjoys a holomorphic/antiholomorphic splitting, 
\be \label{Lzbzbcd}
\L^{\zb \zb cd} X_{cd} =  \L X^{\zb\zb}   
\ee
where $\L$ is a complex (integro-)differential operator with coefficients that are linear in $N^{ab}$.

We now turn to the operator  in \eqref{defKabcd}. Using the fact that $C$ Poisson commutes with itself and with $\Pi^{ab}$, the PB between $F_3$ and itself can be expanded as:
\be \label{F3F31}
  \{F^{ ab}_3  , F^{cd}_3 \}_1 = C \{N^{ab},  F^{cd}_3 \}_1+   \{F^{ ab}_3  , N^{cd} \}_1 C + \frac{1}{4}   \{\Pi^{ ab} , \Pi^{cd} \}_1 .
\ee
The first two terms can then be written in terms of $\Lt$ and $L$, while the last term can be read off from \eqref{hvfPiabone}. One then obtains
\be
\K^{abcd}  = C \tilde{\L}^{abcd} - \L^{abcd}  C - \frac{1}{4} \Db^{\langle a}  \Gb^{b\rangle}_{\phantom{b}mn} \A^{mn pq}   \Gbt_{pq}^{\phantom{ab} \langle c}\Db^{d\rangle}.
\ee
The holomorphic/antiholomorphic splitting property satisfied by $\L^{abcd}$, $\Gb^{a \, bc}$ and $\A^{ab cd}$ ensures $\K^{abcd}$ also satisfies it: 
\be
\K^{\zb \zb cd} X_{cd} =  \K X^{\zb\zb}   
\ee
with
\be
\K= C \L^\dagger - \L  C +\frac{1}{4} q_{z \zb}^2 \Db_z \Db_{\zb}^{-3} \Ab \Db_{z}^{-3} \Db_{\zb},
\ee
where $\L^\dagger \equiv \tilde{\bar{\L}}$ is the adjoint of $\L$.

\subsection{Evaluation of $\L$}
We start by writing \eqref{LabcdXcd} in holomorphic coordinates:
\begin{multline}
 \L^{\zb \zb cd} X_{cd} = \int d^2 w \left( - N^{\zb \zb} \{C  ,  N^{\wb \wb} \}_1+  \frac{1}{2} \{\Pi^{\zb \zb} ,  N^{\wb \wb} \}_1 \right) X_{ \wb \wb}  \\
 + \int d^2 w \left( - N^{\zb \zb} \{C  ,  N^{ww} \}_1+  \frac{1}{2} \{\Pi^{\zb \zb} ,  N^{ww} \}_1 \right) X_{ ww } . \label{LabcdXcdholo}
\end{multline}
The PBs in \eqref{LabcdXcdholo} can be obtained from the HVF of $N^{zz}$ discussed  at the end of section \ref{Nzzsec} and read
\ba
\int d^2 w \ \{C  ,  N^{\wb \wb} \}_1 X_{ \wb \wb} &=&   -2 \Gzero \Db^2_z X^{zz} \label{CNwbwb}  \\  
\int d^2 w \ \{\Pi^{\zb \zb} ,  N^{\wb \wb} \}_1  X_{ \wb \wb} &=&  -4 N^{\zb\zb}  \Gzero \Db^2_z X^{zz} \label{PizzNzbzb} \\
\int d^2 w \ \{\Pi^{\zb \zb} ,  N^{ww} \}_1 X_{ww} &=&  -8  \Db^{\zb} \Db^{-3}_{\zb} \Bb(\Gzero \Db^2_{\zb} X^{\zb\zb}) N_{\zb\zb} \label{PizbzbNww} .
\ea
Using these expressions one finds the first line of \eqref{LabcdXcdholo} vanishes, leading to the form \eqref{Lzbzbcd}. The second line gives the operator 
\be \label{explicitL}
\L X^{\zb\zb}=  2N^{\zb \zb} \Gzero \Db^2_{\zb} X^{\zb\zb}  - 4 \Db^{\zb} \Db^{-3}_{\zb} \Bb(\Gzero \Db^2_{\zb} X^{\zb\zb}) N_{\zb\zb} 
\ee
\\

We finally show how \eqref{explicitL} may be simplified  by using identities satisfied by $\Bb$. We phrase the following discussion in terms of $\bar \L$ rather than $\L$, and reintroduce the notation given in \eqref{defghol},
\be 
\begin{aligned}
g &= -\frac{1}{4}  \Db^{-2}_{\zb} X_{\zb\zb} \\
g^{zz} & =   \frac{1}{2}X^{zz} \\
g^{\zb\zb} & =   \frac{1}{2}  \Db^{-2}_{\zb} \Db^2_{z} X^{zz}.
\end{aligned}
\ee
From the defining equation \eqref{LabcdXcd}
\be
\bar \L X^{zz} = \int d^2 w \left( - N^{zz} \{C  ,  N^{\wb\wb} \}_1+  \frac{1}{2} \{\Pi^{zz} ,  N^{\wb\wb} \}_1 \right) X_{ \wb\wb } ,
\ee
and Eq.  \eqref{HVF1XzzNzz}, one gets
\be \label{barLXzz}
\bar \L X^{zz} = -g N^{zz} + 2  \Db^{z} \Db^{-3}_{z} \Bb(g) N_{zz},
\ee
which is nothing but the complex conjugate of \eqref{explicitL}. On the other hand, if we use the ``unsimplified" form of the $\Pi^{zz} $ component given in \eqref{unsimplifiedPzzNzz} one gets
\be \label{alternativeL}
\bar \L X^{zz} = -\frac{1}{2}(g N^{zz}+N g^{zz}) +q_{z \zb} \Db_{\zb}   \Db^{-3}_{z} \Bb^{\zb \zb}(N,g).
\ee
In this form, the operator manifest a symmetry under the exchange of $N$ with $g$. In particular, it follows that we can also write it as in \eqref{barLXzz} with $g$ and $N$ interchanged:
\ba
\bar \L X^{zz}  &=& -N g^{zz} + 2  \Db^{z} \Db^{-3}_{z} \Bb(N) g_{zz} \\
&=& \frac{1}{2}\left(-N     + 2 q_{z \zb}\Db_{\zb} \Db^{-3}_{z} \Bb(N)  \Db^{-2}_{\zb} \Db^2_{z} \right) X^{zz}. \label{Lbarsimplified}
\ea
The advantage of \eqref{Lbarsimplified} over \eqref{explicitL} is that it allows to factor out $X^{zz}$,  thus giving an explicit operator expression for $\bar \L$.

\section{Physical HVFs} \label{HVF2sec}

In this appendix we discuss the construction of HVFs on $\Gamma_{\phys}=\Gamma_{\kin, 2}$. As we did in section \ref{HVF1sec}, the idea is to apply Dirac formula \eqref{correctedHVFabstract}, this time for $i=1$.  We shall work in general 2d coordinates and parametrize the phase space by
\be \label{indepdirkin2}
\Gamma_{\kin, 2} =   \{  \sigma_{ab}, C, q_{ab} \}.
\ee

From the point of view of  $\Gamma_{\kin, 2} \subset \Gamma_{\kin, 1}$, we are  using  $F_3$ to  determine $\Pi^{ab}$, and the electric part of  $F_5$ to  determine  $N$. The magnetic part of $F_5$, together $F_4$ and $F_6$,  are  constraints on the allowed set of $\sigma_{ab}$'s. Notice there is a subtle difference between the variables used as phase space coordinates in \eqref{indepdirkin2},  and those chosen in subsection \ref{elemPBalgebrasec} to build the elementary PB algebra:  The shear is more convenient than the news when it comes to HVFs expressions.  Conversely, the news is preferable to the shear when considering the PB algebra. \\

Writing the smeared constraint as
\be \label{smeared2ndF}
F[X]=\sum_{i=3}^6 \int_{S^2} F^{ab}_i X_{i \, ab},
\ee
 the corresponding HVF,  along the independent directions in \eqref{indepdirkin2}, can be written as
\begin{multline} \label{hvf2FX}
 \{ \cdot , F[X] \}_1 =  \int_{S^2}  \left(  \delta_X q_{ab} \frac{\delta}{\delta q_{ab}} + \delta_X C \frac{\delta}{\delta C} + \int du \, \delta_X  \sigma_{ a b} \frac{\delta}{\delta  \sigma_{ab}}  \right) \\
  + \{ \cdot, F_6[X_6]\}_1 +\Lambda \{ \cdot, F_6[ X_3]\}_1
 \end{multline}
where 
\ba
\delta_X q_{ab} & = & - \Gbt_{ab}^{\phantom{ab} c}\Db^d X_{3 \langle c d \rangle}  \\ 
\delta_X C & = & 2 \Gzero \Db^{a} \Db^{b }(X_{5 \langle ab \rangle} - C X_{3 \langle ab \rangle})\\
\delta_X \sigma_{ab}(u) & = & \frac{1}{2} u X_{3 \langle ab \rangle} + \delta_+(u) X_{4 \langle ab \rangle}+ X_{5 \langle ab \rangle} + \frac{1}{2}q_{ab} \sigma^{cd}(u)  \delta_X q_{cd} \label{delFX2ndsigma} ,
\ea
and where we left  apart the contributions from  $F_6$. These  come from two places: The explicit one from the $i=6$ term in \eqref{smeared2ndF}, and  one from the $i=3$ term in \eqref{smeared2ndF} that arises  upon using Eq. \eqref{correctedHVFNcalone}  (which also holds  on $\Gamma_{\kin, 1}$). We keep these terms implicit since $F_6$ does not admit a  HVF in the sense of Eq. \eqref{defHVF1}. This will  help  clarify the role of the prescriptions discussed in section \ref{utoinfsec}.

Given Eq. \eqref{hvf2FX} and the inverse Dirac map \eqref{2ndstageInv_DM}, we  now use the general formula \eqref{correctedHVFabstract} to construct the corrected  HVFs on $\Gamma_{\kin, 2}$.  

Let us first consider the smeared 2d metric  as defined in \eqref{smearedkin1}. Its only  non-trivial PB with the constraints is
\be
\{ q[\rho],F_3^{\langle ab \rangle} \}_1= \Db^{\langle a} \Gb^{b \rangle}_{\phantom{b} cd} \rho^{cd}.
\ee
From the inverse Dirac map \eqref{2ndstageInv_DM}  one then finds that the  only non-zero $X^\alpha$ (as defined in \eqref{defXphi}) is
\be \label{X4qrho}
X_{4 ab} =- \Db_{\langle a} \Gb_{b \rangle\, cd} \rho^{cd}.
\ee
Since $\{ \cdot, q[\rho] \}_1$ does not have components along the variables in \eqref{indepdirkin2}, the HVF is just given by substituting \eqref{X4qrho} in  \eqref{hvf2FX}, leading to
\be
\label{HVF2qrho}
\{ \cdot, q[\rho] \}_2 = -\int_{S^2}  \Db_{\langle a} \Gb_{b \rangle\, cd} \rho^{cd} \frac{\delta}{\delta \sigma^+_{ab}} ,
\ee
where  be used the notation
\be \label{defpartialsigmaplus}
 \frac{\delta}{\delta \sigma^+_{a b}}:= \lim_{\Lambda\to + \infty}    \frac{\delta}{\delta \sigma_{ a b}(\Lambda)}.
\ee

We  next consider $C$. This time we have two non-trivial PBs with the constraints:
\be
\{ C(x_0),F_3^{\langle ab \rangle} \}_1= -2 C  \Db^{\langle a} \Db^{b \rangle } \Gzero_{x_0}
\ee
\be
\{ C(x_0),F_5^{\langle ab \rangle} \}_1= 2  \Db^{\langle a} \Db^{b \rangle } \Gzero_{x_0}.
\ee
The  non-trivial $X^\alpha$'s \eqref{defXphi} are then
\be \label{XC2}
\begin{aligned} 
X_{4 ab} & = 2 C  \Db_{\langle a} \Db_{b \rangle } \Gzero_{x_0} \\
X_{6 ab} & = - 2   \Db_{\langle a} \Db_{b \rangle } \Gzero_{x_0}.
\end{aligned}
\ee
Since there are no contributions from $\{ \cdot,C(x_0) \}_1$ along \eqref{indepdirkin2}, the HVF is given by substituting \eqref{XC2} in  \eqref{hvf2FX}:
\be
\{ \cdot,C(x_0) \}_2  =  2 \int_{S^2}  \Db_{\langle a} \Db_{b \rangle } \Gzero_{x_0} C   \frac{\delta}{\delta \sigma_{a b}^+}  -  \{ \cdot, F_6[X_6]\}_1|_{X_{6 ab}=  2\Db_{\langle a} \Db_{b \rangle } \Gzero_{x_0} } . \label{HVF2C}
\ee

We finally discuss the  news tensor, for which we  consider  three type of smearings: First $\dot \sigma[\gamma]$ as in  $\eqref{smearedkin1}$, where  $\gamma_{ab} \stackrel{|u| \to \infty}{\to}$, and then smearings appropriate for the  leading and subelading soft news. In the first case, the only  non-trivial PB with the constraints is:
\be
\{\dot \sigma[\gamma], F_3^{\langle ab \rangle} \}_1 =  -\frac{1}{2} \int du \left( u \dot \gamma^{\langle ab \rangle}  + \Db^{\langle a } \Gb^{b \rangle}_{\phantom{b} cd}(q^{mn}\gamma_{mn} \dot \sigma^{cd})\right) ,
\ee
leading to 
\be 
X_{4 ab}  = \frac{1}{2} \int du u \dot \gamma_{\langle ab \rangle}. 
\ee
Substituting in \eqref{hvf2FX} and adding the contribution from $\{ \cdot,\dot \sigma[\gamma] \}_1$ one gets
\ba 
\{ \cdot,\dot \sigma[\gamma] \}_2 & =&  \frac{1}{2}  \int_{\I}  \left( \gamma_{\langle ab \rangle} \frac{\delta}{\delta \sigma_{a b }}  +\big( u \dot \gamma_{\langle ab \rangle} - \Db_{\langle a} \Gb_{b \rangle\, cd} (q^{mn} \gamma_{mn} \dot \sigma^{cd} )  \big) \frac{\delta}{\delta \sigma^+_{ab}}   \right)   \label{HVF2dotsigma} .
\ea
For the leading soft news we  consider the  smeared version,
\be
\Ncalzero[\chi] = \int_{S^2} \Ncalzero^{ab}\chi_{ab}.
\ee
Using the prescriptions given in section \ref{utoinfsec}, its non-trivial PB with the constraints are found to be
\be
\begin{aligned} 
\{\Ncalzero[\chi], F_3^{\langle ab \rangle} \}_1 &=  -\frac{1}{2}  \Db^{\langle a } \Gb^{b \rangle}_{\phantom{b} cd}(q^{mn}\chi_{mn} \Ncalzero^{cd}) \\
\{\Ncalzero[\chi],  F_6^{\langle ab \rangle} \}_1 & = -\chi^{\langle ab \rangle},
\end{aligned}
\ee
leading to
\be
\begin{aligned} 
X_{4 ab} & = - \L_{ab }^{\phantom{ab} cd} \chi_{cd}, \\
X_{5 ab} & =-\chi_{\langle ab \rangle}.
\end{aligned}
\ee
Substituting in \eqref{hvf2FX} and adding the contribution from $\{ \cdot,\Ncalzero[\chi]  \}_1$ one gets
\be
\{ \cdot,\Ncalzero[\chi] \}_2  =   \int_{S^2} \left( -2 \Gzero \Db^{\langle a} \Db^{b \rangle } \chi_{ab}  \frac{\delta}{\delta C} - \Big( L \chi_{\langle a b \rangle} + \frac{1}{2}\Db_{\langle a} \Gb_{b \rangle\, cd} (q^{mn} \chi_{mn} \Ncalzero^{cd} ) \Big) \frac{\delta}{\delta \sigma^+_{ab}}   \right),  \label{HVF2Ncalzero}
\ee
where we used the notation
\be \label{Lshortnotation}
L \chi_{\langle ab \rangle} \equiv \L_{ab }^{\phantom{ab} cd} \chi_{cd}  .
\ee

Finally, for the subleading soft news we consider
\be
\Ncalone[\chi] = \int_{S^2} \Ncalone^{ab}\chi_{ab}.
\ee
Using again the prescriptions of section \ref{utoinfsec}, its non-trivial PBs with the constraints are
\be
\begin{aligned} 
\{\Ncalone[\chi], F_3^{\langle ab \rangle} \}_1 &=  -\frac{1}{2}\Db^{\langle a } \Gb^{b \rangle}_{\phantom{b} cd}(q^{mn}\chi_{mn} \Ncalone^{cd}) \\
\{\Ncalone[\chi], F_4^{\langle ab \rangle} \}_1 & =  - \chi^{\langle ab \rangle}  
\end{aligned}
\ee
leading to
\be \label{Xsmearednews2}
\begin{aligned} 
X_{3 ab} & = - \chi_{\langle ab \rangle} \\
X_{4 ab} & = - K \chi_{\langle ab \rangle}  \\
X_{6 ab} & =  \Lt \chi_{\langle ab \rangle} ,
\end{aligned}
\ee
where we used the notation \eqref{Lshortnotation} for   $K$ and $\Lt$. Substituting in \eqref{hvf2FX} and adding the contribution from $\{ \cdot,\Ncalone[\chi]  \}_1$
we get\footnote{The term proportional to $\Lambda$ coming from $\{ \cdot,\Ncalone[\chi]  \}_1$ cancels with the one coming from the last term in \eqref{hvf2FX}.}
\be
\{ \cdot,\Ncalone[\chi] \}_2  =   \int_{\I} \delone_{\chi} \sigma_{ab} \frac{\delta}{\delta \sigma_{a b }} + \int_{S^2} \left( \delone_\chi q_{ab} \frac{\delta}{\delta q_{ab}} + \delone_\chi C \frac{\delta}{\delta C} \right) + \{ \cdot, F_6[X_6]\}_1|_{X_{6 ab}=\Lt \chi_{\langle ab \rangle}},  \label{HVF2Ncalone}
\ee
where
\ba
\delone_\chi q_{ab}  &= & \Gbt_{ab}^{\phantom{ab} \langle c}\Db^{d \rangle} \chi_{cd} \\
\delone_\chi C &= &2 \Gzero \Db^{\langle a} \Db^{b \rangle }( C \chi_{ab}) \\
\delone_\chi \sigma_{ab}(u) &=&  - \delta_+(u) \big( K \chi_{\langle ab \rangle} + \frac{1}{2}  \Db_{\langle a} \Gb_{b \rangle\, cd} (q^{mn} \chi_{mn} \Ncalone^{cd} )\big)    +\frac{1}{2}q_{ab} \sigma^{cd}(u)  \delone_\chi q_{cd}.
\ea

\noindent{\emph{Comments}}

\begin{itemize}
\item The HVF of both $C$ and $\Ncalone$ exhibit the a priori undetermined  term $\{ \cdot, F_6\}_1$ . This is not a problem for evaluating the PB algebra, if one uses the prescriptions given in section \ref{utoinfsec}, prior to Eq. \eqref{prescriptionnewsF6}, and treats $\dot \sigma[\gamma]$,  $\Ncalzero[\chi]$ and $\Ncalone[\chi]$ separately, as we did above. A unified treatment of these three quantities,  however, requires the identification of distributional terms at infinity in the unintegrated bracket $\{ \dot \sigma, F_6\}_1$. The prescription given in Eq. \eqref{prescriptionnewsF6} implies
\be
\int du \{ C, \dot \sigma^{\langle a b \rangle}(u) \}_2 =  \{ C, \Ncalzero^{\langle a b \rangle} \}_2 ,
\ee
as can be checked from Eqs. \eqref{HVF2C} and \eqref{HVF2Ncalzero}. This is how  the analogue of the kinematical discontinuity between \eqref{shearsoftnewsPB} and \eqref{shearnewsPB} is resolved.

\item The HVFs share some of the structural aspects of their kinematical counterparts discussed in appendices \ref{kinHVFsapp} and \ref{HVF1sec}. In particular, potential ambiguities associated to Green's function are associated to null directions of the symplectic structure, see the next appendix.

\end{itemize}

\section{Null directions of the symplectic form} \label{nullappendix}
The  variables  $N$ and $C$ introduced in Eq. \eqref{defCNscalars} are defined  modulo the kernel of $ \Db_{\langle a} \Db_{b \rangle}$ (itself identical to the kernel of $\D$). Denoting by $\tb$ a general element of this kernel, we then have an intrinsic redundancy in our description, in which 
\be\label{shiftCN}
C \sim C+ \tb, \quad    N \sim N+ \tb, \quad \quad  \tb \in \ker  \Db_{\langle a} \Db_{b \rangle} \equiv \ker \D,
\ee
where the shifts may also include  changes in other variables, see below. 

In this appendix we show that, for all the three  spaces involved in our discussion, $\Gamma_{\kin, i}, i=0,1,2$, the shifts \eqref{shiftCN} span the null directions $\delta^i_{\text{null}}$ of the corresponding symplectic structures, 
\be \label{defnulldirection}
\Omega_{\kin, i}(\delta, \delta^i_{\text{null}}) =0 \quad \forall \ \delta \in \Gamma_{\kin, i} .
\ee
Thus, the phase spaces should actually be understood as being defined modulo such null directions.

One way to  construct the  null shifts is as follows. Consider on $\Gamma_{\kin, i}$ the HVFs associated to $C$ and $N$,
\be \label{HVFsCNi}
 \{ \cdot, C(x_0)\}_i , \quad \{ \cdot, N(x_0)\}_i.
\ee
They all involve the Green's function $\Gzero$ of $\D$. We then consider
the formal replacement 
\be
\Gzero_{x_0}(x) \to \tb(x)
\ee
in  \eqref{HVFsCNi}. This gives, for every $\tb \in \ker \D$, a pair of two null directions that we denote by $\delta^{N}_{\tb,i}$ and $\delta^{C}_{\tb,i}$: 
\be \label{defnullCNi}
\begin{aligned}
\delta^{N}_{\tb,i} &:= - \{ \cdot, C\}_i|_{\Gzero \to \tb} \\ 
\delta^{C}_{\tb,i} &:=  \{ \cdot, N\}_i|_{\Gzero \to \tb} ,
\end{aligned}
\ee
where the  sign choices correspond to those in \eqref{shiftCN}.

For $i=0$, Eq. \eqref{defnullCNi} leads to
\be \label{nullCN0}
\begin{aligned}
\delta^{N}_{\tb,0} &=  \int_{S^2}  \tb \frac{\delta}{\delta N} ,\\ 
\delta^{C}_{\tb,0} &=  \int_{S^2} \left( \tb \frac{\delta}{\delta C} + \tb N^{ab} \frac{\delta}{\delta \Pi^{ab}}  - 2 \Delta^{ab}_{\phantom{ab} cd}(\tb) N^{cd}\frac{\delta}{\delta p^{ab}} \right) .
\end{aligned}
\ee
One can explicitly check these satisfy the null condition \eqref{defnulldirection}. Notice the shift in $C$ comes with additional shifts in $p_{ab}$ and $\Pi_{ab}$. Such terms appear due the fact that  $\ker \D$ depends on  $q_{ab}$ and $T_{ab}$ respectively. We also notice that terms involving  $\Gzero^{ab}_{x_0} \equiv -2\Db^{\langle a} \Db^{b \rangle} \Gzero$ in $ \{ \cdot, N\}_0$ do not contribute since $\Db_{\langle a} \Db_{b \rangle} \tb=0$.

For  $i=1$, one finds 
\be \label{nullCN1}
\begin{aligned}
\delta^{N}_{\tb,1} &=  \int_{S^2}  \tb \frac{\delta}{\delta N} ,\\
\delta^{C}_{\tb,1} & =  \int_{S^2} \left( \tb \frac{\delta}{\delta C}  +  \left( \tb N^{ab}+  2 \Db^{a}  \Gb^{b }_{\phantom{b}cd} \Bb^{cd}_{\phantom{cd}mn}(\tb) N^{mn}  \right)    \frac{\delta}{\delta \Pi^{\langle ab \rangle}} \right).
\end{aligned}
\ee
where we have only displayed the  independent directions on $\Gamma_{\kin, 1}$, \eqref{parametrizationGamma1}. The absence/presence of corrections to the $i=0$ expressions can be understood from the fact that $\delta^{N}_{\tb,0}F_1=\delta^{N}_{\tb,0}F_2=0$  while $\delta^{C}_{\tb,0} F_{1} \neq 0$.  One can again verify  that \eqref{nullCN1} are null directions: For $\delta^{N}_{\tb,1} $ this is direct while for $\delta^{C}_{\tb,1}$ it involves a non-trivial cancellation of terms.

Notice that, so far,  the null directions only acted on soft-sector variables:
\be \label{null01equalsoftnull}
\Omega_{\kin, i}(\delta, \delta^{N/C}_{\tb,i}) = \Omega_{\soft, i}(\delta, \delta^{N/C}_{\tb,i}) = 0, \quad i=0,1.
\ee
This situation  changes for $i=2$. Using the parametrization \eqref{indepdirkin2} one finds
\be \label{nullCN2}
\begin{aligned}
\delta^{N}_{\tb,2} &=  0 \\ 
\delta^{C}_{\tb,2} &=  \int_{S^2} \left( \tb \frac{\delta}{\delta C} + \left(\tfrac{1}{2}\tb \Ncalzero^{ab} - \Db^{a}  \Gb^{b }_{\phantom{b}cd} \Bb^{cd}_{mn}(\tb) \Ncalzero^{mn} \right)\frac{\delta}{\delta \sigma^+_{\langle a b\rangle}}\right) . 
\end{aligned}
\ee
The fact that $\delta^{N}_{\tb,2}$ vanishes can be  understood from the fact that $N$ is no longer an independent variable in $\Gamma_\phys=\Gamma_{\kin, 2}$. Rather, it is defined through the leading soft news by $N=-2\Gzero \Db^{\langle a} \Db^{ b\rangle} \Ncalzero_{ab}$. The parametrization \eqref{indepdirkin2}, however, treats $C$ as an independent variable, that comes with its corresponding null direction. To show that $\delta^{C}_{\tb,2}$ satisfies \eqref{defnulldirection}, we notice that its action on the soft variables is identical to that of $\delta^{C}_{\tb,1}$ (for the same reasons leading to \eqref{PBunaltered})
\be \label{delC2eqdelC1onsoft}
 \varphi \in (C,N, \Pi^{ab},q_{ab} ) \implies  \delta^{C}_{\tb,2} \varphi =\delta^{C}_{\tb,1} \varphi .
\ee
From \eqref{null01equalsoftnull} this in turn implies 
\be
\Omega_{\soft}(\delta, \delta^{C}_{\tb,2}) = 0.
\ee
Unlike the previous cases, however, there could be  contributions from the hard part of the symplectic structure. These  yield terms proportional to $\lim_{U \to \infty} \delta \dot \sigma_{ab}(U)$ which vanish on $\Gamma_\phys$ and thus
\be
\Omega_{\hard}(\delta, \delta^{C}_{\tb,2}) =0.
\ee

We finally argue that \eqref{nullCN0}, \eqref{nullCN1} and \eqref{nullCN2} span all null directions of   $\Gamma_{\kin, i}$ for $i=0,1,2$ respectively. For $i=0$, this follows from the ``block diagonal'' form of $\Omega_{\kin, 0}$.\footnote{Modulo crossed terms due to the trace-free condition which do not affect the argument.} The hard piece is known to be  non-degenerate \cite{AS}. The soft piece is a sum of three terms, with only the $\delta (\D N) \wedge \delta C$ piece  exhibiting degeneracies. These are the ones captured by  $\delta^{N}_{\tb,0}$ and $\delta^{C}_{\tb,0}$. Since the first and second stage constraints are second class, there cannot be additional null directions on  $\Gamma_{\kin, 1}$ and  $\Gamma_{\kin, 2}$.

\end{document}